%% file: PGC_RF_v210303b_arxiv.tex
\documentclass[12pt]{article}
\usepackage{lscape}
\usepackage{geometry}
\usepackage[onehalfspacing]{setspace}
\usepackage{amssymb}
\usepackage{amsfonts}
\usepackage{graphicx}
\usepackage{amsmath}
\usepackage{epstopdf}
\usepackage{natbib}
\usepackage{bibunits}
\usepackage{morefloats}
\usepackage{float}
\usepackage{comment}
\usepackage{pdflscape}
\usepackage{adjustbox}
\usepackage{subcaption}
\usepackage{longtable}
\usepackage{apalike}
\usepackage{xfrac}
\usepackage[dvipsnames,table]{xcolor}
\usepackage{qtree}
\usepackage{multicol}
 \usepackage{booktabs}
\usepackage{footnote}
\usepackage{algorithm,algorithmic}

\usepackage[super]{nth}

\usepackage{threeparttable}
\usepackage{tabularx}
\usepackage{dcolumn}
\newcolumntype{d}[1]{D..{#1}} 

\usepackage{mathpazo} 
\usepackage[scaled]{helvet} 
\usepackage{courier} 
\normalfont
\usepackage[T1]{fontenc}
\usepackage{listings}

\usepackage{siunitx}
\usepackage[pdftex,bookmarks,colorlinks]{hyperref}
\hypersetup{
  colorlinks,
  citecolor=PineGreen,
  linkcolor=red,
  urlcolor=blue}
\usepackage{cleveref}

\usepackage{tikz}
\usetikzlibrary{positioning}

\usepackage{threeparttable}
\usepackage{tabularx}

\begin{document}
\title{\vspace{-0.1cm} \Huge The Macroeconomy as a Random Forest}
\author{Philippe Goulet Coulombe\thanks{%
\noindent Department of Economics, \href{mailto:gouletc@sas.upenn.edu}{{gouletc@sas.upenn.edu}}. For many helpful discussions, I would like to thank Karun Adusumilli, Edvard Bakhitov, Francesco Corsello, Frank Diebold, Maximilian Göbel, Magnus Reif, Frank Schorfheide, Dalibor Stevanovic, David Wigglesworth and Boyuan Zhang. For excellent research assistance during different eras of this project, I am grateful to Tony Liu and JiWhan Moon. MRF is now available as an \href{https://philippegouletcoulombe.com/code}{\textbf{\texttt{R} package}}.}}
\date{\vspace{-0.4cm}
University of Pennsylvania\\[2ex]%
\small
First Draft: November 15, 2019 \\
This Draft: \today \\ 
\href{https://drive.google.com/file/d/14lGhUPa9tSRbSngTz4HfeBp2TXX6CzG-/view?usp=sharing}{Latest Draft Here} \\ 
\vspace{0.4cm}
\large
  }
\maketitle

\vspace{-0.5cm}

\begin{abstract}

\noindent I develop \textit{Macroeconomic Random Forest} (MRF), an algorithm adapting the canonical Machine Learning (ML) tool to flexibly model evolving parameters in a linear macro equation. Its main output, \textit{Generalized} Time-Varying Parameters (GTVPs), is a versatile device nesting many popular nonlinearities (threshold/switching, smooth transition, structural breaks/change) and allowing for sophisticated new ones. The approach delivers clear forecasting gains over numerous alternatives, predicts the 2008 drastic rise in unemployment, and performs well for inflation. Unlike most ML-based methods, MRF is directly interpretable --- via its GTVPs. For instance, the successful unemployment forecast is due to the influence of forward-looking variables (e.g., term spreads, housing starts) nearly doubling before every recession. Interestingly, the Phillips curve has indeed flattened, \textit{and} its might is highly cyclical.

\end{abstract}







\thispagestyle{empty}

\clearpage 
\setcounter{page}{1}

\section{Introduction}

The rise of Machine Learning (ML) led to great excitement in the econometrics community. In applied macroeconomics, a first wave of papers took ML algorithms off the shelf and went hunting for forecasting gains. With the emerging consensus that some ML offerings can appreciably increase predictive accuracy, a question emerges: what is the place of economics in all that?

The conditional mean is the most basic input to any empirical macroeconomic analysis. Anything else that follows (e.g., structural analysis) depends on it. Thus, getting it right is not merely useful, it is \textit{necessary}. Clearly, in that regard, ML can help. However, while the latter gladly delivers prediction accuracy gains (and ergo a conditional mean closer to the truth), it is much more reluctant to disclose its inherent model. Consequently, ML is currently of great use to macroeconomic forecasting, but of little help to macroeconomics. I propose a simple remedy: shifting the focus of the algorithmic arsenal away from predicting $y_t$ into modeling $\beta_t$, which are economically meaningful coefficients in a time-varying macroeconomic equation. The newly proposed algorithm, \textit{Macroeconomic Random Forest} (MRF) kills two coveted birds with one stone. First, in most instances, MRF forecasts better than off-the-shelf ML algorithms and traditional econometric approaches. Second, its main output, \textit{Generalized Time-Varying Parameters} (GTVPs), can be interpreted. Their versatility comes from nesting many popular specifications (structural breaks/change, threshold effects, regime-switching, etc.) and letting the data decide whichever combination of them is most suitable.  Ultimately, we get a new methodology leveraging the power of ML and big data to provide a modern take on the decades-old challenge of estimating latent states driving linear macroeconomic equations. 

\vskip 0.15cm

{\sc \noindent \textbf{The State of Empirical Macro Affairs}.}  Answering positively two questions guarantees a viable conditional mean: "are all the relevant variables included in the model?" and at a higher level of sophistication, "is linearity a valid approximation of reality?". The first one led to the successful development of factor models and large Bayesian Vector Autoregressions (VARs) over the last two decades. To address the second, applied macroeconomic researchers have proposed many non-linear time series models based on reasonable economic intuition. Most of them amount to have regression coefficients $\beta_t$ in 
$$y_t = X_t \beta_t +\varepsilon_t$$
evolving through time. The $\beta_t$ process can take many forms, and a choice must be made \textit{a priori} out of many equally plausible alternatives. Notable members of the vast time-variations catalog are threshold/switching regressions \citep{hansen2011TAR}, smooth transition \citep{terasvirta1994}, structural breaks \citep{perron2006,stock1994unit}, and random walk time-varying parameters \citep{sims1993,cogley2001evolving,primiceri2005}. While it is uncontroversial that factor models and large Bayesian VARs have gone a long way in meeting their original goals, less victorious statements are available for the various time-variation proposals. Why?





More often than not, nonlinear time series models use little data and/or restrict stringently the shape of $\beta_t$'s path.  While the consequences for forecasting are direct and obvious, those for analysis of macroeconomic relationships are equally problematic. Is the evolving Taylor rule characterized by switching regimes \citep{simszha2006}, a Volker structural break \citep{clarida2000}, or gradually evolving parameters \citep{boivin2005has,primiceri2005}?  This discordance interferes with our understanding of the past while impacting our expectations for tomorrow's $\beta_t$. I now divide popular time-variation approaches into two strands, discuss their shortcomings, and complete by explaining how MRF addresses them.




\vskip 0.15cm
{\sc \noindent \textbf{Observable Time-Variation Via Interaction Terms}.} Using interaction terms and related refinements is a parsimonious way to create time variation in a linear equation. For instance, switching regimes based on an observed regressor can be obtained by interacting the linear equation with the indicator function $I(q_t>c)$, where $c$ is some value, and $q_t$ is a threshold variable chosen by the researcher. However, using the FRED-QD US macro data set \citep{mccrackenng} reveals an overwhelmingly large number of candidates for $q_t$.  Additionally, there may be multiple regimes interacting together. Or the "true" $q_t$ could be an unknown function of available regressors. And structural breaks or slow exogenous variation could get in the way. The list goes on. This renders a credible exploration of the threshold structures' space impossible and the enterprise of manually specifying the model very much compromised.

Here is an empirical example. \cite{auerbach2012measuring} and \cite{rameyzubairy2018} use a GDP/unemployment indicator to let the effects of fiscal stimulus (potentially) vary with the state of the economy. \cite{batini2012successful} allow for additional dependence on the origin of the impulse (revenue or spending). Such honorable explorations could go on endlessly. MRF provides a hammer solution to the problem. First, the near-universe of threshold structures can be characterized by regression trees  --- see section \ref{macroastrees}. Second, MRF embeds, among other things, a powerful greedy algorithm designed to explore such "structure" spaces. 


   
\vskip 0.15cm 

{\sc \noindent \textbf{Latent Time-Variation}.} Some methods with an aura of greater flexibility are labeled as "latent change". In this line of work, $\beta_t$ either follows a law of motion (random walk, Markov process) or could be subject to discrete breaks.\footnote{Simpler derivatives are often used in applied work. In forecasting, rolling-window estimation drops early observations. In empirical macro, pre-defined subsamples are popular \citep{clarida2000,del2020s}.} At first glance, this appears to solve many of the problems of interaction terms approaches. By treating $\beta_t$ as a state to be filtered/estimated within the model, the complexity of characterizing its path correctly out of abundant data seems to vanish. Alas, estimating $\beta_t$'s path implies a great number of parameters \citetext{in fact, often greater than the number of observations, \citealt{GC2019}} which inevitably necessitates strong regularization. That regularization is the law of motion itself, a choice far from innocuous -- and akin to that of $q_t$ in "observable" change models. Accordingly, whether it is latent regime-switching, exogenous breaks, or slow change, none can easily accommodate for the additional presence of the other. Yet, these models are routinely fitted \textit{separately} on the \textit{same data}. Consequently, methods often detect what they are designed to detect, in near-complete abstraction of imaginable interference from other nonlinearities.
 


Additionally, while "latent" approaches may sometimes rationalize the data well in-sample, many of them will struggle to outperform a simple benchmark \textit{out-of-sample}. Often, the very nature of $\beta_t$'s law of motion creates forecasting headaches. Classical TVPs imply a two-sided vs one-sided filtering problem. Analogously, detecting a structural break is much harder without a great amount of data on both sides of it. Moreover, there is the obvious problem of statistical efficiency. If the Phillips curve flattened because an economy became increasingly open, including an interaction term with imports/exports is wildly more efficient than obtaining the whole $\beta_t$ path non-parametrically. Thus, exogenous structural change should be, in some sense, a time variation of last resort. The advantage of MRF is that it algorithmically search for "observable" low-hanging fruits, and turn to split the sample with $t$ only if necessary. Further, it implicitly creates a forecasting function for $\beta_t$ which is an RF in its own right. This is, almost in any case, much more powerful than existing alternatives -- like random walks. 

\vskip 0.15cm

{\sc \noindent \textbf{Mechanics}.} 
The key difference when adding the M to MRF is the inclusion of a linear part within each of the tree leaves, rather than just an intercept. Motivated in cross-sectional applications to improve the efficiency of nonparametric estimation (in the spirit of local linear regression), trees with linear parts have been considered (among others) in \cite{treed} and \cite{modeltrees}. \cite{friedberg2019llf} expand on this by considering an ensemble of them (i.e., a forest) and focusing on the problem of treatment effect heterogeneity. Of course, the difference here is that a linear part is much more meaningful when one can look at $\beta_t$ as a process of its own -- and as a synthesis of nonlinear time series models. Finally, it is noteworthy that the approach may come in semiparametric partially linear clothing, yet it makes no compromise on the range of nonlinearities it captures. This is a virtue of time-varying coefficients models being able to approximate any nonlinear function \citep{granger2008}.


The paper also introduces new devices enhancing MRF's predictive and interpretability potential. First, I propose Moving Average Factors (MAFs) as a simple way to compress ex-ante the information contained in the lags of a regressor entering the RF part of MRF. They boost the meaningfulness of tree splits and helps avoid running out of them quickly. The transformation is motivated by the literature on constraining/regularizing lag polynomials \citep{shiller1973}. Precisely, MAFs' contribution is to induce similar shrinkage when there are no explicit coefficients to shrink. When it comes to GTVPs themselves, I provide a regularization scheme better suited for time series which procures a desirably smoother path with respect to time. It is inspired by the random walk shrinkage of the classical TVP literature and is implemented within the tree procedure by weighted least-squares. Finally, a variant of the Bayesian Bootstrap provides credible regions that are instrumental for the interpretation of GTVPs.

\vskip 0.15cm
{\sc \noindent \textbf{Results}.} In simulations, the tool does comparably well to traditional nonlinear time series models when the data generating process (DGP) matches what the latter is designed for. When the time-variation structure becomes out of reach for classical approaches, MRF wins. Additionally, it supplants plain RF whenever persistence is pervasive. In a forecasting application, the MRFs gains are present for almost all variables and horizons under study, a rarity for nonlinear forecasting approaches. For instance, the Autoregressive Random Forest (ARRF) almost always supplant its resilient OLS counterpart. Also, an MRF where the linear part is a compact factor-augmented autoregression generates very accurate forecasts of the 2008 downturn for both GDP and the unemployment rate (UR). Inspection of resulting GTVPs reveals they behave differently from random walk TVPs. For instance, in the UR equation, the contribution of forward-looking variables nearly doubles before every recession --- including 2008 where the associated $\beta_t$ is forecasted  to do so out-of-sample. This reinforces the view that financial indicators and other market-based expectations proxies can rapidly capture downside risks around business cycle turning points \citep{adrian2019}. MRF learned and applied it to great success.

Inflation is subject to a variety of time-variations, detection of which would be compromised by approaches lacking the generality of MRF. The long-run mean and the persistence evolved slowly and in an exogenous fashion --- this has been repeatedly found in the literature \citetext{e.g., \citealt{cogley2001evolving}}. More novel is the finding that the real activity factor's effect on the price level depends positively on the strength of well-known leading indicators, especially housing-related. Following this lead, I complete the analysis by looking at a traditional Phillips' curve specification. I report that the inflation/unemployment trade-off coefficient decreased significantly since the 1980s \textit{and} also varies strongly along the business cycle. Among other things, it is extremely weak following every recession. This nuances current evidence on the flattening Phillips curve, which, by design, focused almost entirely on long-run exogenous change \citep{blanchard2015inflation,galigambetti2019, del2020s}. Overall, MRF suggests inflation can rise from a positive unemployment gap, but it goes down much more timidly from economic slack.  These findings are made possible by combining different tools within the new framework, such as credible intervals for the GTVPs, new variable importance measures specifically designed for MRF, and surrogate trees as interpretative devices for $\beta_t$.

\vskip 0.15cm

{\sc \noindent \textbf{Outline}.} Section \ref{sec:MRF} introduces MRF, motivates its use, considers practical aspects, and discusses relationships with available alternatives. Sections \ref{sec:sim} and  \ref{forecasting} report simulations and forecasting results, respectively. Section \ref{sec:anal} analyzes various GTVPs of interest. Section \ref{sec:con} concludes. 


\section{Macroeconomic Random Forests}\label{sec:MRF}

This section introduces MRF. I first motivate the use of trees as basis functions by casting  standard switching structures for autoregressions as special cases. Second, I detail the MRF mechanics and how it yields GTVPs. Third, I discuss how the approach relates to both standard RF and traditional random walk TVPs. Fourth, I discuss interpretability potential and provide a way to assess parameter uncertainty.

\subsection{Traditional Macro Non-Linearities as Trees}\label{macroastrees}
Within the modern ML canon, Random Forest (RF) is an extremely popular algorithm because it allows for complex nonlinearities, handles high-dimensional data, bypasses overfitting, and requires little to no tuning. This is in sharp contrast with, for example, Neural Networks, whose ability to fail upon a bad choice of hyperparameters is largely unmatched. Thus, RF is a reasonable device to look into for constructing GTVPs. But there is more: many common time series nonlinearities fit within a tree structure. Hence, it will be all the more natural to think of MRF as a generalization of previous nonlinear offerings.  Overall, it eliminates the arbitrary search for a specification. By creating a unified view, the myriad of time-variations suggested separately can now be tackled jointly.



I now present two examples displaying how common time series nonlinearities imply a tree structure for an AR process. Let us consider the inflation process in a country where inflation targeting (IT) was implemented at a publicly known date (like in Canada). Let $\pi_t$ be inflation at time $t$ and $t^*$ is the onset date of IT. Additionally, $g_t$ is some measure of output gap. A plausible model  is reported in the tree graph below. The story is straightforward. Inflation behaved differently before vs after IT. After IT, it is a simple AR process. Before IT, it was a switching AR process which dynamics and mean depended on the sign of the output gap.\footnote{Note that a standard regression tree would set all $\phi$'s to 0.}

\vspace{0.5cm}

\Tree[.{Full Sample} 
[.{$t<t^*$} 
[.{$g_{t-1}<0$} {$\pi_t = c_1 + \phi_1 \pi_{t-1}+\epsilon_t$} ]
[.{$g_{t-1} \geq 0$} {$\pi_t = c_2+ \phi_2 \pi_{t-1} +\epsilon_t$} ]
 ]
[.{$t \geq t^*$} {$\pi_t = c_3 + \phi_3 \pi_{t-1} +\epsilon_t$} ]
 ]
   
\vspace{0.5cm}

This is one story out of many that trees can characterize. In practice, none of the above is known. The structure, the splitting variables, and the splitting points could be different. This is both good and bad news. It highlights the flexibility of trees. It also suggests that designing the "true" one from economic deduction is a daunting task --- equally plausible alternatives are easily imaginable. Fortunately, algorithms can point out which trees in better agreement with the data.
 
A global grid search is computationally unfeasible if either $S_{t}$ is large or if we want to consider more than a few splits (examples above included 2 and 3, respectively). A natural way forward is recursive partitioning of the data set via a \textit{greedy} algorithm \citep{breiman1984classification}.\footnote{A single autoregressive tree was proposed in \cite{meek2002art}.} A greedy algorithm optimizes functions by iteratively doing the best local update, rather than directly solving for a global optimum. As a result, it is prone to high variance \citep{ESL}.  Hence, considering a diversified portfolio of trees appears as the most sensible route. To achieve that, it is highly effective to use Bootstrap Aggregation  \citetext{\textit{Bagging}, \citealt{breiman1996bagging}} of many de-correlated trees.  This is the famous Random Forest proposition of \cite{breiman2001}.

\subsection{Generalized Time-Varying Parameters}\label{sec:gtvp}
The general model is
\begin{align*}
y_t &= X_{t}\beta_t +\epsilon_t \\
\beta_t &= \mathcal{F}(S_t) 
\end{align*}
where $S_{t}$ are the state variables governing time variation and $\mathcal{F}$ a forest. $S_t$ is oberved macroeconomic data which composition is motivated in section \ref{MAF} and laid out explicitly in section \ref{forecasting}.  $X$ determines the \textit{linear} model that we want to be time-varying. Typically, ${X_t}\subset S_t$ is rather small (and focused) compared to $S_t$. For instance, an autoregressive random forests (ARRF) -- which generalizes the cases of the previous section --  uses lags of $y_t$ for $X_t$. The tree fitting procedure underlying \textit{plain} RF is not adequate, as it sets $X_t=\boldsymbol{1}$ by default. Thus, analogously to \cite{friedberg2019llf}, it is modified to
\begin{equation}\label{mrf_algo}
\begin{aligned}
\min _{j \in \mathcal{J^-}, \enskip c \in \rm I\!R}  \Bigg[ & \min _{\beta_{1}} \sum_{\{ t \in l | S_{j,t} \leq c  \}}\left(y_{t}-X_t\beta_{1}\right)^{2}+\lambda \Vert \beta_1 \Vert_2  \\
+ & \min _{\beta_{2}} \sum_{\{ t \in l | S_{j,t} > c  \}}\left(y_{t}-X_t \beta_{2}\right)^{2}+\lambda \Vert \beta_2 \Vert_2 \Bigg].
\end{aligned}
\end{equation}
The purpose of this problem is to find the optimal variable $S_j$ (so, finding the best $j$ out of the random subset of predictors indexes $\mathcal{J}^{-}$) to split the sample with, and at which value $c$ of that variable should we split.\footnote{Note that, unlike \cite{friedberg2019llf}, $S_t$ and $X_t$ will differ, which is natural when motivated from a TVP perspective (but not so much from local linear regression one). Forcing their equivalence is not feasible nor desirable in a macro environment.} It outputs $j^*$ and $c^*$ which are used to split  $l$ (the parent node) into two children nodes, $l_1$ and $l_2$. We start with the leaf $l$ being the full sample. Then, we perform a split according to the minimization problem, which procures us with 2 subsamples.  Within each of these two newly created subsamples, we run \eqref{mrf_algo} again. Repeating this process recursively constructs an ever-growing set of $l$'s which are of ever-shrinking size.  Doing so until a stopping criteria is met generates a tree.



\vskip 0.15cm
{\sc \noindent \textbf{Let the Trees Run Deep}.}  Recursively splitting $\beta_0$ into $\beta_1$ and $\beta_2$ eventually leads to $\beta_t$. However, $\beta_t$, by construction, has very little company within its terminal node/leaf. As result, a single tree has low bias, but also very high variance for $\beta_t$. When fitting a single tree, the (early) stopping point must be tuned to avoid overfitting. However, this is not necessary when a sufficiently diversified ensemble of trees is considered. Originally, \cite{breiman2001} himself provided a bound on the generalization error that grows with the correlation between trees.\footnote{Also, \cite{duroux2016impact} derive a formula (for a "median" forest) linking tuning parameters related to the depth of the trees and that of diversification.} In \cite{MSoRF}, I go further by showing that RF's out-of-sample prediction is equivalent to the optimally "stopped" or "pruned" one, provided sufficiently diversified trees.  The desirable property is attributed to the peculiar behavior of "randomized greedy algorithms", which are often overlooked as mere computational necessities. Those insights are of even greater use when it comes to time series since dependence and structural change pose challenges to hyperparameter tuning. Given a large enough $B$, a reasonable \texttt{mtry} (see "De-Correlation" below on this) and standard subsampling rate, we can be confident that the out-of-bag prediction and $\beta_t$'s exclude fitted noise. In our specific context, it means the sample will not be over-split, and we are not going to see time variation when it is not there. Naturally, the credible regions proposed in section \ref{bayes} will also help in that regard. The property will be illustrated in section \ref{sec:newsimul}.





(M)RF prediction is the \textit{simple average} from those of its single trees. Same goes for $\beta_t$. RF is a clever diversification scheme which generates sufficient randomization for that average to inherit the above properties. To achieve that, it mixes elements of re-sampling and model averaging: Bagging and de-correlated trees.\footnote{See \cite{MSoRF} for a discussion on how RF compares and contrast with the forecast combinations/averaging literature.}
 


\vskip 0.15cm
{\sc \noindent \textbf{Bagging}.} Each tree is "grown" on a bootstrapped sample (or a random subsample) \citep{breiman1996bagging}.\footnote{This does not preclude from obtaining $\beta_t$ for all $t$'s since $\beta_t$'s attached to the excluded observations are simply generated by applying the tree on the "out-of-bag" data.} When the base learner is highly nonlinear in observation and/or unstable, gains from Bagging can be large \citep{breiman1996bagging,grandvalet2004bagging}. Nonparametric \citetext{or "pairs" \citealt{mackinnon2006bootstrap}} bootstrap is being used --- i.e., we are \textit{not} shuffling residuals.\footnote{Nonetheless, Bagging in itself is not estranged to macro forecasting  \citep{inoue2008useful,hillebrand2010benefits,hillebrand2020bagging}. However, nearly all studies consider the more common problem of variable selection via hard-thresholding rules -- like t-tests \citep{lee2020bootstrap}.} Rather, we are randomly selecting many observations triples $[y_t \enskip  X_t \enskip S_t]$ (or pairs $[y_t \enskip S_t]$ for Plain RF), and then fit a tree on them. To deal with the dependence inherent to time series data and other reasons detailed in section \ref{bayes}, a slightly more sophisticated bootstrapping/subsampling procedure (involving blocks) will be used for MRF.




\vskip 0.15cm
{\sc \noindent \textbf{De-Correlation}.} The second ingredient, proposed in \cite{breiman2001}, is to consider "de-correlated" trees. RF is an average of many trees, and any averaging scheme reduces variance at a much faster rate if its components are uncorrelated. In our context, this is obtained by growing trees semi-stochastically. In equation \eqref{mrf_algo}, this is made operational by using $\mathcal{J}^- \subset \mathcal{J}$ rather than $\mathcal{J}$. In words, this means that at each step of the recursion, a different subsample of regressors is drawn to constitute candidates for the split. This prevents the greedy algorithm (which, as we know, only "thinks" locally) to always embark on the same optimization route. As a result, trees are further diversified and computing time, reduced. The fraction of randomly selected predictors is a tuning parameter typically referred to as \texttt{mtry} in the literature (and all software), with a default value of $\frac{1}{3}$ for regression settings. This, other algorithmic parameter settings, and some practical aspects are discussed in appendix \ref{CV_section}.





Plain RF has many qualities readily transferable to MRF. It is easy to implement and to tune. That is, it has few tuning parameters that are usually of little importance to the overall performance -- robustness. It is relatively immune to the adverse effects of including many irrelevant features \citep{ESL}. Given the standard ratio of regressors to observations in macro data, this is a non-negligible advantage. Furthermore, with a sufficiently high \texttt{mtry}, it can adapt nicely to sparsity and discard useless predictors \citep{olson2018making}. Finally, its vanilla version already shows good forecasting performance for US inflation \citep{medeiros2019} and macro data in general \citep{chen2019off,GCLSS2018}. 

\subsection{Random Walk Regularization}\label{RWR}


Equation (\ref{mrf_algo}) uses Ridge shrinkage  which implies that each time-varying coefficient is implicitly shrunk to 0 at every point in time. $\lambda$ and the prior it entails can exert a significant influence. For instance, if a process is highly persistent (AR coefficient lower than 1 but nevertheless quite high) as it is the case for SPREAD (see section \ref{forecasting}), shrinking the first lag heavily to 0 could incur serious bias. Fortunately, this can easily be refined to a Minnesota-style prior if $X_t$ corresponds to a Bayesian VAR equation. If $X_t$ is low-dimensional (as it will often be), a simpler alternative consists in using OLS coefficients as prior means. Nonetheless, the specification of previous sections implies that if $\lambda$ grows large, $\forall t \enskip \beta_t=0$ (or whatever the prior mean is). $\beta_i=0$ is a natural stochastic constraint in a cross-sectional setting, but its time series translation $\beta_t=0$ can easily be suboptimal. The traditional regularization employed in macro is rather the random walk 
$$\beta_t=\beta_{t-1}+u_t.$$
Thus, it is desirable to transform (\ref{mrf_algo}) so that it implements the prior that coefficients evolve smoothly (at least, to minimal extent), which is just shrinking $\beta_t$ to be in the neighborhood of $\beta_{t-1}$  and $\beta_{t+1}$ rather than 0. This is in line with the view that economic states (as expressed by $\beta_t$ here) last for at least a few consecutive periods. Note that unlike traditional TVP methods which rely extensively on smoothness regularization -- as it is the sole regularizer, MRF makes only an very mild use of it to get rid of high-frequency noise that may be left in $\beta_t$. The main benefit is to facilitate the interpretation of resulting GTVPs.

%



I implement the desired regularization by taking the "rolling-window view" of time-varying parameters, which has been exploited recently to estimate large TVP-VARs \citep{giraitis2018inference,petrova2016quasi}. That is, the tree, instead of solving a plethora of small ridge problems, will rather solve many weighted least squares problems (WLS) which includes close-by observations. The latter are in the neighborhood (in time) of observations within current leaf. They are included in estimation, but are allocated a smaller weight. 

For simplicity and to keep computational demand low, the kernel used by WLS is rather rudimentary: it is a symmetric 5-step Olympic podium. Informally, the kernel puts a weight of 1 on observation $t$, a weight of $\zeta<1$ for observations $t-1$ and $t+1$ and a weight of $\zeta^2$ for observations $t-2$ and $t+2$. Since some specific $t$'s will come up many times (for instance, if both observations $t$ and $t+1$ are within the same leaf, podiums overlap), I take the maximal weight allocated to $t$ as the final weight ${w(t;\zeta)}$.  

Formally, define $l_{-1}$ as the "lagged" version of leaf $l$. In other words, $l_{-1}$ is a set  containing each observation from $l$, with all of them lagged one step. $l_{+1}$ is the "forwarded" version. $l_{-2}$ and  $l_{+2}$ are two-steps equivalents. For a given candidate subsample $l$, the podium is
\begin{align*}
{w(t;\zeta)}&=
\begin{cases}
1 ,& \text{if } t \enskip  \in  l\\
\zeta ,& \text{if } t \enskip  \in  (l_{+1} \cup l_{-1})/l \\
\zeta^2 ,& \text{if } t \enskip  \in  (l_{+2} \cup l_{-2})/\left(l \cup (l_{+1} \cup l_{-1})\right) \\
0 ,& \text{otherwise}
\end{cases}
\end{align*}
where $\zeta<1$, a tuning parameter guiding the level of time-smoothing. Then, it is only a matter of how to include those additional (but down weighted) observations in the tree search procedure. The usual candidate splitting sets 
$$l_1(j,c) \equiv \{ t \in l | S_{j,t} \leq c  \} \quad \text{  and} \quad l_2(j,c) \equiv \{ t \in l| S_{j,t}>c \}$$
are expanded to include all observations of relevance to the podium
$$\text{for } i=1,2: \quad l_i^{RW}(j,c) \equiv l_i(j,c) \enskip \cup \enskip l_i(j,c)_{-1}
\enskip \cup \enskip l_i(j,c)_{+1} \enskip \cup \enskip l_i(j,c)_{-2} \enskip \cup \enskip l_i(j,c)_{+2}. $$
\noindent The splitting rule becomes
\begin{equation}\label{mrf_algo_rw}
\begin{aligned}
\min _{j \in \mathcal{J^-}, \enskip c \in \rm I\!R}  \Bigg[ & \min _{\beta_{1}}\sum_{t \in  l_1^{RW}(j,c)} w(t;\zeta) \left(y_{t}-X_t\beta_{1}\right)^{2}+\lambda \Vert \beta_1 \Vert_2  \\
+ & \min _{\beta_{2}} \sum_{t \in  l_2^{RW}(j,c)} w(t;\zeta)\left(y_{t}-X_t \beta_{2}\right)^{2}+\lambda \Vert \beta_2 \Vert_2 \Bigg].
\end{aligned}
\end{equation}
Note that the Ridge penalty is kept in anyway, so the final model has in fact two sources of regularization. With $\zeta\rightarrow0$, we are heading back to pure Ridge. 

Although not considered in the main applications of this paper, models with a larger linear part $X_t$ are possible. For instance, one could estimate, equation by equation, a high-dimensional VAR. In practice, this simply requires harsher regularization via higher values of $\lambda$, $\zeta$ and a larger minimum leaf size. Nevertheless, the forecasting benefits from this strategy could prove limited: MRF is "high-dimensional" whenever $S_t$ is large. The time-varying constant in MRF is a RF in its own right. It can be seen as a complex misspecification function (in the deep learning jargon, it is effectively called the bias) that adaptively controls for omitted variables in a way that is both non-linear and strongly regularized via randomization. Consequently, the cost from omitting a regressor of minor importance in $X_t$ is low since it can be picked up by the time-varying intercept. 

Of course, the small $X_t$ strategy treats the extra regressors as exogenous, which could be at odds with some researchers' will to investigate a large web of impulse response functions. Anyhow, both approaches are possible. The dynamic coefficients of a (large) GTVP-VARs can be estimated by either fitting MRF equation by equation, or modifying the splitting rule in \eqref{mrf_algo_rw} to be multivariate so that each tree is fitted jointly for all equation -- pooling time-variation across equations. Finally, elements of the covariance matrix of residuals can be fitted separately with a plain RF, which is very fast.


\subsection{Relationship to Random Walk Time-Varying Parameters}
			 		 
GTVPs have many advantages over classical TVPs. While it is known that any nonlinear model can be approximated by a linear one with TVPs \citep{granger2008}, nothing is said about how efficient that estimation is going to be. As it turns out, efficiency crucially matters in a macro context, and random-walk TVPs can be quite inefficient \citep{aruoba2017dsge}. For example, if the true $\beta_t$ follows a recurrent switching mechanism, random walk parameters already have two strikes against them. Some dimensionality reduction  techniques -- like reduced-rank restrictions \citep{dewindgambetti2014,stevanovic2016common,chan-ftvp2018,GC2019} -- can help, but nothing in that paradigm can come close to the parsimony of simply interacting $X_t$ with relevant variables. In contrast, MRF considers all time-variations options, and choose the "obvious thing", which may or may not be splitting on $t$. Also, it is absolutely possible that the resulting $\mathcal{F}$  pools \textit{both} latent and observable time variation. 

Even though MRF is remarkably flexible, its variance remains low thanks to the diversified portfolio of trees. The variance of classical TVPs can be controlled by cross-validation \citep{GC2019} or via an elaborate hierarchical prior \citep{amir2018choosing}. A number of applications opt for a "manual" approach \citep{AAG2013}. However, it is understood that no tuning, however careful it may be, can overcome the hardship of fitting random-walks when the true $\beta_t$'s look nothing like it. 






Econometrically, one way to more formally connect this paradigm to recent work on TVPs is to adopt the view that RF are adaptive kernel estimators  \citep{meinshausen2006,athey2019grf,friedberg2019llf}. That is, the tree ensemble is a machine generating kernel weights. Once those are obtained, estimation amounts to weighted least squares (WLS) problem with a Ridge penalty. By running (\ref{mrf_algo}) recursively, one obtains terminal nodes/leaves $L_{b}()$ to construct kernel weights
$$
\alpha_{t}\left(\mathrm{x}_{0}\right)=\frac{1}{B} \sum_{b=1}^{B} \frac{1\left\{\mathrm{X}_{\mathrm{t}} \in L_{b}\left(\mathrm{x}_{0}\right)\right\}}{\left|L_{b}\left(\mathrm{x}_{0}\right)\right|}
$$ 
to use in 
\begin{align}\label{mrf_kernel}
\forall t \enskip : 
\operatorname{argmin}_{\beta_t}\left\{\sum_{\tau=1}^{T} \alpha_{t}\left(\mathbf{s}_{\tau}\right)\left(Y_{\tau}-X_\tau \beta_{\tau})\right)^{2}+\lambda\Vert \beta_t \Vert_2 \right\}.
\end{align}
As shown in \cite{GC2019}, standard random walk TVPs are in fact a smoothing splines problem, and for those, a reproducing kernel exists \citep{dagum2009}. \cite{giraitis2014inference} drop the random walk altogether and proposed to use kernels directly. Anyhow, in both cases, the only variable entering the kernel is $t$. In other words, only proximity in time is considered for the clustering of observations. This makes the seemingly flexible estimator in fact quite restrictive -- and dependent on its inherent smoothness prior. Moreover, standard kernel methods are known to break down even in medium dimensions (say <10 variables) \citep{friedberg2019llf}. Therefore, augmenting $t$ -- the sole variable in the kernel (implicit of explicit) of traditional TVP methods -- with additional regressors is not an option. No such constraints bind on the RF approach.

\subsection{Relationship to Standard Random Forest}

The standard RF is a restricted version of MRF where $X_t=\iota$, $\lambda=0$ and $\zeta=0$. In words, the only regressor is a constant and there is no within-leaf shrinkage.  Previous sections motivated MRF as a natural generalization of non-linear  time-series models. At this point, a reasonable question emerges from a ML standpoint. Why should we prefer the partially linear MRF to the fully nonparametric RF? One reason is statistical efficiency. The other is potential for interpretation. 

\subsubsection{Smooth Relationships are Hard Relationships (to estimate)}\label{smoothishard}
In finite samples, plain RF can have a hard time learning smooth relationships -- like a AR(1) process. This is bad news for time series applications. For prediction purposes, estimating 
$$y_t = \phi y_{t-1} +\varepsilon_t$$
by OLS implies a single parameter. However, approximating the same relationship with a tree (or an ensemble of them) is far more consuming in terms of degrees of freedom. To get close to the straight line once parsimoniously parametrized  by $\phi$, we now need a succession of many step functions.\footnote{In a standard regression setup, nobody would model a continuous variable as an ordinal one unless some wild nonlinearities are suspected.} With short time series, modeling smooth/linear relationships in such a way is a luxury one rarely can afford. The mechanical consequence is that RF will waste many splits on capturing the linear part, and may run out of them before it gets to focus more subtle nonlinear phenomena.\footnote{One necessary (but not sufficient) symptom is AR terms being flagged as really important by typical RF variable importance measures (one example is \cite{borup2020now}).} In a language more familiar to economists, this is simply running out (quickly) of degrees of freedom. MRF provides a workaround. Modeling the linear part concisely leaves more room to estimate the nonlinear one. By its more strategic budgeting of degrees of freedom, the resulting (estimated) partially linear model could be, in fact, more non-linear than the fully nonparametric one.





This paper is not the first to recognize the potential need for a linear part in tree-based models. For instance, both \cite{treed} and \cite{modeltrees} proposed linear regressions within a leaf of a tree, respectively denominated "Treed Regression" and "Model Trees". More focused on real activity forecasting, \cite{woloszko2020adaptive} and \cite{wochner2020dynamic} blend insights from macroeconomics to build better-performing tree-based models.\footnote{Specifically, \cite{wochner2020dynamic} also note that using trees in conjunction with factor models can improve GDP forecasting. An analogous finding will be reported in section \ref{forecasting}.} On a different end of the econometrics spectrum, \cite{friedberg2019llf}  proposed to improve the nonparametric estimation of treatment effect heterogeneity by combining those ideas developed for trees into a forest.\footnote{More broadly, this is extending to trees and ensemble of trees the "classical" non-parametrics literature's knowledge that local linear regression usually has much better properties (especially at the sample boundaries) than the Naradaya-Watson estimator.} To my knowledge, this paper is the first to exploit the link between this strand of work and the sempiternal search for the "true" state-dependence in empirical macroeconomic models. 


\subsubsection{A Note on Interpretability}

The interpretation of ML outputs is now a field of its own \citep{molnar2019interpretable}. RF is widely regarded as a black box model which needs to be interpreted using an external device. Indeed, it usually averages over 100 trees of substantial depth, which makes individual inspection impossible. MRFs partially circumvent the problem by providing time series $\beta_t$ which can be examined, and have a meaning as time-varying parameters for the linear model. Thus, whatever one may do with TVPs, it can be done with GTVPs. There are also some new avenues. For instance, Variable Importance (VI) measures usually deployed to dissect RF's prediction can be used to inspect what is driving $\beta_t$'s. Those will be used in section \ref{surro}.



A popular approach to dissect a standard RF is to use interpretable surrogate tree models  to partially replicate the black box model's fit. The idea can be transferred to MRF \citep{molnar2019interpretable}. In fact, partial linearity facilitates such an exercise. The linear part in MRF splits the nonparametric atom into different pieces ($X_{t,k}\beta_{t,k}$) which can be analyzed separately.  Each time series $\beta_{t,k}$ can be dissected with its own surrogate model, and meaningful combination/transformations of coefficients can be considered.





\subsection{Engineering $S_t$}\label{MAF}

This section discusses principles guiding the composition of $S_t$, which is the raw material for $\mathcal{F}$ in both MRF and plain RF. Macroeconomic data sets \citetext{e.g. FRED, \citealt{mccracken2020fred}} typically contains many regressors and few observations. After incorporating lags for each variable, it can easily be the case that predictors outnumber observations. The curse of dimensionality has both computational and statistical ramifications. The former is mostly avoided in RF since it does not rely on inverting a matrix. However, the statistical curse of dimensionality, a feature of the regressors/observations ratio, remains a difficulty to overcome. 


There are two extreme ways of reducing dimensionality: sparse or dense. The former selects a small number of features out of the large pool in a supervised way (e.g. LASSO), the latter compresses the data in a set of latent factors that should span most of  the original regressors space. This is often seen as a necessity to choose \textit{one} of them.\footnote{In macro forecasting work using RF, \cite{GCLSS2018} follow a dense approach by only including factors while \cite{borup2020targeting} opt for sparsity by proposing a Lasso pre-selection step.} However, in a regularized model, both can be included, and we can let the algorithm select an optimal combination of original features and factors.\footnote{A more detailed discussion of this can be found in Appendix \ref{sec:morest}.} This is useful --- it is not hard to imagine a situation where opting for one or the other would prove suboptimal to a more nuanced solution.

{\sc \noindent \textbf{Lag Polynomials}.} From a predictive standpoint, residuals autocorrelation  implies there is forecasting power left on the table. To get rid of it, many lags might be necessary. In multivariate contexts (like that of a VAR), doing so quickly pushes the model to overfit. A standard solution is Bayesian estimation and the use of priors in the line of \cite{litterman1984}, which are specially designed for blocks of lags structures. Outside of the VAR paradigm, there is an older literature estimating restricted/regularized lag polynomials in Autoregressive Distributed Lags (ARDL) models \citep{almon,shiller1973}. More recently, these methods have found new applications in mixed-frequency models \citep{ghysels2007midas} where the design of the model leads to an explosion of lag parameters. 

(M)RF experiences an analogous situation.  A tree may waste many splits trying to efficiently extract information out of a lag polynomial: for instance, splitting on the first lag, then the 7th one, then the 3rd one. In linear parametric models, the above methods can extract the relevant information out of a lag polynomial without sacrificing many degrees of freedom. A significant roadblock to this enterprise in the RF paradigm is that there are no explicit lag polynomials to penalize. An alternative route is to exploit the insight that RF can choose for itself relevant restrictions. We just have to construct regressors that embodies those, and include them in $S_t$.





\vskip 0.2cm

{\sc \noindent \textbf{Moving Average Factors}.}  To extract the essential information out of the lag polynomial of a specific variable, a linear transformation can do the job. Consider forming a panel of $P$ lags of variable $j$:
$$X_{t,j}^{1:P}\equiv [X_{t-1,j} \enskip ... \enskip X_{t-P,j}] \enskip.$$
We want to form weighted averages of the $P$ lags so that it summarizes most efficiently the temporal information of the feature indexed by $j$.\footnote{$P$ is a tuning parameter the same way the set of included variables in a standard factor model is one.} The weighted averages with that property will be the first few factors (extracted by PCA) of $X_{t,j}^{1:P}$.\footnote{While I work directly with the latent factors, a related decomposition called singular spectrum analysis works with the estimate of the \textit{summed} common components. Since this decomposition naturally yields a recursive formula, it has been used to forecast macroeconomic and financial variables \citep{hassani2009forecasting, hassani2013predicting}.} This can be seen as the time-dimension analog to the traditional cross-sectional factors. The latter are defined such as to maximize their capacity to replicate the cross-sectional distribution of  $X_{t,j}$ fixing $t$ while the Moving Average Factors (MAFs) proposed here seek to represent the temporal distribution of $X_{t,j}$ for a fixed $j$ in a lower-dimensional space.\footnote{In the spirit of the Minnesota prior, one can assign decaying (in $p$) weights to each lag before running PCA. This has the analogous effect of shrinking more heavily the distant lags and less so the recent ones.}  By doing so, our goal to summarize the information of $X_{t,j}^{1:P}$ without modifying the RF algorithm (or any other) is achieved: rather than using the numerous lags as regressors, we can use the MAFs which compress information ex-ante. As it is the case for standard factors, MAF are designed to maximize the explained variance in $X_{t,j}^{1:P}$, not the fit of the final target. It is the RF part's job to select the relevant linear combinations among $S_t$ so to maximize the fit. Finally, it is noteworthy that MAFs facilitate interpretation. As these are moderately sophisticated averages of a single time series, they can be viewed as a smooth index for a specific (but tangible) economic indicator. This is arguably much easier to interpret than a plethora of lags coefficients.

The take-away message from this subsection can be summarized in three points. First, there is no need to choose ex-ante between sparse and dense when the model performs selection/regularization. We can let the algorithm find the optimal balance. Second, to make the inclusion of many lags useful, we need to regularize the lag polynomial. Third, such compression can be achieved most easily by generating MAFs and using those as regressors in RF -- or any algorithm. 


\subsection{Quantifying Uncertainty of $\beta_t$'s Estimates}\label{bayes}
 \cite{taddy2015bayesian} and \cite{taddy2016nonparametric} interpret RF's prediction as the posterior mean of a tree functional $\mathcal{T}$ (the splitting algorithm) obtained by an approximate Bayesian bootstrap.\footnote{The connection between \cite{breiman1996bagging}'s bagging and \cite{rubin1981bayesian}'s Bayesian Bootstrap was acknowledged earlier in \cite{clyde2001bagging}.} Through those lenses, each tree is a posterior draw. Seeing $\mathcal{T}$ as a Bayesian nonparametric statistic (independently of the DGP) is of even greater interest in the case of MRF.\footnote{An alternative (frequentist) inferential approach is that of \cite{friedberg2019llf}. However, their asymptotic argument requires estimating the linear coefficients and the kernel weights on two different subsamples. This is hard to reconciliate with our goal of modeling time-variation and different regimes throughout the entire sample. Furthermore, when the sample size is small, splitting the sample in such a way carries binding limitations on the complexity of the estimated function.} It provides inference for meaningful time-varying parameters $\beta_t$ rather than an opaque conditional mean function. Such techniques, originating from \cite{ferguson1973bayesian}, have seldomly found applications in econometrics, such as \cite{chamberlain2003nonparametric} for instrumental variable and quantile regressions. 
 
While the Bayesian Bootstrap desirably does not assume many things about the data, it yet makes the assumption that $Z_t = [y_t \enskip X_t \enskip S_t]$ is an \textit{iid} random variable. Thus, it cannot be used directly as a proper theoretical motivation for using the bag of trees directly to conduct inference. I propose a block extension to make \cite{taddy2015bayesian}'s convenient approach amenable to this paper's setup. 

Block Bayesian Bootstrap (BBB) is a simple redefinition of $Z$ so that it is plausibly \textit{iid}. Hence, in the spirit of traditional frequentist block bootstrap \citep{mackinnon2006bootstrap}, blocks of a well-chosen size will be exchangeable. Thus, a new variable can be defined $Z_\mathfrak{b} \equiv [y_{\underline{\mathfrak{b}}:\bar{\mathfrak{b}}} \enskip X_{\underline{\mathfrak{b}}:\bar{\mathfrak{b}}} \enskip S_{\underline{\mathfrak{b}}:\bar{\mathfrak{b}}}]$. There will be a total of $\mathfrak{B}=\sfrac{T}{\text{block size}}$ fixed and non-overlapping blocks. Under covariance stationarity, $\tilde{Z}_\mathfrak{b} = vec(Z_\mathfrak{b})$ are \textit{iid}, for a properly chosen block length.\footnote{In practice, I will use block of two years for both quarterly or monthly data.}  Analogously to \cite{taddy2015bayesian}, block-subsampling is preferred to BBB in implementations since it is faster and gives nearly identical results. Details of BB and BBB are available in Appendix \ref{sec:bbb_details}.


It is reasonable to wonder how the above procedure deals with the possible presence of heteroscedasticity.  Fortunately, the nonparametric bootstrap/subsampling that RF uses is in fact the "pairs" bootstrap of \cite{pairsbootstrap} which is valid under general forms of heteroscedasticity \citep{mackinnon2006bootstrap}.  From a Bayesian point of view, \cite{lancaster2003} show that the obtained variance for OLS from using such a bootstrap is asymptotically equivalent to that of White's  sandwich formula.\footnote{\cite{poirier2011bayesian} propose better priors and \cite{karabatsos2016dirichlet} incorporate such ideas into a generalized ridge regression.} Hence, in the spirit of heteroscedasticity-robust estimation, no attempt will be made at directly evolving volatility (which is a GLS approach). Rather, it will be reflected in larger bands for periods of smaller signal-to-noise ratio.

\section{Simulations}\label{sec:sim}

Simulations are divided in two parts. The first shows that Autoregressive Random Forest (ARRF) delivers forecasting gains over standard nonlinear time series model when the true DGP mixes both endogenous and exogenous time-variation. Moreover, the former is very resiliant against traditional approaches, even when the DGP matches the latter's restrictive assumptions. Additionally, those simulations will numerically document the superiority of ARRF over RF when the AR part is pervasive (as discussed in section \ref{smoothishard}). Overall, this helps rationalizing forecasting results from section \ref{forecasting}, where ARRF supplants $\sim$TARs for the vast majority of targets. 

The second simulations section considers simpler linear parts and look at how the algorithm behaves when $S_t$ is large. Further, I focus on $\beta_t$ itself and its credible regions. The main point is to visually show that (i) GTVPs adapts nicely to a wide range of DGPs and (ii) are not prone to discover inexistent time-variation.

\subsection{Comparison of ARRF to Traditional Nonlinear Autoregressions}



I consider 3 DGPs: Autoregression (AR), Self-Exciting Threshold ARs (SETAR), and a SETAR model that collapse to an AR (via a structural break).  Those DGPs include two types of time variations, endogenous ($y_{t-1}$) and exogenous ($t$).\footnote{Since a structural break is just a threshold effect with respect to variable $t$, one can conclude without loss of generality that similar results would be obtained using different additional switching variables.} They are meant to encapsulate compactly the usual nonlinearities considered in empirical studies, like dependence on the state of the business cycle \citep{AG2012,rameyzubairy2018} and exogenous time variation \citep{clarida2000}.

For all DGPs, $X_{t}=[1 \enskip {y_{t-1}} \enskip {y_{t-2}}]$. The simulated series sample size is either $T=150$ or $T=300$. The last 40 observations of each sample consist the hold-out sample for evaluation. I forecast 4 different horizons: $h=1,2,3, 4$. Models are estimated once at the last available data point. 

\vspace{0.2cm}
{\sc \noindent \textbf{Models.}} SETAR, Rolling-Window (RW) AR, Random Forest (RF) and Autoregressive Random Forest (ARRF) are included. Iterated SETAR forecasts are obtained via the standard bootstrap method \citep{clements1997setar} and all the others are generated via direct forecasting. That is, in the latter case, I fit the model directly on $y_{t+h}$ rather than iterating forward the one-step ahead forecast. To certify that the observed differences between SETAR and other models is not merely due to the choice of iterated vs direct forecasts -- a non-trivial choice in many environments \citep{chevillon2007direct} --, I also include SETAR-d where "d" means its forecasts were alternatively obtained by direct forecasting. 

In all simulations, MRF's $S_{t}$ includes 8 lags of $y_t$ and a time trend, which match what will be referred to in section \ref{forecasting} as "Tiny ARRF". Thus, unlike $\sim$TARs, it is "allowed" to split on what we know (by the DGP choices) to be useless regressors (especially at horizon $h=1$). The specified linear part for all models matches that of the true DGP ($X_{t}=[1 \enskip {y_{t-1}} \enskip {y_{t-2}}]$).

\vspace{0.2cm}
{\sc \noindent \textbf{Performance Metric.}} Performance is evaluated using the mean squared prediction error (MSPE). In simulation $s$, for the forecasted value at time $t$ made $h$ steps ahead, I compute 
$$RMSE_{h,m}= \sqrt{ \frac{1}{40 \times 100}\sum_{s=1}^{100} \sum_{t \in \text{OOS}} (y_{t}^s-\hat{y}_{t-h}^{s,h,m})^2}.$$
100 different simulations are considered, which means the total number of squared errors being averaged for a given horizon and model is 100$\times$40=4000. To provide a visually useful normalization, bar plots report $RMSE_{h,m}$'s relative to that of the oracle, who knows perfectly the law of motion of time-varying parameters $\beta_t$.\footnote{Precisely, if the model has a break and a switching variable, it knows exactly the break points, thresholds and AR parameter values in each regime. The only things the oracle does not know are the future shocks ($\epsilon_{t+h}$), and the out-of-sample evolution of parameters  ($\beta_{t+h}$) -- unless the latter is purely deterministic.} Formally, the metric is 
$\Delta_o {RMSE}_{h,m}=\sfrac{{RMSE}_{h,m}}{{RMSE}_{h,o}} -1.$

\vskip 0.2cm
{\sc \noindent \textbf{SETAR Morphing Instantly into AR(2).}} The two sources of time-variation are combined to display MRF's edge in this not so implausible situation. Further, 
\begin{align*}
\text{DGP 1}=
\begin{cases}
\text{SETAR} ,& \text{if } t<T/2  \\
\text{AR}  ,& \text{otherwise}
\end{cases} 
\end{align*}
can rightfully be hypothesized for some economic time series: complex dynamics up until the mid-1980's followed by a very simple autoregressive structure during the Great Moderation.\footnote{The AR has ${\beta}= [0 \enskip 0.7 \enskip -0.2]$ and the SETAR has $\beta_{t}=I(y_{t-1}\geq 1)[2 \enskip 0.8 \enskip -0.2] + I(y_{t-1}<1)[0.25 \enskip 0.4 \enskip -0.2]$. Results for the latter in isolation are in Appendix \ref{sec:moresims}.} In Figure \ref{v1_graph}, MRF comes out as the best model for all horizons in the smaller sample. RF fails particularly at short horizons because it attempts to model all dynamics nonparametrically. Doubling the sample size helps, but its disimprovement with respect to the oracle remains at least twice as large as that of MRF. SETAR and AR both focus on dynamics but are misspecified. Their increase in relative RMSE is about thrice that of MRF at longer horizons for the shorter sample. For horizon 1, RW-AR does equally well as MRF, which is expected in this DGP since it discards earlier observations we only know ex-post to be harmful. Thus, in this DGP much akin to that of the hypothetical inflation tree of section \ref{macroastrees}, MRF comes out as the clear winner.





\vskip 0.2cm
{\sc \noindent \textbf{Persistent SETAR.}}  DGP 2, with $\beta_{t}=I(y_{t-1}\geq 0)[2 \enskip 0.8 \enskip -0.2] + I(y_{t-1}<0)[0.25 \enskip 1.1 \enskip -0.4]$,
represents an endogenous switching process which may suit well real activity variables: it includes high/low regimes, and mildly different dynamics in each of them. Can MRF match traditional nonlinear times series model when the world is nonlinear, yet simple? The broad answer from Figure \ref{v2_graph} is yes. For all horizons and sample sizes considered, MRF is practically as good as SETAR, the optimal model in this context. Because of its capacity to control overfitting, MRF will be competitive even if nonlinearities turn out to be as simple as often postulated in the empirical macroeconomics literature. With the relative importance of changing persistence, RF cannot match MRF's performance and is trailing behind with RW-AR.\footnote{In Appendix \ref{sec:moresims}, the case where the changing persistence is less important is considered.} Nevertheless, RF improves substantially at shorter horizons when the sample size increase. Finally, AR is resilient at longer horizons but is much worse than MRF and SETAR at shorter ones. 


\begin{figure}[p!]
  \begin{subfigure}[b]{\textwidth}
  \begin{center}
         \includegraphics[trim={1.5cm 0cm 0.25cm 0cm},clip,width=0.94\textwidth]{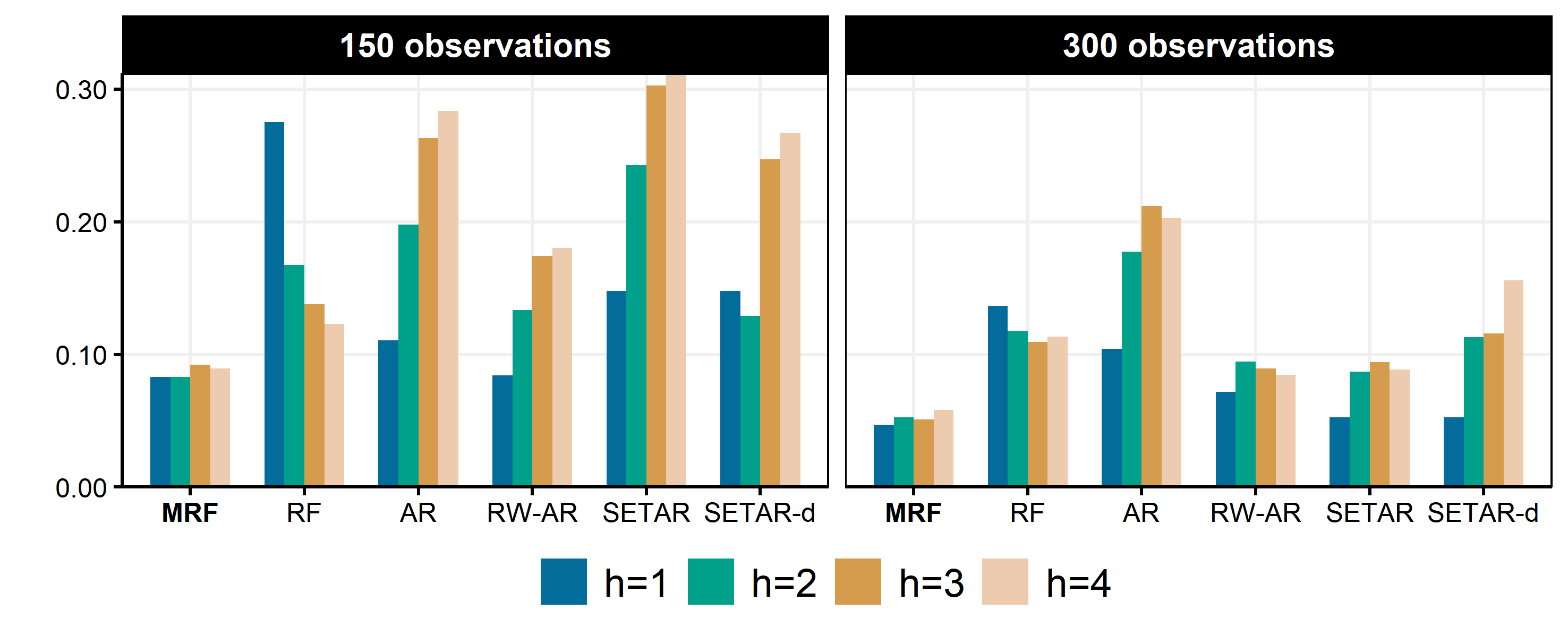}
    \caption{DGP is SETAR morphing into AR(2).}
        \label{v1_graph}
                    \end{center}
  \end{subfigure}
  \hspace{2em}
  \begin{subfigure}[b]{\textwidth}
    \begin{center}
         \includegraphics[trim={1.5cm 0cm 0.25cm 0cm},clip,width=0.94\textwidth]{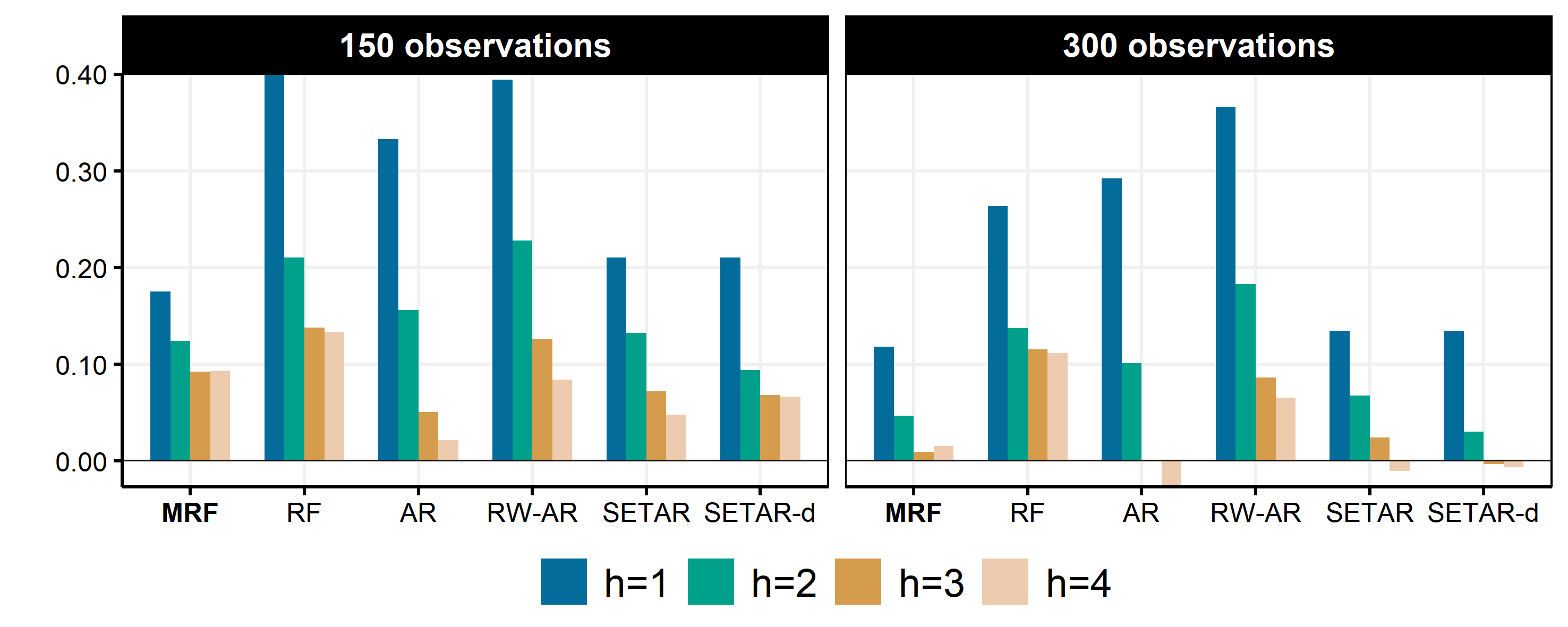}
    \caption{DGP is Persistant SETAR.}
        \label{v2_graph}
                    \end{center}
      \end{subfigure}
  \begin{subfigure}[b]{\textwidth}
    \begin{center}
         \includegraphics[trim={1.5cm 0cm 0.25cm 0cm},clip,width=0.94\textwidth]{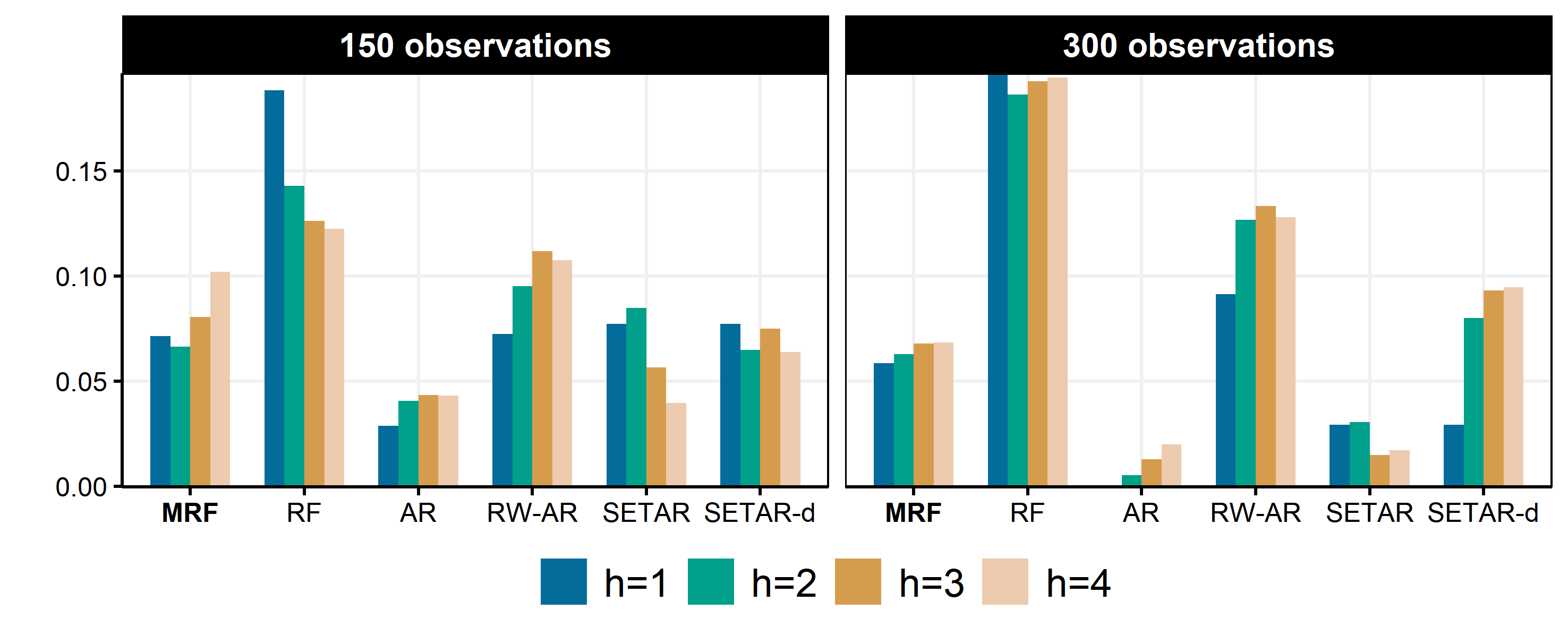}
    \caption{DGP is Plain AR(2).}
        \label{v3_graph}
                    \end{center}
      \end{subfigure}
  \caption{Displayed are increases in relative RMSE with respect to the oracle.}
    \label{oldsimul_all1}
\end{figure}

\vskip 0.2cm
{\noindent \sc \textbf{No Time-Variation.}}  Given the widespread worry that ML-based algorithms can overfit, a time-invariant DGP is a natural check.\footnote{Also, the incredible resilience of linear AR models is well documented in the macroeconomic forecasting literature (see \cite{KLS2019} and references therein).} Can MRF still deliver competitive performance if reality equates simple linear dynamics? Results for DGP 3, an AR(2) process with ${\beta}= [0 \enskip 0.7 \enskip -0.2]$, 
are reported in Figure \ref{v3_graph}. As expected, AR is the best model for all horizons and both sample sizes. The RW-AR suffers from high variance and it is assumed that tuning the window length in a data-driven way would help. Plain RF struggles irrespective of the sample size. For the smaller sample, MRF performs as well as the tightly parametrized SETARs. Their marginal increases in RMSE with respect to the oracle are typically less than 10\%, which is small in contrast to that of previous DGPs. More observations generally helps AR, the iterated SETAR, and MRF especially at longer horizons.


\vskip 0.2cm
{\sc \noindent \textbf{About Misspecification of $X_t$.}}  Most of the reported gains from using MRF come from avoiding misspecification when a more complex DGP arises. What happens if the arbitrary linear part in MRF, $X_t$, is itself misspecified? Figure \ref{MISS_graph} in the appendix report corresponding results. For all DGPs under consideration, a "Bad" MRF, where $X_t$ is composed of two white noise series (instead of the first two lags of $y_t$), performs similarly well (or bad) as plain RF.\footnote{This result may not hold, however, when the law of motion for the intercept is highly complex and requires a great number of split (unlike what is considered here). This is due to the linear part restricting the depth of trees (with to what plain RF could allow for), especially if observations are scarce. In that regard, increasing the ridge penalty (via $\lambda$) will help. Nevertheless, in practice, it is a safer bet to use a small linear part if uncertainty around its composition is high.  More on this and the effect of hyperparameters can found in appendix \ref{CV_section}.}

\vskip 0.2cm

{\sc \noindent \textbf{Summary.}}  First, when the true DGP is \textit{not} that of the tightly parametrized classical nonlinear time series model, the more flexible MRF does better. Second, when classical nonlinear time series model are fitted on their corresponding DGPs, those perform better than MRF -- but only marginally. Third, when there are pervasive linear autoregressive relationships, plain RF struggles. Fourth, MRF and RF relative performance both increase with the number of observations but MRF's one increases faster if the linear part is well-chosen. In Appendix \ref{sec:moresims}, results for 3 additional DGPs are reported: another SETAR, AR with a structural break, and SETAR morphing in another SETAR (through a break). Again, MRF is shown to have an edge when other models are misspecified and almost as good when those are not. 


\subsection{A Look at GTVPs when $S_t$ is Large}\label{sec:newsimul}


A notable difference between the simulations presented up to now and the applied work being carried in later sections is the size of $S_t$. In many macro applications, there is no shortage of variables to include in MRF's $\mathcal{F}$. For instance, the FRED-QD data base \citep{mccracken2020fred} contains over 200 potential predictors that can join lags of $y$ and a time trend within $S_t$. As a result, there is now considerable interest in allowing for time variation in empirical models using large information sets. For instance, \cite{koop2013large} propose large TVP-VARs while \cite{abbate2016changing} extend \cite{bernanke2005measuring}'s factor-augmented VAR to be time-varying. Interestingly, those papers (and the corresponding literature) almost exclusively focus on a setup where, in MRF notation, $X_t$ is large and $S_t$ is extremely small (usually just $t$). Of course, MRF could deal with this case (as discussed in section \ref{RWR}), but its edge will be more apparent when we let the RF part deal with large data and keep $X_t$ concise. Indeed, in addition to lessened misspecification concerns, RFs also benefits from more data through increased randomization -- which prevents fully grown trees from overfitting \citep{breiman2001,MSoRF}. 

\begin{figure}[p!]
  \begin{subfigure}[b]{\textwidth}
   \includegraphics[trim={1.6cm 1.8cm 0.3cm 0.5cm},clip,width=\textwidth]{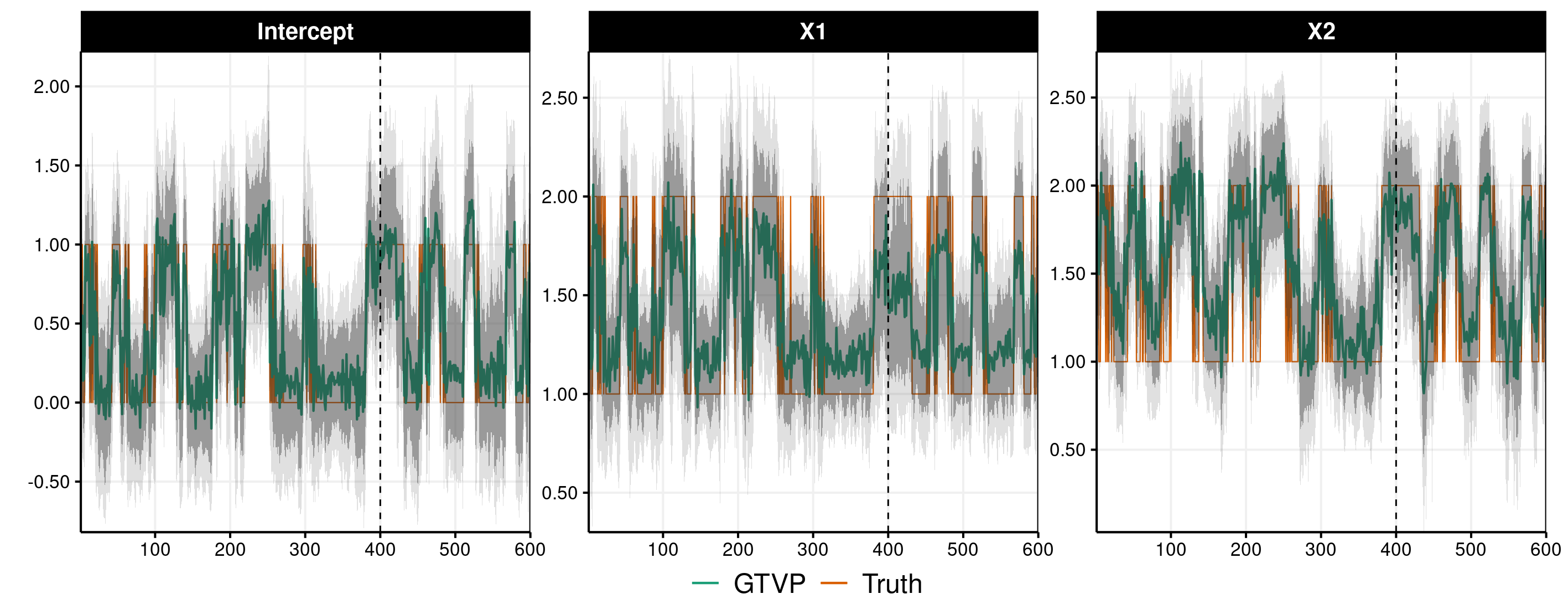}
    \caption{DGP 1}
        \label{newsimul_v1}
  \end{subfigure}
  \hspace{2em}
  \begin{subfigure}[b]{\textwidth}
   \includegraphics[trim={1.6cm 1.8cm 0.3cm 0.5cm},clip,width=\textwidth]{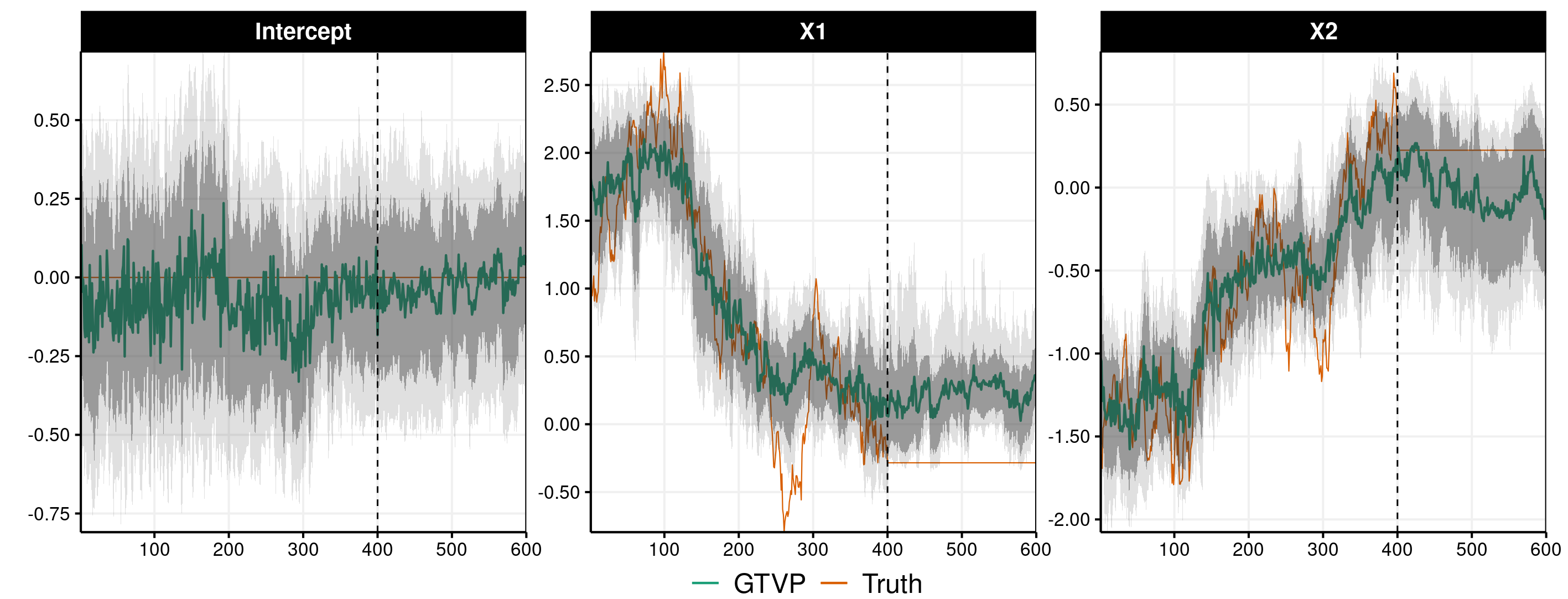}
    \caption{DGP 2}
        \label{newsimul_v2}
      \end{subfigure}
  \begin{subfigure}[b]{\textwidth}
   \includegraphics[trim={1.6cm 1.8cm 0.3cm 0.5cm},clip,width=\textwidth]{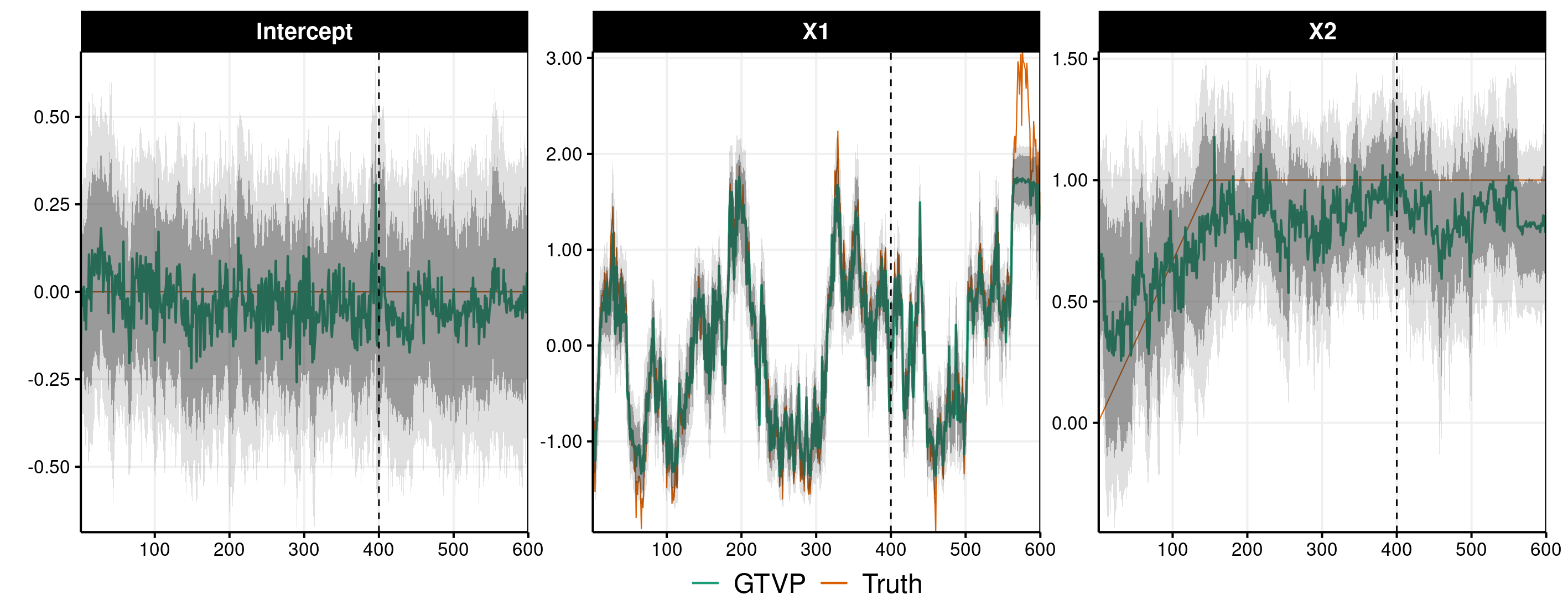}
    \caption{DGP 3}
        \label{newsimul_v3}
      \end{subfigure}
  \caption{The grey bands are the 68\% and 90\% credible region. After the blue line is the hold-out sample. Green line is the posterior mean and orange is the truth. The plots include only the first 400 observations for visual convenience.}
    \label{newsimul_graph_main}
\end{figure}

The additional simulations go as follow. First, I simplify the analysis by looking at a static model with mutually orthogonal but autocorrelated regressors $X_1$ and $X_2$, both driving $y_t$ according to some process. I simulate each of them for 1000 periods and estimate the models with the first 400 observations. The remaining 600 are used to evaluate the out-of-sample performance. The signal to noise ratio is calibrated to $\sfrac{2}{3}$ which is about what is found (out-of-sample) for most models in the empirical section.

The only remaining questions are that of the constitution of $S_t$ and the generation of $\beta_t$'s. I create two autocorrelated (but not cross-correlated) factors. Out of each of them, I create 50 series with a varying amount of additional white noise.\footnote{To be precise, their standard deviation is $U[0.5,3]$\% that of the original factor standard deviation.} Joining those 100 series with lags of $y_t$ and a time trend, the final size of $S_t$ is slightly above 100. Finally, $\beta_t$'s are functions of the underlying \textit{first} factor which (like the second) is not directly included in the data set. In certain DGPs, some $\beta_t$'s will also be a pure function of $t$ (like random walks, structural breaks).\footnote{To clarify, the second factor and underlying series are completely useless to the true DGP -- arguably mimicking the inevitable when using a data base of the size of FRED-QD.} Table \ref{NS_dgps} summarizes the six DGPs in words. 

\begin{table}[!h]
\caption{Summary of Data-Rich Simulations DGPs}
\vspace{-0.3cm}
\footnotesize
\centering
\renewcommand{\arraystretch}{1.15}
\rowcolors{2}{white}{gray!15}
\label{NS_dgps}
\begin{tabular}{llllll}
\toprule
\toprule

DGP \#   & \multicolumn{1}{l}{Intercept} & \multicolumn{1}{l}{$\beta_t^{X_1}$}  & \multicolumn{1}{l}{$\beta_t^{X_2}$} & \multicolumn{1}{l}{Residuals Variance} \\
\midrule
1   & Switching & Switching & Switching & Flat\\
2 & Flat          & Random Walk  & Random Walk & Flat \\
3 & Flat          & Latent factor directly  & Slow Change (function of $t$) & Flat \\
4 & Flat          & Switching  & Slow Change (function of $t$) & Flat \\
5 & Flat          & Switching  & Structural Break & Flat \\
6 & Flat          & Flat & Flat & Stochastic Volatility \\
\bottomrule
\bottomrule
\end{tabular}
\end{table}

More illustratively, Figures \ref{newsimul_graph_main} and \ref{newsimul_graph} (appendix) plots one example of each DGPs as well as the estimated GTVPs and their credible region (as discussed in section \ref{bayes}). It is visually obvious that GTVPs are adaptive in the sense that it can discover which kind of time-variation is present in the data while estimating it. In Figure \ref{newsimul_v1}, MRF successfully estimate the rather radical switching regimes present in all coefficients. In Figure \ref{newsimul_v2}, MRF realizes that almost all of $S_t$ is useless because true $\beta_t$ follow random walks.  Rather, it manages to fit $\beta_t$'s nicely by relying on a multitude of $t$ splits. In Figure \ref{newsimul_v3}, things get "easier" for the true $\beta_{X_1,t}$ as it is driven directly by the first latent factor. MRF discovers that and leverage it to have a very tight fit for it, both in-sample and out-of-sample. This is merely a reflection that if time variation can be constructed by simple interaction terms, this is certainly the easiest statistical route to by -- and MRF chooses it thanks to its inherent ability to perform "time variation selection".

Figure \ref{ns_rmse} reports distributions of RMSE differentials with respect the oracle (the forecast that knows the $\beta_t$'s law of motion). MRF performance is compared to OLS, Rolling-Window OLS (RW-OLS) and plain RF.  As expected, MRF outperforms all alternatives by wide margins for most DGPs. By construction, RW-OLS and OLS also perform well for DGP 5 (random walks) and DGP 6 (constant parameters). Nonetheless, it is reassuring to see that MRF either performs much better than OLS or worse by a thin margin (in cases with no time-variation). 



\section{Macroeconomic Forecasting}\label{forecasting}

In this section, I present results for a pseudo-out-of-sample forecasting experiment at the quarterly frequency using the dataset FRED-QD \citep{mccracken2020fred}. The latter is publicly available at the Federal Reserve of St-Louis's  web site and contains 248 US macroeconomic and financial aggregates observed from 1960Q1. The forecasting targets are real GDP, Unemployment Rate (UR), CPI Inflation (INF), 1-Year Treasury Constant Maturity Rate (IR) and the difference between 10-year Treasury Constant Maturity rate and Federal funds rate (SPREAD). These series are representative macroeconomic indicators of the US economy which is based on \cite{GCLSS2018} exercise for many ML models, itself based on \cite{KLS2019} and a whole literature of extensive horse races in the spirit of \cite{SW1998comparison}. The series transformations to induce stationarity for predictors are indicated in \cite{mccracken2020fred}. For forecasting targets, GDP, UR, CPI and IR are considered $I(1)$ and are first-differenced. For the first two, the natural logarithm is applied before differencing. SPREAD is kept in "levels".  Forecasting horizons are 1, 2, 4, 6 and 8 quarters.

The pseudo-out-of-sample period starts in 2003Q1 and ends 2014Q4.  I use expanding window estimation from 1961Q3. Models are estimated (and tuned, when applicable) every two years. For all models except SETAR and STAR, I use direct forecasts, meaning that $\hat{y}_{t+h}$ is obtained by fitting the model directly to $y_{t+h}$ rather than iterating one-step ahead forecasts.  $\sim$TAR iterated forecasts are calculated using the block-bootstrap method which is standard in the literature \citep{clements1997setar}.

Following standard practice, the quality of point forecasts is evaluated using the root Mean Square Prediction Error (MSPE). For the out-of-sample (OOS) forecasted values at time $t$ of variable $v$ made $h$ steps ahead, I compute 
$$RMSE_{v,h,m}= \sqrt{ \frac{1}{\#\text{OOS}}\sum_{t \in \text{OOS}} (y_{t}^v-\hat{y}_{t-h}^{v,h,m})^2}.$$
The standard \cite{dieboldmariano} (DM) test procedure is used to compare the predictive accuracy of each model against the reference AR(4) model. $RMSE$ is the most natural loss function given that all models are trained to minimize the squared loss in-sample.

It has been argued in section \ref{MAF} that feature engineering matters crucially when the number of regressors exceeds the sample size. $S_{t}$, the set of variables from which RF can select, is motivated by such concerns. Its exact composition is detailed in Table \ref{S_t_comp}. Among other things, it includes both cross-sectional and moving average factors, which are compressing information along their respective dimensions. The usefulness of MAFs is further studied in \cite{MDTM} and found to help, mostly with tree-based algorithms. However, it is supplanted by a more computationally demanding (but more general) transformation of the raw data that \cite{MDTM} propose specifically for ML-based macroeconomic forecasting.

\begin{table}[!h]
\caption{Composition of $S_t$}
\vspace{-0.3cm}
\footnotesize
\centering
\setlength{\tabcolsep}{1.1em}
\renewcommand{\arraystretch}{1.3}
\rowcolors{2}{white}{gray!15}
\label{S_t_comp}
\begin{tabular}{lll}
\toprule
\toprule

What   & \multicolumn{1}{l}{Why} & \multicolumn{1}{l}{How} \\
\midrule
8 lags of $y_t$   & Endogenous SETAR-like dynamics & -- \\
$t$   & Exogenous "structural" change/breaks & -- \\
2 lags of FRED   & Fast switching behavior & -- \\
8 lags of 5 traditional factors $F$   & Compress cross-sectional information ex-ante & Usual PCA \\
2 MAFs for each variable $j$  & Compress lag polynomial information ex-ante & PCA on 8 lags of $j$\\
\bottomrule
\bottomrule
\end{tabular}
\vspace{-0.3cm}
\end{table}

\vspace{0.2cm}
{\noindent \sc \textbf{Models}.} To better understand where the gains from MRF are coming from, I include models that use different subsets of ideas developed in earlier sections. Those are summarized in Table \ref{fcst_models}. The competitive data-rich models are in the benchmarks group. Non-linear time series models are also included as they share an obvious familiarity with ARRF. "Tiny" versions of both ARRF and RF are considered to gauge the effect from only having access only to a small $S_t$ --- this could be the case for many non-US applications. Conversely, this helps quantify how a data-rich environment contributes to the success of ARRF versus its plain flexibility. Indeed, Tiny ARRF corresponds to what was shown in the "data-poor" simulations (section \ref{sec:sim}) to be a generalization of $\sim$TARs and related models.

Here are some remarks motivating some inclusions and specifications choices. To assess the marginal effects of MAFs alone, Lasso, Ridge and RF are considered using $S_t$ --- those are known to handle high-dimensional feature space. When it comes to FA-ARRF, I opt for a parsimonious linear specification including one lag of the first two factors. First, concise models make interpretation easier. Second, considering compact linear specifications within MRF is usually the better strategy. Parameters (including the intercept) are all RFs in their own right and can palliate to the omission of marginally important features, if need be. Consequently, it is desirable to fix a humble linear part and let $\beta_t$'s take care of the rest.\footnote{Further backing a parsimonious choice (with MRF), \cite{mccracken2020fred} report that the first two factors account for 30\% of the variation in the data while adding two more only bumps it up to 41\%, making the last two presumably more disposable in our context.} Finally, as discussed in \cite{mccracken2020fred}, the first factor mostly loads on real activity variables while the second is a composite of forward-looking indicators like term spreads, permits and inventories. They are baptized and interpreted accordingly.

  \begin{savenotes}
  \begin{table}[!th]
    \begin{threeparttable}
\caption{Forecasting Models}\label{fcst_models}
\footnotesize
\centering
\rowcolors{2}{white}{gray!15}
\setlength{\tabcolsep}{0.76em}
\renewcommand{\arraystretch}{1.3}
\begin{tabular}{llll}
\toprule
\toprule

 Name & Acronym   & \multicolumn{1}{l}{Linear Part ($X_t^m$)} & \multicolumn{1}{l}{RF part}  \\
\midrule
Autoregression & \textbf{\textcolor{black}{AR}}   & $[1, \hspace*{0.1cm} y_{t-\{1:4\}}]$ & $\emptyset$ \\

Factor-Augmented Autoregression & \textbf{FA-AR}   & $[1, \hspace*{0.1cm} y_{t-\{1:4\}}, \hspace*{0.1cm} F_{1,t-\{1:2\}},  \hspace*{0.1cm} F_{2,t-\{1:2\}}]$ & $\emptyset$ \\

Plain Random Forest & \textbf{RF}   & $\emptyset$ & Raw data\footnote{Precisely, this means 8 lags of FRED-QD, after usual transformations for stationarity have been applied.} \\
Low-Dimensional Plain RF & \textbf{Tiny RF}  & $\emptyset$ &  $[ y_{t-\{1:8\}}, \hspace*{0.1cm} t ]$ \\
\midrule
Plain RF but using $S_t$  & \textbf{RF-MAF}   & $\emptyset$ & $S_t$ \\
RF-MAF on de-correlated $y_t$ & \textbf{AR+RF}  & Filter $y_t$ first with an AR(2), then RF &  $S_t$ \\
Autoregressive Random Forest & \textbf{ARRF} & $[1, \hspace*{0.1cm} y_{t-\{1:2\}}]$   & $S_t$ \\
Low-Dimensional Autoregressive RF & \textbf{Tiny ARRF} & $[1, \hspace*{0.1cm} y_{t-\{1:2\}}]$         & $[ y_{t-\{1:8\}}, \hspace*{0.1cm} t ]$  \\
Factor-Augmented Autoregressive RF & \textbf{FA-ARRF} & $[1,  \hspace*{0.1cm} y_{t-\{1:2\}},  \hspace*{0.1cm} F_{1,t-1},  \hspace*{0.1cm}F_{2,t-1}]$     & $S_t$  \\
Vector Autoregressive RF\footnote{Note that the VAR appellation refers to the linear equation consisting of typical "small monetary VAR". The model remains univariate and direct forecasts are used.}  & \textbf{VARRF} &$[1,  \hspace*{0.1cm} y_{t-\{1:2\}}, \hspace*{0.1cm} {GDP}_{t-1},  \hspace*{0.1cm} {IR}_{t-1},  \hspace*{0.1cm} {INF}_{t-1}]$ & $S_t$ \\
\midrule
Self-Exciting Threshold AR & \textbf{SETAR} & $[1, \hspace*{0.1cm} y_{t-\{1:2\}}]$   &  $\emptyset$ \\
Smooth Transition AR\footnote{The state variable is $y_{t-1}$, as in SETAR.} & \textbf{STAR} & $[1, \hspace*{0.1cm} y_{t-\{1:2\}}]$   &  $\emptyset$ \\
10 years Rolling-Window AR & \textbf{RW-AR} & $[1, \hspace*{0.1cm} y_{t-\{1:2\}}]$   &  $\emptyset$ \\
Time-Varying Parameters AR\footnote{Estimated and tuned via the Ridge approach proposed in \cite{GC2019}.}  & \textbf{TV-AR} & $[1, \hspace*{0.1cm} y_{t-\{1:2\}}]$   &  $\emptyset$ \\
LASSO using $S_t$ & \textbf{LASSO-MAF} & $S_t$   &  $\emptyset$ \\
Ridge using $S_t$ & \textbf{Ridge-MAF} & $S_t$   &  $\emptyset$ \\
\bottomrule
\bottomrule
\end{tabular}
\begin{tablenotes}[para,flushleft]
\footnotesize  Notes: models are classified in 3 categories: benchmarks, MRFs (and related prototypes), and misc (non-linear time series models, other reasonable additions). The main analysis in section \ref{RF_Q_main} omits the \nth{3} club for parsimony. 
  \end{tablenotes}
  \end{threeparttable}
\end{table}
\end{savenotes}

\subsection{Main Quarterly Frequency Results}\label{RF_Q_main}


Violin plots are used throughout to summarize dense RMSEs tables (like Table \ref{Q_table}). I report the distribution of $RMSE_{v,h,m}/RMSE_{v,h, AR}$. This is informative about the overall ranking and versatility of considered models. Of course, being ranked first does not imply being the best model for every $h$ and $v$. Rather, it means that it performs better on average, over all targets.



\begin{figure}[h!]
\begin{center} 
\hspace*{-.3cm}\includegraphics[trim={1.5cm 1.5cm 0.25cm 0cm},clip,scale=.3]{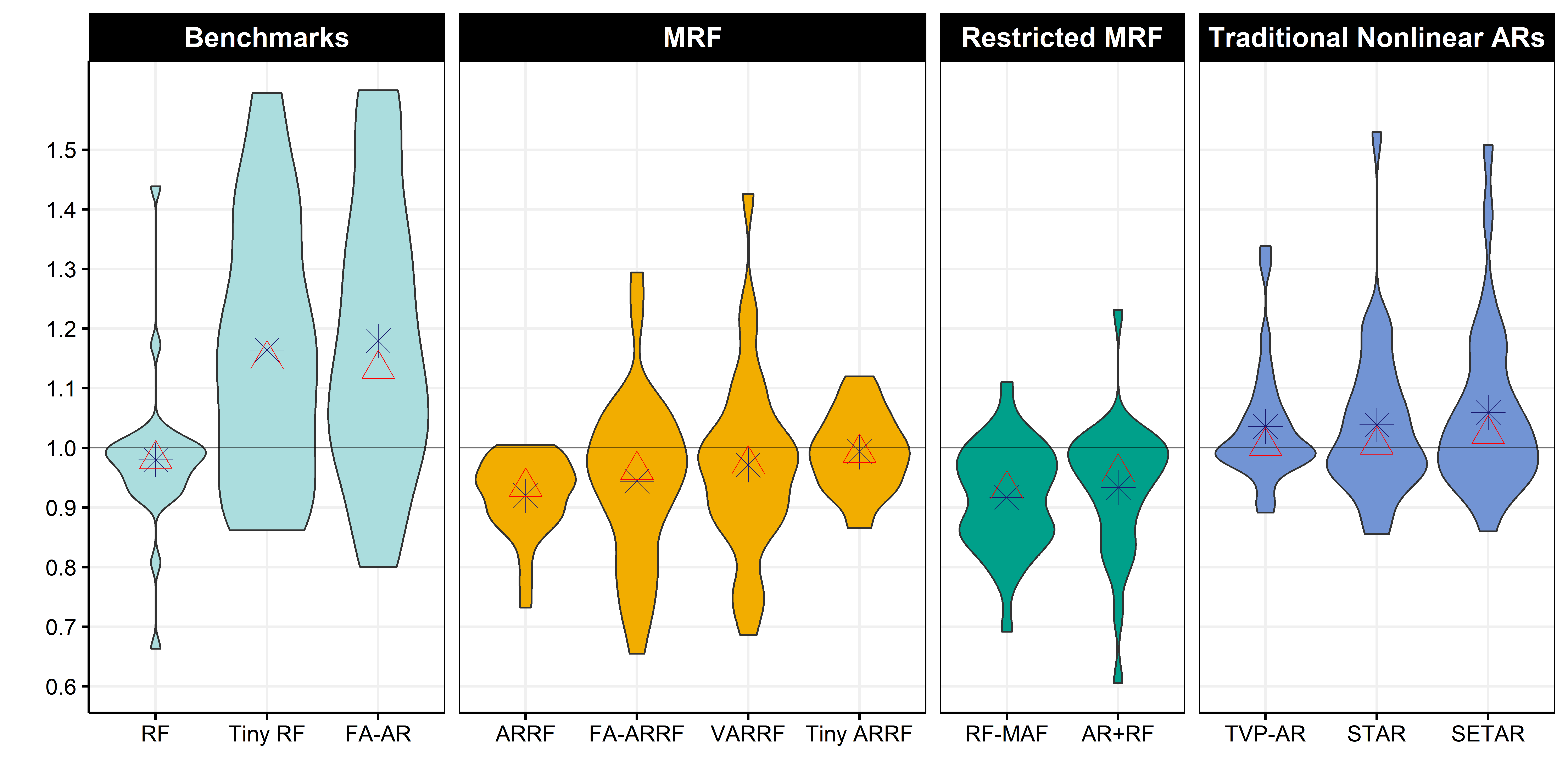}
\caption{\footnotesize The distribution of $RMSE_{v,h,m}/RMSE_{v,h, AR}$. The star is the mean and the triangle is the median.}  
\label{violin_main}
\end{center}
\end{figure} 

Here are interesting observations from Figure \ref{violin_main}. Clearly, MRFs deliver important gains over both the AR and FA-AR benchmarks (the latter is second to last). ARRF has a noticeably small mass above the 1 line. In other words, there are no targets for which ARRF does significantly worse than its OLS counterpart, which makes it atypically adaptable among nonlinear autoregressions. A look at Table \ref{Q_table} confirms this observation also extends to FA-ARRF vs FA-AR. The simplification AR+RF, ranks third with a performance that is much more volatile. This suggests that imposing time-invariant dynamics can sometimes help (see one example in Figure \ref{v2h1_betas}), but can also be highly detrimental (as reported for inflation). Of course, that we do not know ex-ante, and it is why AR+RF does not inherit ARRF's "off-the-shelf" quality.

MAFs are useful: RF-MAF does much better than RF which uses the raw data. The latter only exhibits conservative gains over the benchmark. Thus, it is understood that a fraction of MRFs' forecasting gains emanates from considering more sensible transformations of time series data -- and which are trivially implementable. The relevance of MAFs is studied more systematically in \cite{MDTM}. 

FA-ARRF provides very substantial improvements, but can also fail. This is the linear part's doing: FA-AR will mostly work well for real activity variables while AR is a jack of all trades. Thus, it is not surprising to see FA-ARRF inherit some of these uneven properties, albeit to a much milder extent. For instance, in Table \ref{Q_table}, FA-AR is noticeably worse than AR for all inflation horizons, while FA-ARRF beats AR for all of them. This phenomenon is well summarized by FA-AR being second to last \textit{overall}, well behind FA-ARRF. VARRF has a behavior similar to that of FA-ARRF, but with less highly noticeable gains.

Does a large $S_t$ pay off? Most of the time, yes. It is worth re-emphasizing that restricting $S_t$ restricts the space of time-variations possibilities as well as the potential for trees diversification. Nonetheless, if the restrictions are "true", gains are possible.\footnote{An interesting specific case is Tiny ARRF being close behind ARRF for inflation. This is intuitive given that INF has often been associated with exogenous time variation.} This is reported to be a rare occurrence, with ARRF $\succ$ Tiny ARRF (and RF $\succ$ Tiny RF) for almost any target. Thus, we can safely conclude that a rich $S_t$ is desirable, with $\mathcal{F}$ being tasked with selection of relevant items. 

As discussed in earlier sections, ARRF connects to the wider family of nonlinear autoregressive models. It clearly does better on average than SETAR and Smooth-Transition TAR. This advantage is attributable to both a more flexible law of motion and a large $S_t$. Tiny ARRF is better than the $\sim$TAR group, while ARRF is \textit{much} better. Linking this result to those of simulations, this means that no $\sim$TAR is likely the true model. 






\begin{figure}[h!]
\begin{center}
  \begin{subfigure}[b]{\textwidth}
  \begin{center}
\includegraphics[trim={1.5cm 0cm 0.25cm 0cm},clip,scale=.28]{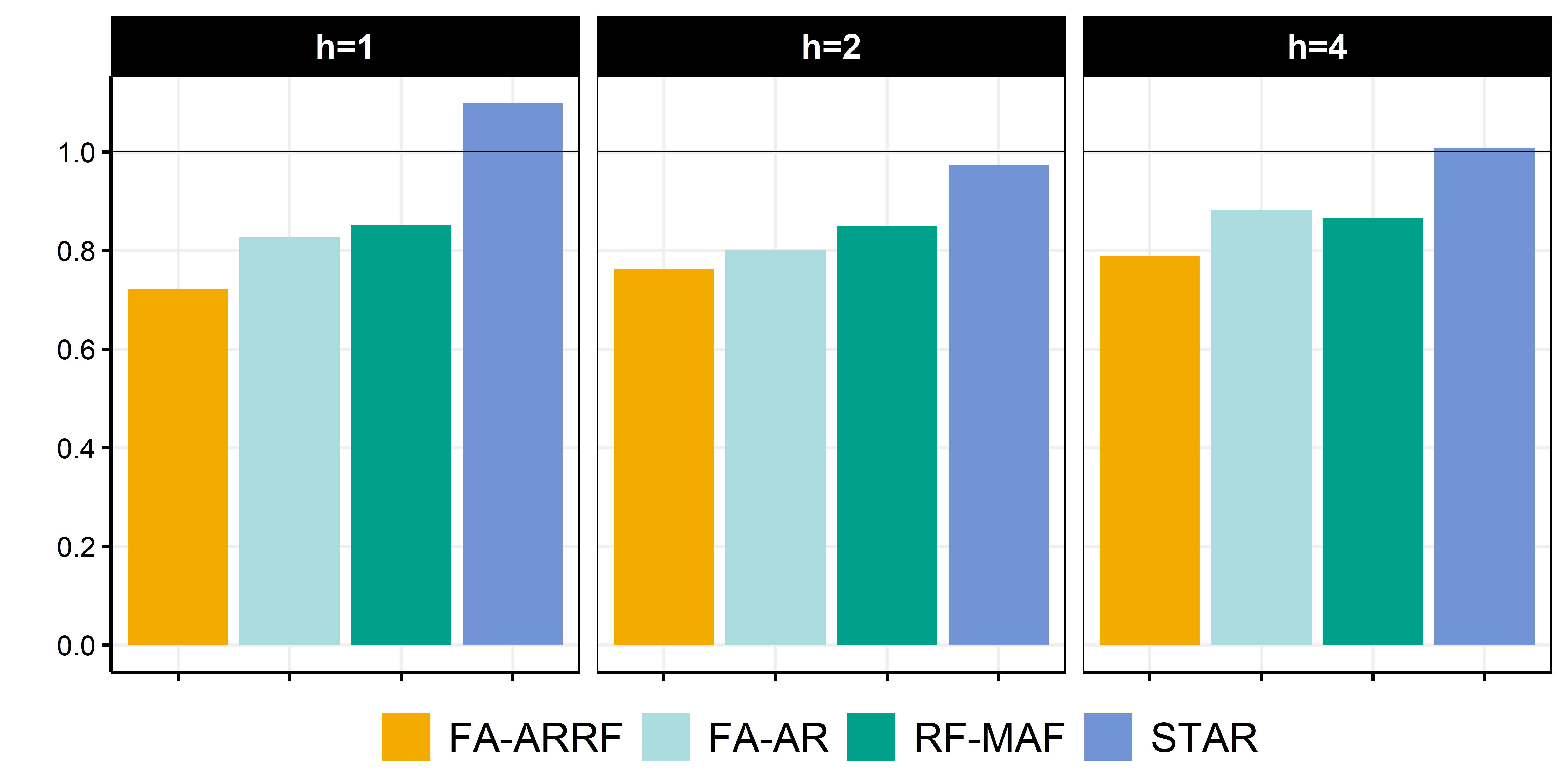}
\caption{$RMSPE_{\textcolor{black}{UR},h,m}/RMSPE_{\textcolor{black}{UR},h, AR(4)}$ }  
\end{center}
  \end{subfigure}
  \begin{subfigure}[b]{\textwidth}
    \begin{center}
\includegraphics[trim={1.5cm 0cm 0.25cm 0cm},clip,scale=.28]{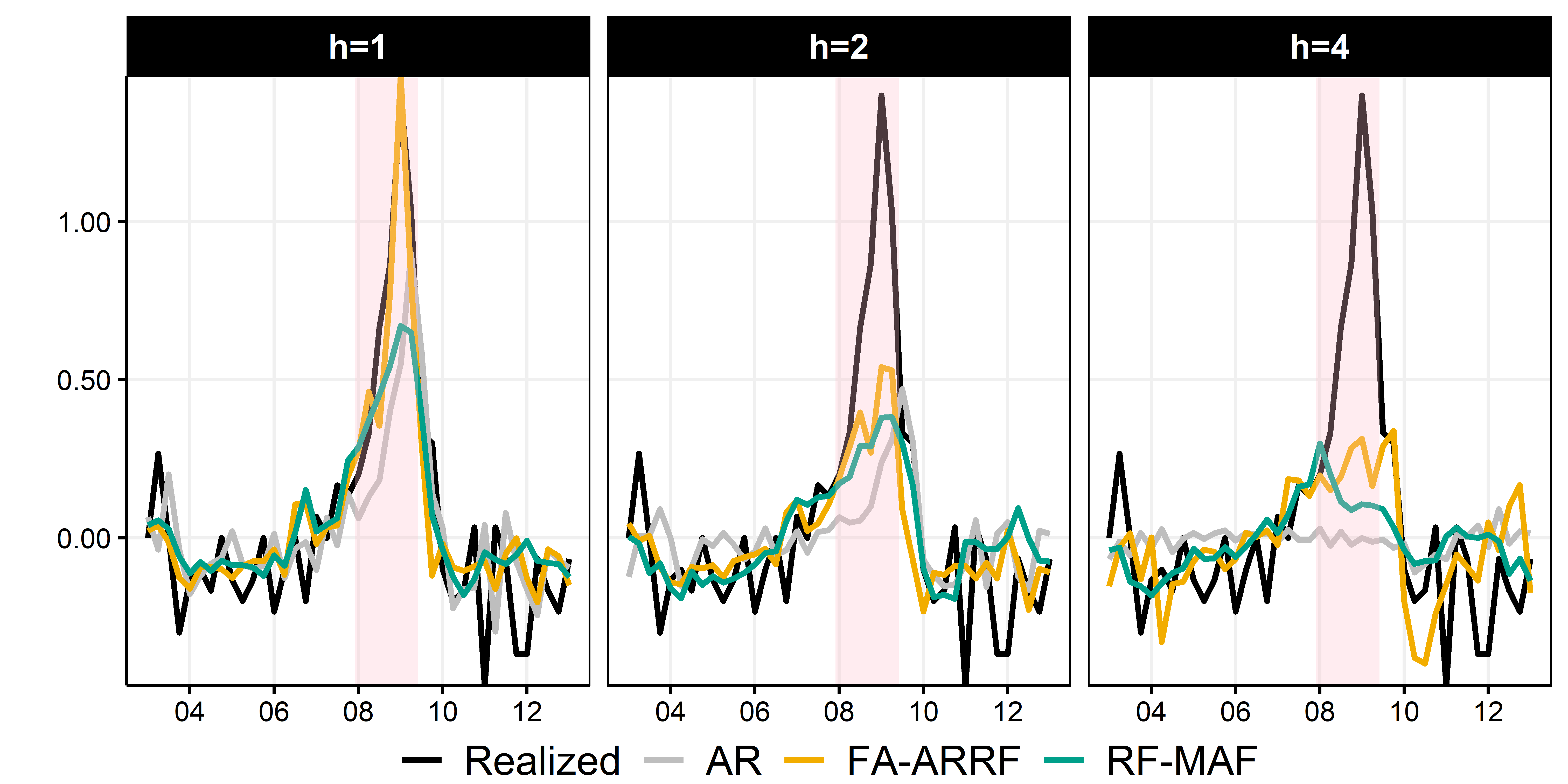}
\caption{A look at some forecasts}  
\label{UR_fpaths}
\end{center}
  \end{subfigure}
    \vspace*{-0.5cm}
  \caption{Zooming on best model within each group for UR (change)}  
\label{UR_detail}
\end{center}
\vspace*{-1cm}
\end{figure}


\vspace{0.2cm}
{\sc \noindent \textbf{Real Activity Targets.}} Figure \ref{UR_detail} reports results for UR. FA-ARRF dominates strongly. Table \ref{Q_table} confirms it is the best model for all horizons but the last one (8 quarters ahead, where the encompassed RF-MAF is the best). Clearly, at an horizon of one quarter, the preferred model successfully predicts the drastic rise in unemployment during the Great Recession. Rather than responding with a lag to negative shocks (which is what we observe from AR and ARRF), the model visibly predicts them. As a result, improvements in RMSE are between 25\% and 30\% over AR for all horizons. Specifically, predicting UR (change) with FA-ARRF at $h=1$ yields an unusually high out-of-sample $R^2$ of about 80\%.  The nearly perfect overlap of the yellow and black lines highlight the absence of a one-step ahead shock around 2008. Note that FA-AR and STAR forecasts are omitted from Figure \ref{UR_fpaths} to enhance visibility. STAR forecasts are either similar or worse than the benchmark (as often found for nonlinear time series models). FA-AR forecasts at $h=1$ follows a proactive pattern similar to the yellow line, but with a 1 to 2 quarter delay -- hence the inferior results.

For $h=2$, the quantitative rise is nowhere near the realized one, but it reveals 6 months ahead the arrival of a significant economic downturn. Additionally, ARRF and FA-ARRF both flag one year ahead the arrival of a rise in unemployment, which is a quality shared by very few models. The barplot in Figure \ref{UR_detail} (and Table \ref{Q_table}) provides a natural decomposition of FA-ARRF's gains. Adding the MAFs to an otherwise plain RF procures an improvement of roughly 15\% across all horizons (RF-MAF $\succ$ RF, in Table \ref{Q_table}). The linear FA-AR part and the rest of algorithmic modifications discussed in section \ref{sec:MRF} provide an additional reduction of 10\% to 15\% depending on the forecasted horizon (FA-ARRF $\succ$ RF-MAF and FA-ARRF $\succ$ FA-AR). It is noteworthy that good results for $h=1$ are mechanically close to impossible with a plain RF since it cannot extrapolate -- i.e., predict values of $y_t$ that did not occur in-sample. In contrast, this is absolutely feasible within MRF thanks to the linear part.

GDP is known to have a lower signal-to-noise ratio. In Figure \ref{GDP_detail}, FA-ARRF exhibits a bit less than a 20\% drop in RMSE over the AR and nicely grasp the 2008 drop one quarter ahead.\footnote{\cite{diebold1994measuring} proposed an empirically sucessful regime-switching factor model. Given that line of work and more recent results in \cite{wochner2020dynamic}, the FA-ARRF's success is not an anomaly.} However, FA-ARRF performance does not stand apart as much as it did for UR. One reason can be traced visually to predicting higher post-recession growth than its competitors. Finally, RF-MAF closing in on ARRF will be investigated on its own in section \ref{whyfail}. In short, this occurs because once the time-varying intercept is flexibly modeled by RF, there is very little room left for autoregressive behavior (at the quarterly frequency).



\vspace{0.2cm}
{\sc \noindent \textbf{Spread and Inflation.}} VARRF shines for SPREAD (Figure \ref{SPREAD_detail}) by capturing key movements, even up to a year ahead. The simpler AR+RF also does remarkably well. FA-ARRF provides successful one-year ahead forecasts.  Overall, these results highlight the common importance of the autoregressive part, which is no surprise given SPREAD's persistence. For INF, Table \ref{Q_table} displays that RF-MAF is the leading model (with ARRF close behind) reducing RMSEs by 12-15\% for all horizons. I investigate this with GTVPs in section \ref{whyfail}.





\section{Analysis}\label{sec:anal}
Based on forecasting results, I concentrate on FA-ARRF's GTVPs. Additionally, its parameters are easier to interpret (given factors are labeled) since regressors are mechanically orthogonal.
First, I look at $\beta_t$ and analyze their behavior around the Great Recession. Second, I compare GTVPs to random walk TVPs, ex-post vs ex-ante, with a focus on the recessionary episode. Finally, I use a surrogate model approach to explain of the parameters' paths in terms of observed variables.  

\subsection{Forecast Anatomy}\label{interpretation}


$\beta_t$'s characterize completely MRF's forecasts. Thus, we can investigate GTVPs to understand results from the previous section. The FA-ARRF forecasting equation is 
$$ y_{t+h}=\mu_t + \phi_{1,t}y_{t}+\phi_{2,t}y_{t-1}+\gamma_{1,t} F_{1,t}+\gamma_{2,t} F_{2,t}+u_{t+h}.$$
and naturally $\beta_t = \left[\mu_t \enskip \phi_{1,t} \enskip \phi_{2,t} \enskip \gamma_{1,t} \enskip \gamma_{2,t}\right]$. To avoid overfitting, $\hat{\beta}_t$'s are (as in section \ref{sec:newsimul}) the mean over draws that did not include observations $t-4$ to $t+4$ (a two-year block) in the tree-fitting process. Intuitively, this mimics in-sample the real out-of-sample experiment that starts here in 2007Q2.\footnote{Note that this is partially different from what gave the results reported in section \ref{RF_Q_main}, where the model was re-estimated every 2 years. Here, estimation occurs once in 2007Q2.}

\begin{figure}[ht!]
\begin{center} 
\hspace*{-1cm}\includegraphics[scale=.25]{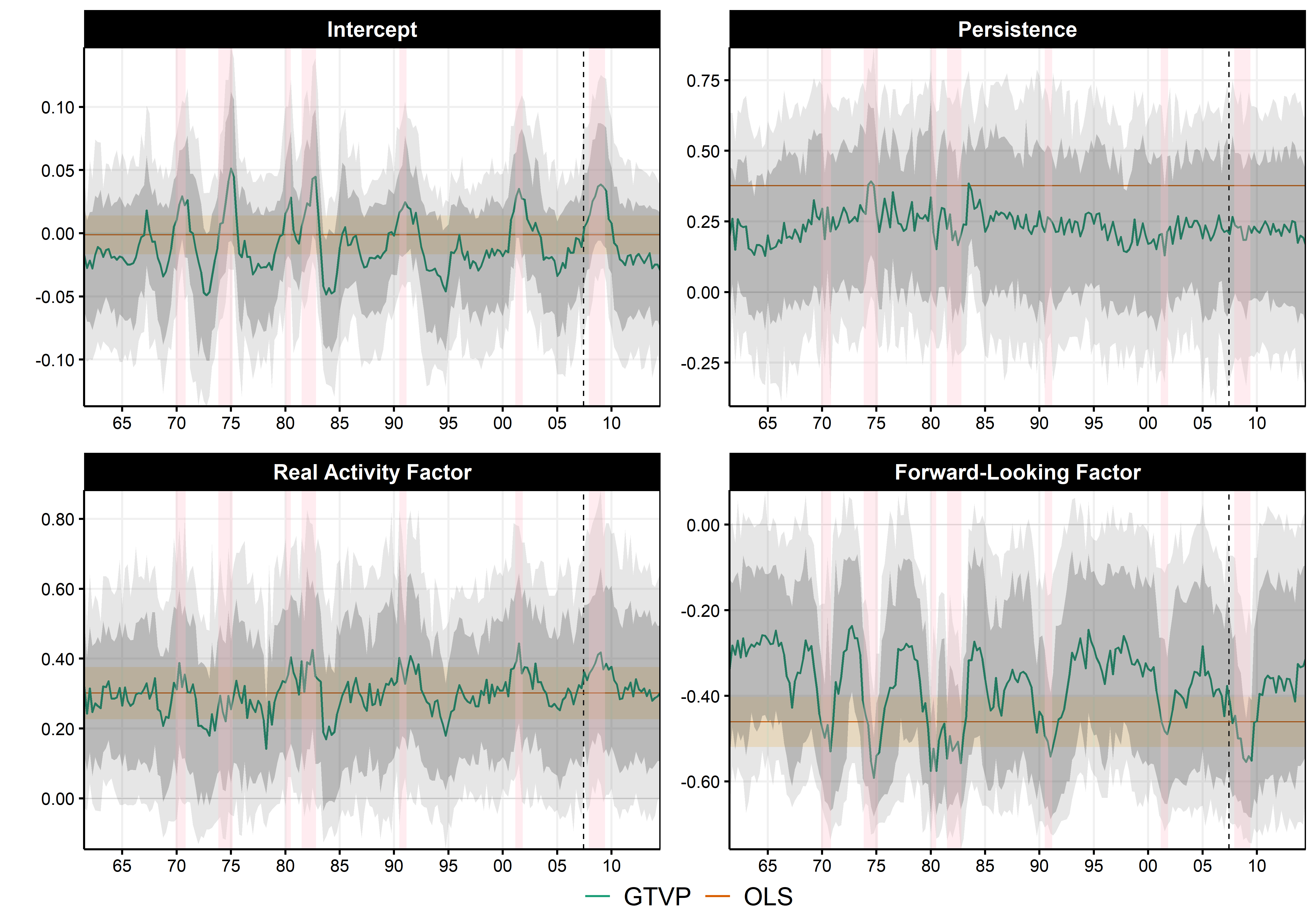}  
\caption{\footnotesize GTVPs of the one quarter ahead UR forecast. Persistence is defined as $\phi_{1,t}+\phi_{2,t}$. The gray bands are the 68\% and 90\% credible region. The pale orange region is the OLS coefficient $\pm$ one standard error. The vertical dotted blue line is the end of the training sample. Pink shading corresponds to NBER recessions. }
\label{v2h1_betas}
\end{center}
\end{figure}

Figure \ref{v2h1_betas} displays GTVPs underlying the successful one-step ahead UR \textit{change} forecast. The intercept clearly alternates between at least two regimes and the "increasing UR" one is in effect circa 2008. In levels, this translates to UR alternating between a positive and negative (albeit small) trend.  Overall persistence is strikingly time-invariant, and marginally smaller than for OLS estimates. The effect of $F_1$, the real activity factor, is generally within OLS confidence intervals, suggesting that while $\gamma_{1,t}$ almost doubles around recessions, this is subject to great uncertainty.

What is less uncertain, however, is the magnified contribution of the forward-looking factor $F_2$ during recessionary episodes, which stands out as the key difference with OLS. $\gamma_{2,t}$ smooth-switching behavior can be best interpreted by remembering that $F_2$ is highly correlated with capacity utilization, manufacturing sector indicators, building permits and financial indicators (like spreads) \citep{mccracken2020fred}. Many of those variables are considered "leading" indicators and have often been found to increase forecasting performance, mostly before and during recession periods \citep{stock1989new, estrella1998predicting,leamer2007housing}. Recently, there has been renewed attention on the matter, with financial indicators highlighted as capable of capturing economic activity downside risk \citep{adrian2019,delle2020modeling}.
This brand of nonlinearity can translate to a more active $\gamma_{2,t}$ around business cycle turning points. MRF learns that, while OLS provides a clumsy average of two regimes. In Figure \ref{v2h1_betas}, the obvious consequence of OLS' rigidity is being over-responsive to leading indicators during tranquil economic times, and under-responsive when it matters.

Section \ref{surro} will investigate formally the underlying variables driving this time variation. Figure \ref{v1h1_betas} displays equivalent $\beta_t$ for GDP one quarter ahead. The pattern  $\gamma_{2,t}$ is also visible for GDP, but it is quantitatively weaker and more uncertain -- which is is no surprise given GDP being generally noisier than UR.  Additionally, slow and relatively mild long-run change is observed. Interestingly, $\gamma_{1,t}$ has been shrinking since the mid 1980s, and its regime dependence exhibited in the first four recessions is no more.

\subsection{Comparing Generalized TVPs with Random Walk TVPs}\label{tvp_comp}

The relationship between random walk TVPs and GTVPs was evoked earlier. I compare them for the small factor model. I estimate standard TVPs using the ridge regression technique developed \cite{GC2019}. Conveniently, the procedure incorporates a cross-validation step that determines the optimal level of time variation in the random walks.\footnote{I show with simulations that this much easier approach performs similarly well (and sometimes better) to traditional Bayesian TVP-VAR, for model sizes that the latter is able to estimate.} 

\begin{figure}[ht!]
\begin{center} 
\hspace*{-1cm}\includegraphics[scale=.25]{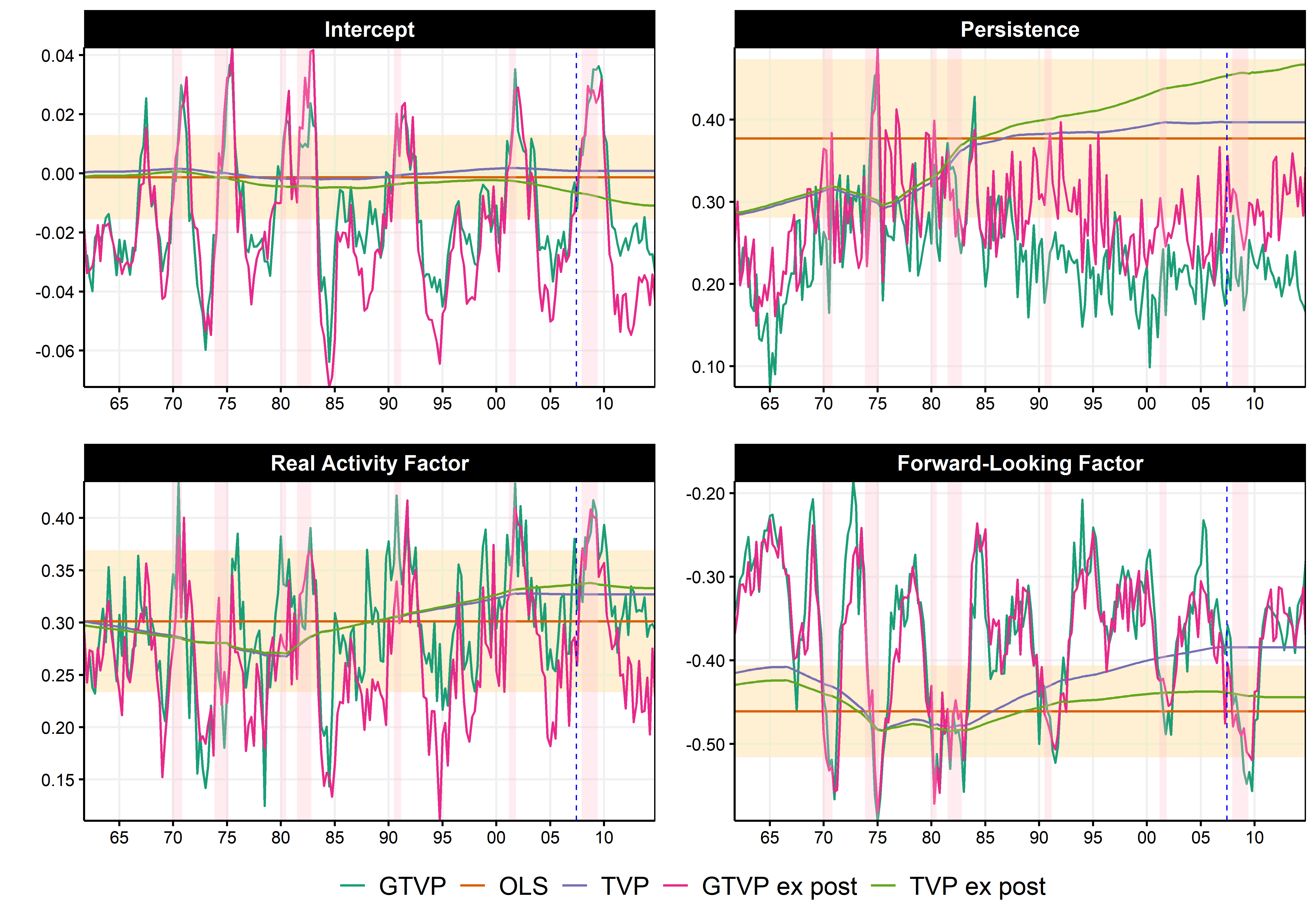}  
\caption{\footnotesize UR equation $\beta_t$'s obtained with different techniques. Persistence is defined as $\phi_{1,t}+\phi_{2,t}$. TVPs estimated with a ridge regression as in \cite{GC2019} and the parameter volatility is tuned with k-fold cross-validation --- \textbf{see Figure \ref{v2h1_tvpcomp_lambdatresdoux} for a case where TVP parameter volatility is forced to be higher}. Ex Post TVP means using the full sample for estimation and tuning as opposed to only using pre-2002 data as for GTVPs. The pale orange region is the OLS coefficient $\pm$ one standard error. Pink shading corresponds to NBER recessions.}
\label{v2h1_tvpcomp}
\end{center}
\end{figure}

As Figure \ref{v2h1_betas} suggested for $\mu_t$ and $\gamma_{2,t}$, parameters can be subject to recurrent, rapid and statistically meaningful shifts. Such behavior creates difficulties for random-walk TVPs, which put the accent on smooth and slow structural change. Figure \ref{v2h1_tvpcomp} confirms this conjecture. Standard TVPs look for long-run change when regime-switching behavior is the main driving force. As a result, they are flat and within OLS confidence bands, as often reported in the literature \citep{AAG2013}. Of course, more action will mechanically be obtained for TVPs when considering a smaller amount of smoothness than what cross-validation proposed. In appendix \ref{sec:addigraphs}, I report the same figures, but using the optimal smoothing parameters (as picked by CV) divided by 1000. This provides much more volatile random walk TVPs that are inclined, at certain specific moments, to follow the GTVPs. However, it is clear in Figure \ref{v2h1_tvpcomp} that the end-of-sample/revision problem is worsen by the forced lack of smoothing. 

It is known in the traditional TVP literature that there is a balance between flexible (but often erratic) $\beta_t$ paths and very smooth ones where time variation may simply vanish.\footnote{In the case of ridge regression-based TVPs, cross-validation is just a data-driven way of backing this necessary empirical choice.}  Since random-walk TVPs are unfit for many forms of the time-variation present in macroeconomic data, high bias estimates are usually reported as only them can keep variance at a manageable level. This can have serious implications. Relying too much on time-smoothing can create a mirage of long-run change and/or dissimulate parameters that mostly (but not solely) vary according to expansions/recessions.



Another concern, particularly consequential for forecasting, is the boundary problem. As discussed earlier, random-walk TVP models forecasts can suffer greatly from it because by construction, forecasts are always made at the boundary of the variable on which the kernel is based -- i.e., time. One can deploy a 1-sided kernel, but this only alleviate a few pressing symptoms without attacking the heart of the problem. In sharp contrast, GTVPs use a large information set $S_t$ to create the kernel, which implies that the likelihood of making a forecast at the boundary is rather low, unless the RF part constantly selects $t$ as splitting variable. 

Figures \ref{v2h1_tvpcomp} and  \ref{v1h1_tvpcomp} show, for both random walk and generalized TVPs, their full-sample versions (up to the end of 2014, "ex post") and their version with a training sample ending in 2007Q2 (the dashed blue line). There are two main observations. First, GTVPs are much less prompt to rewrite recent history than random-walk TVPs. Indeed, the green line and the magenta one closely follows each other all the way up to the end of the training sample.  Second, while GTVPs can change many quarters after 2007Q2 (like the GDP constant), they are generally very close to each other at the boundary -- especially when the time variation is statistically meaningful (like that of $\mu_t$ and $\gamma_{F_2,t}$), which is what matters for forecasting. This is much less true of random walk TVPs as there are clear examples where the two version differ for a long period of time (for instance, the intercept and the coefficient on $F_2$ in the GDP equation), and this often culminates at the boundary.\footnote{In (real) practice, all models would be re-estimated each quarter. However, it is worth pointing out that re-estimating every period is much more important for random-walk TVP than it is for GTVPs. For such reasons, the TV-AR in section \ref{forecasting} was the sole model estimated every period rather than every two years.}

\subsubsection{Why and When MRF Can Fail to Deliver Better Forecasts}\label{whyfail}
MRF can sometimes be outperformed by simpler alternatives, like standard RF that incorporate MAFs. When that occurs, it is usually due to the inadequacy of the linear part rather than GTVPs themselves. Unlike traditional TVPs, GTVPs rarely provides a model worse than OLS. 

\begin{figure}[ht!]
  \begin{subfigure}[b]{\textwidth}
\begin{center} 
\hspace{-1cm}\includegraphics[trim={0cm 1.5cm 0.1cm 0cm},clip,scale=.25]{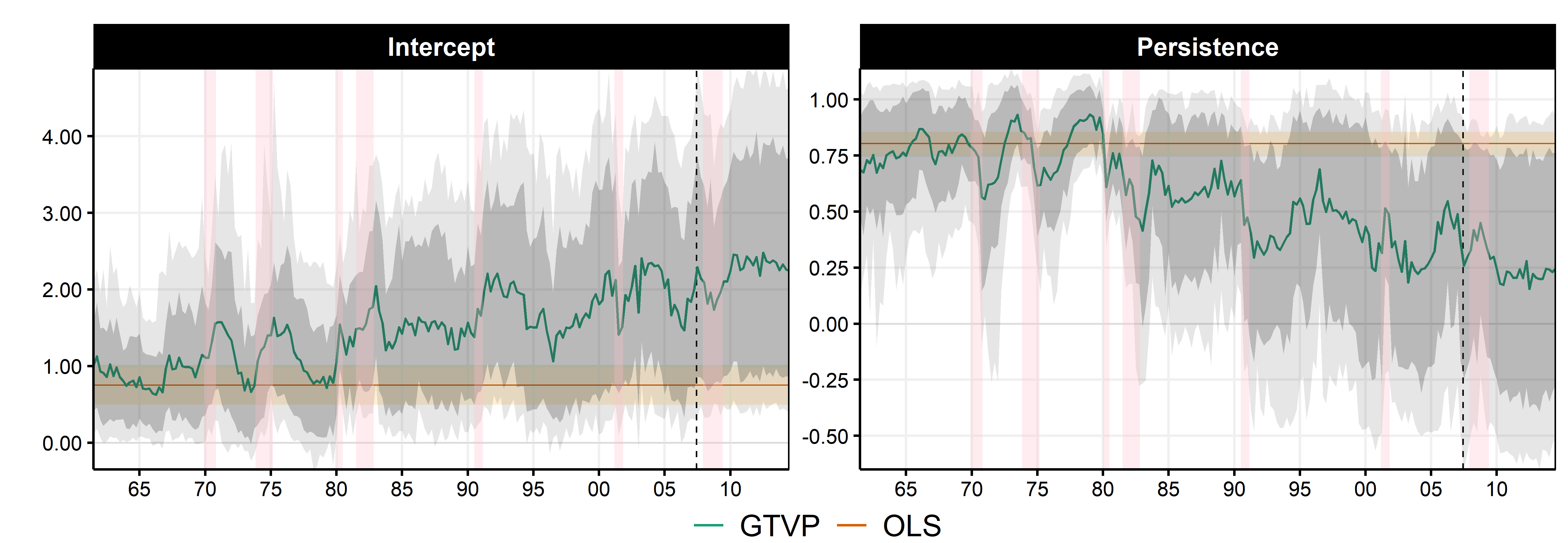}  
\caption{Inflation}
\label{v4h1_betas_arrf}
\end{center}
  \end{subfigure}
  \hspace{2em}
  \begin{subfigure}[b]{\textwidth}
\begin{center} 
\hspace{-1cm}\includegraphics[scale=.25]{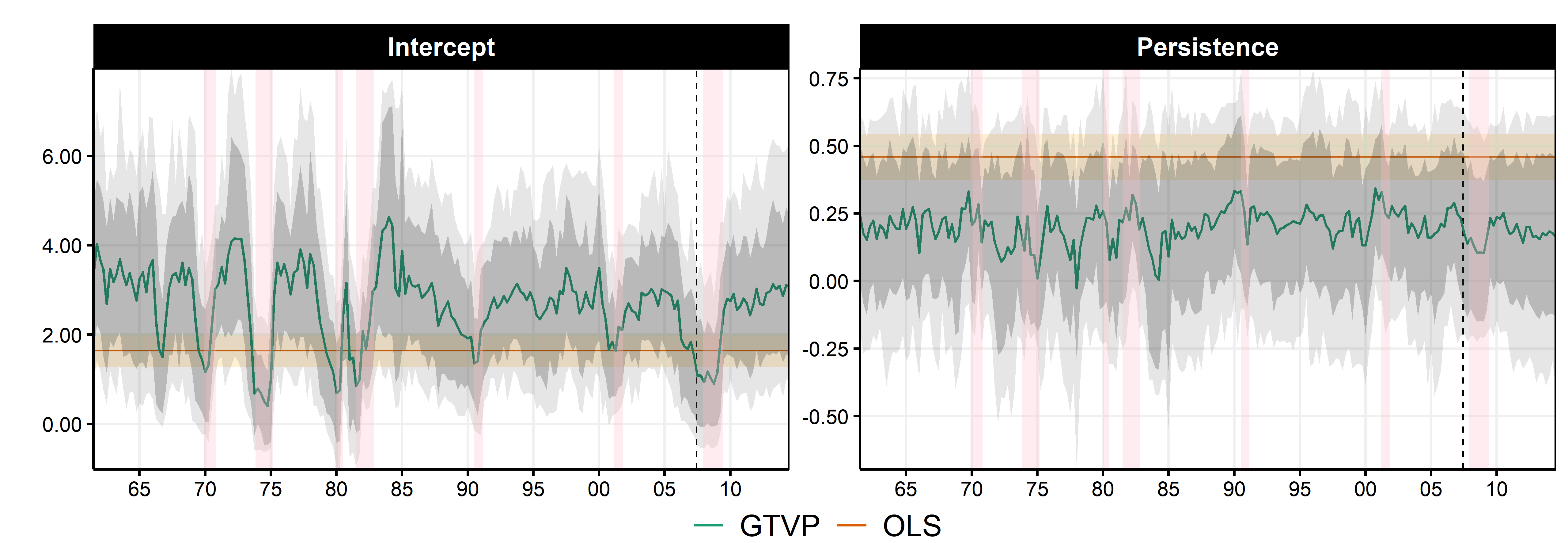}  
\caption{GDP}
\label{v1h1_betas_arrf}
\end{center}
  \end{subfigure}
\caption{\footnotesize GTVPs of the one-quarter ahead forecasts using ARRF. Persistence is defined as $\phi_{1,t}+\phi_{2,t}$. The gray bands are the 68\% and 90\% credible regions. The pale orange region is the OLS coefficient $\pm$ one standard error. The vertical dotted line is the end of the training sample. Pink shading corresponds to NBER recessions.}
\end{figure}

Trivially, $\beta_t$ helps understanding relative performance. For instance, in the case of forecasting inflation with the \textit{quarterly} data set, ARRF does not supplant RF-MAF. The critical difference between ARRF (reported in Figure \ref{v4h1_betas_arrf}) and its restricted analog is that the two autoregressive coefficients of the former are shut to 0.\footnote{Of course, lags of INF can still enter the forest part for $\mu_t$, so RF-MAF does not suppress entirely the link between current and recent inflation.} In Figure \ref{v4h1_betas_arrf}, the estimates of ARRF broadly agree with the view that inflation persistence has substantially decreased during and following Volker disinflation  \citep{cogley2001evolving,cogley2010inflation}. 

In terms of anticipated forecasting performance, such decline in persistence suggests a constrained version simply including $\mu_t$ may do better. The OOS evaluation period corresponds to the region of Figure \ref{v4h1_betas_arrf} where $\phi_{1,t}+\phi_{2,t}$ is the nearest to 0. Given that observation, RF-MAF mildly improving upon ARRF is less surprising. An analogous finding emerges for GDP at many horizons. ARRF does not outperform RF-MAF like FA-ARRF and larger VARs versions of MRF do. GTVPs showcased in Figure \ref{v1h1_betas_arrf} provide a simple explanation. There is only a limited role for persistence when allowing for a forest-driven $\mu_t$. $\phi_{1,t}+\phi_{2,t}$ is below the OLS counterpart most of the time and the credible 68\% credible region frequently includes 0. The ensuing forecast is essentially a time-varying constant, which is what RF-MAF does.\footnote{This result is largely in accord with the reported sufficiency of a switching intercept (without additional autoregressive dynamics) to model US GDP in \cite{camacho2007jump}. However, Figure \ref{v1h1_betas_arrf} suggests that there are rather 3 regimes: recession, expansion before 1985 (growth rate $\approx$ 3.5\%), expansion after 1985 (growth rate slightly below 3\%). The sufficiency of the switching intercept has also been documented in Markov-switching dynamic factor models for Norway \citep{aastveit2016identification} and Germany \citep{carstensen2020predicting}.} In sum, unlike many ML offerings, MRF successes and failures can be understood via a time-varying parameter interpretation. The helpfulness of this attribute cannot be overstated when thinking about future model improvements.  


\subsection{Cutting Down the Forest, One Tree at a Time }\label{surro}

Evolving $\beta_t$ can limit macroeconomists in their ability to use the model for counterfactuals. Complementarily, policy-makers will complain about the limited use for a model in which tomorrow's parameters are unknown (random walks). Fortunately, GTVPs may be the result of an opaque ensemble of trees, but they are made out of observables rather than a multiplicity of latent states. That is, they change, but according to a \textit{fixed} structure. Hence, the reduced-form coefficients could easily change, and yet remain completely predetermined as long as $\mathcal{F}$ itself is stable. In this paradigm, a changing $\beta_t$ is not necessarily  empirical evidence supporting \cite{lucascritique}'s critique -- rather, a changing $\mathcal{F}$ could be. Hence, dissecting $\mathcal{F}$ is inherently interesting.  One way to get started on this is to use well-established measures of Variable Importance (VI), originally proposed in \cite{breiman2001}.  Those extract features driving the \textit{prediction}. Conveniently, they can be adapted to inquire $\beta_t$. Then, one can capitalize on VI's insights to build interpretable small trees parsimoniously approximating $\beta_{t,k}$'s path.


The construction of upcoming graphs consists in two steps. I start by computing 3 different VI measures: $VI_{OOB}$ (out-of-bag predictive performance), $VI_{OOS}$ (out-of-sample predictive performance) or $VI_{\beta}$ (for a specific coefficient rather than the whole prediction).  Appendix \ref{surro_plus} contains a detailed explanation those and a discussion on how the current approach relates to recent work in the ML interpretability literature. As a potential data set for the construction of a surrogate tree, I consider the union of the 20 most potent predictors as highlighted by any of the three VIs.  The tree is pruned with a cost-complexity factor (usually referred to as \texttt{cp}) of 0.075. That tuning parameter is set such as to balance its capacity to mimic the original GTVP and potential for interpretation.





\subsubsection{Unemployment Equation}
\begin{figure}[ht!]
  \begin{subfigure}[b]{0.55\textwidth}
   \includegraphics[width=\textwidth]{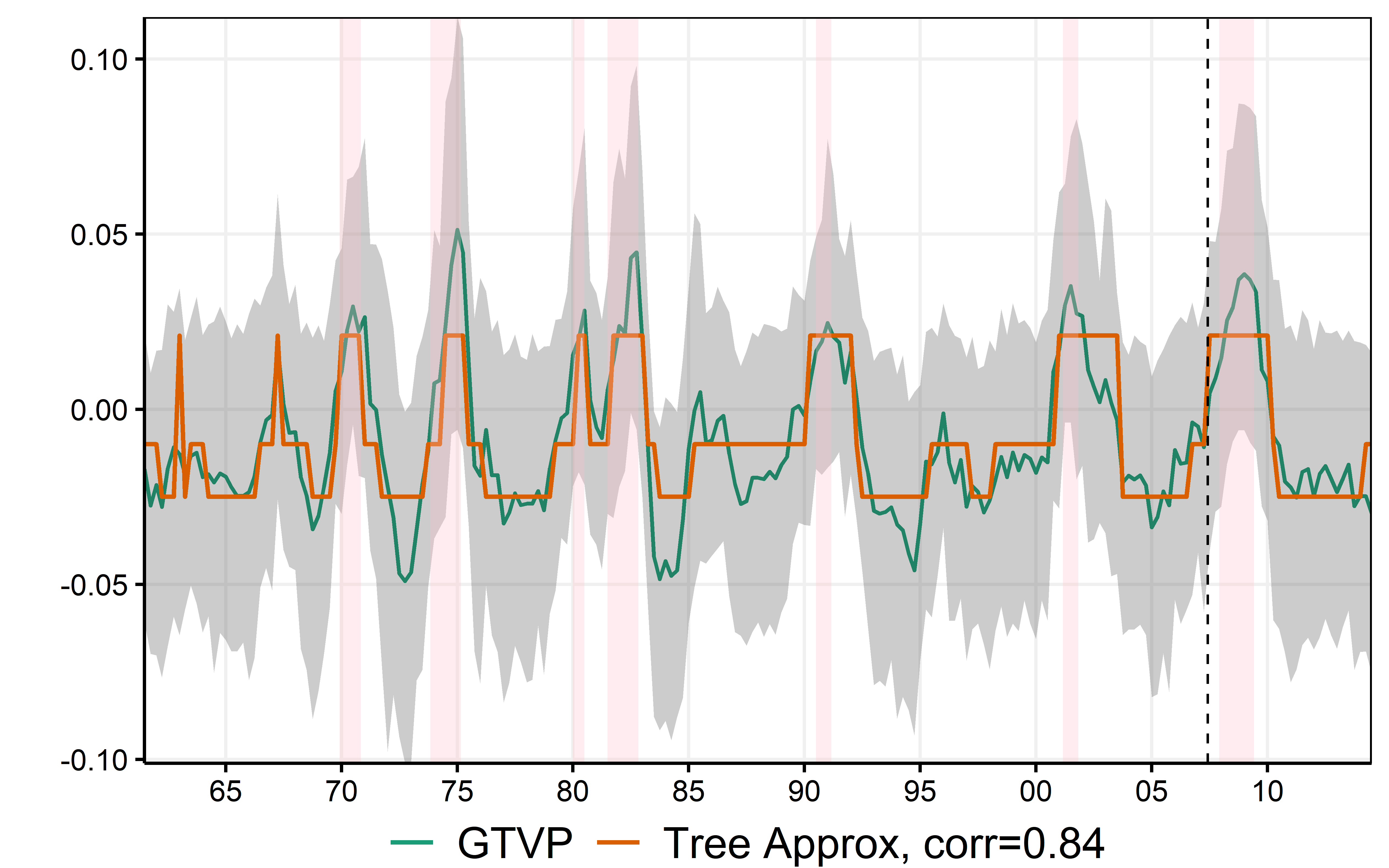}
\caption{$\mu_{t}^{UR,h=1}$: Surrogate Model Replication} 
\label{surro_v2b1_ts} 
      \end{subfigure}
  \hspace{1em}
  \begin{subfigure}[b]{0.45\textwidth}
   \includegraphics[trim={1.5cm 2.5cm 0.5cm 2.5cm},clip,width=\textwidth]{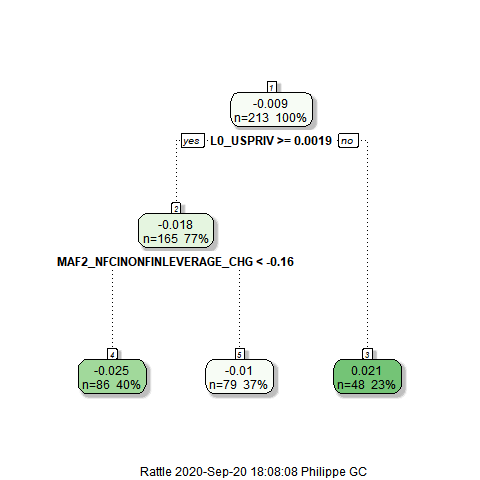}
\caption{$\mu_{t}^{UR,h=1}$: Corresponding Tree}  
\label{surro_v2b1} 
      \end{subfigure}
  \begin{subfigure}[b]{0.55\textwidth}
   \includegraphics[width=\textwidth]{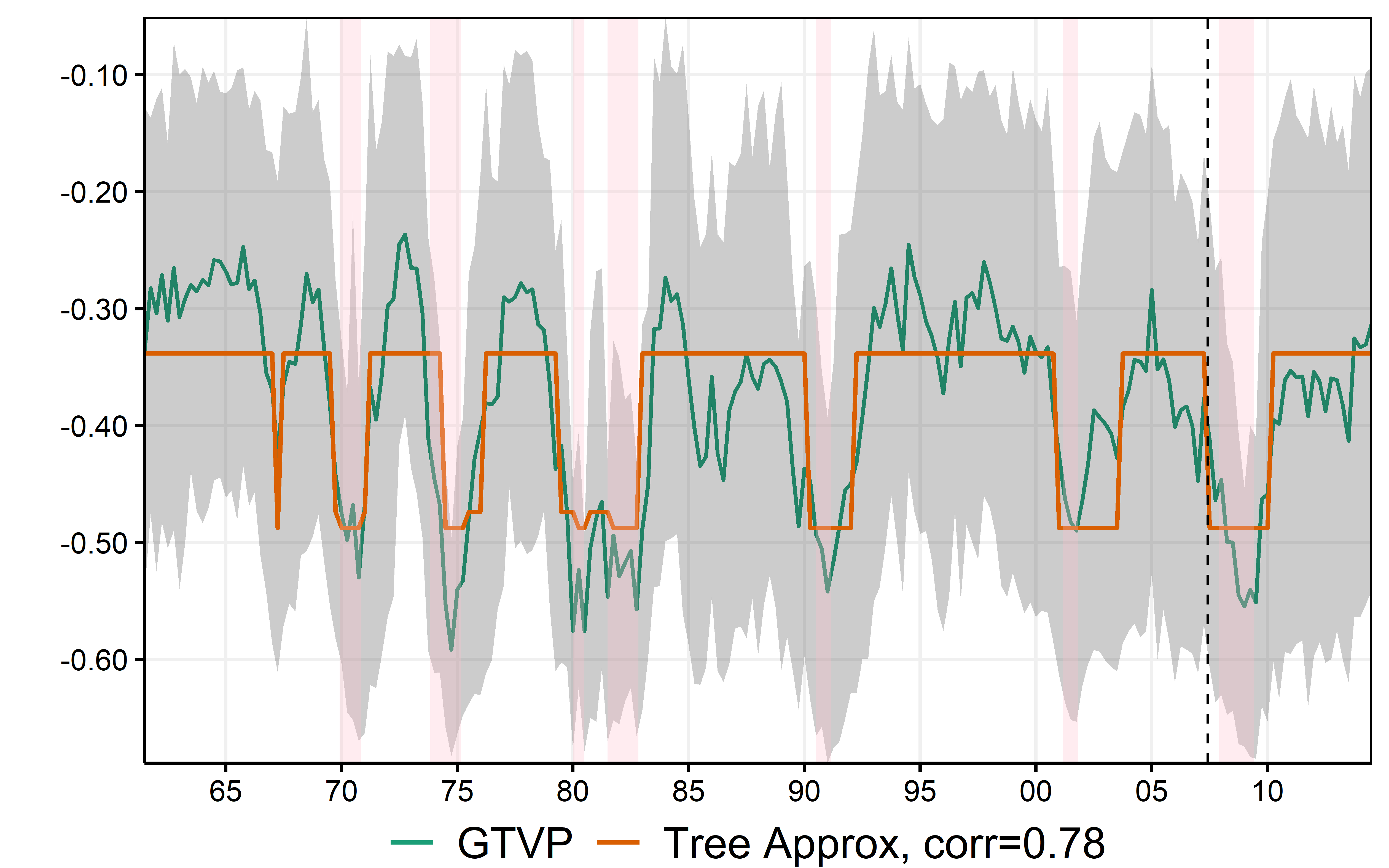}
\caption{$\gamma_{t,F_2}^{UR,h=1}$: Surrogate Model Replication} 
\label{surro_v2b5_ts}  
      \end{subfigure}
  \hspace{1em}
  \begin{subfigure}[b]{0.45\textwidth}
   \includegraphics[trim={1.5cm 2.5cm 0.5cm 2.5cm},clip,width=\textwidth]{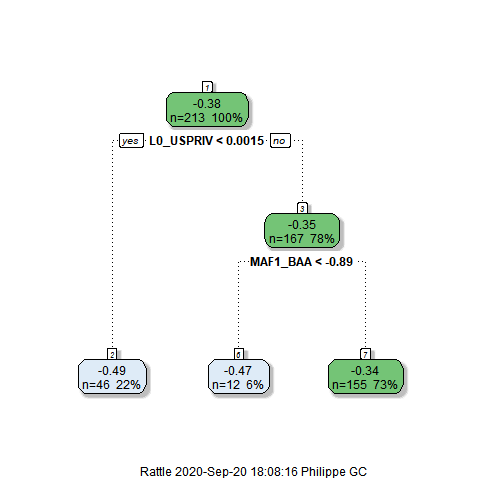}
\caption{$\gamma_{t,F_2}^{UR,h=1}$: Corresponding Tree} 
\label{surro_v2b5}   
  \end{subfigure}
  \caption{\footnotesize Surrogate $\beta_{t,k}$ Trees. Shade is 68\% credible region. Pink shading is NBER recessions.}
    \label{surro_Q}
\end{figure}

I limit the attention $\mu_t$ and $\gamma_{2,t}$ paths, which were argued of greater importance to FA-ARRF's success in forecasting UR. Also, the nature of their variation is easier to characterize with a single tree (ex-post). Figures \ref{surro_v2b1} and \ref{surro_v2b5} show that paths can sometimes be summarized succinctly using a handful of predictors. 


Most of $\mu_t$ can be captured by two states which are determined by a cut-off on total private sector employees (USPRIV): 0.021 (increasing unemployment) and -0.018 (decreasing). This first layer basically classifies recession vs expansions in a very parsimonious way, which is inevitably crude and imperfect. The additional split on a MAF of non-financial leverage provides a more refined classification: there are more of less three states. The time series plot shows the alternation between two symmetrically opposed states of 0.021 and -0.025 (respectively entering and exiting a recession) and a transitory (and seldomly visited) middle ground around 0. 

The impact of $F_2$ on UR switches significantly, and most of the action can be summarized by a private sector employees dummy (USPRIV).  The indicator's movement downwards -- which usually commence from the \textit{onset} of a recession -- can double the effect of $F_2$ on UR in absolute terms. However, some high (absolute) $\gamma_{2,t}$ episodes would be left behind when merely using USPRIV. Those are retrieved by an additional split with a MAF of average corporate bonds yield with a BAA rating (lower medium grade).

The GTVP (green line) often plunges earlier than the ex-post surrogate tree's replica (orange). This is important, especially from a forecasting perspective. In Figure \ref{VI_Q_v2h1}, it is clear that leading indicators (especially financial ones) play a prominent role in driving the GTVP $\gamma_{2,t}$ -- well before USPRIV starts showing signs of an imminent downturn.  Since $F_2$ is already composed mostly of forward-looking variables, this hints at a convex effect of market-based expectations proxies.



Lastly, a word of caution. Given the points raised earlier in section \ref{macroastrees}, it is more appropriate to see these surrogate trees as suggestive of one potential explanation. It is an open secret that their exact structure is sensitive to small changes in the estimated path. For instance, little variation in $\beta_t$ is needed to observe a change in the exact choice of variables itself. As a result, some of them may rightfully seem exotic when singled out in such a simple tree. GTVPs, as the product of a forest, will more often than not rely on a multitude of indicators from a specific group (which we observe in Figure \ref{VI_Q_v1h1}) rather than a single indicator. 

\subsubsection{Monthly Inflation Equation}\label{PCanal}

As detailed in Appendix \ref{sec:month}, FA-ARRF is a very competitive model for \textit{monthly} inflation at all horizons. By its use of $F_1$, the real activity factor, it has the familiar flavor of a Phillips' curve (PC).\footnote{As noted in \cite{stock2008phillips}, the plethora of output gap indicators used in literature makes the use of a common statistical factor a credible alternative.} This is of interest given PCs have at best a very uneven forecasting track record \citep{atkeson2001phillips,stock2008phillips,faust2013forecasting}. For instance, simple autoregressive/random walk/historical mean benchmarks often do much better. 

Given its paramount importance within New Keynesian models, many explanations have been proposed for PC forecasts failures.  The curve could be time-varying in a way that annihilates its forecasting potential \citep{stock2008phillips}. Closely related, some have stipulated the PC is nonlinear \citep{dolado2005monetary,doser2017inflation,linde2019resolving,mineyama2020downward}. If that were to be true, this should be exploitable. Lastly, an adjacent point of view, which became increasing popular following the Great Recession, is that the PC has irreversibly flattened to the point of predictive desuetude \citep{blanchard2015inflation,blanchard2016phillips,del2020s}. Unlike the first two propositions, this one is, by nature, terminal.


Of course, all those explanations amount to hypotheses on the nature of $\gamma_{1,t}$'s time variation, of which MRF provides a very flexible account. It is worth emphasizing that MRF is estimated up to 2007Q2, unlike many of the above models explaining the "missing disinflation" after observing that it took place.\footnote{Indeed, they do so either by fitting the post-2008 data directly, or by choosing a specification (or building a theoretical model) directly inspired by the experience of the Great Recession.}  The variable importance measures reported in Figure \ref{VI_M} showcase a  "consensus" subset of variables that matters for inflation time variation. Three popular ones are the trend, MAF of building permits and MAF of housing starts.  The leading role for the trend suggests that exogenous time variation is important to explain inflation -- to no one's surprise \citep{cogley2001evolving}. Studying $\beta_t$-specific VI's suggest that this is mostly a feature of the intercept and persistence.

\begin{figure}[h!]
  \begin{subfigure}[b]{0.55\textwidth}
   \includegraphics[trim={0.5cm 0cm 0cm 0cm},clip,width=\textwidth]{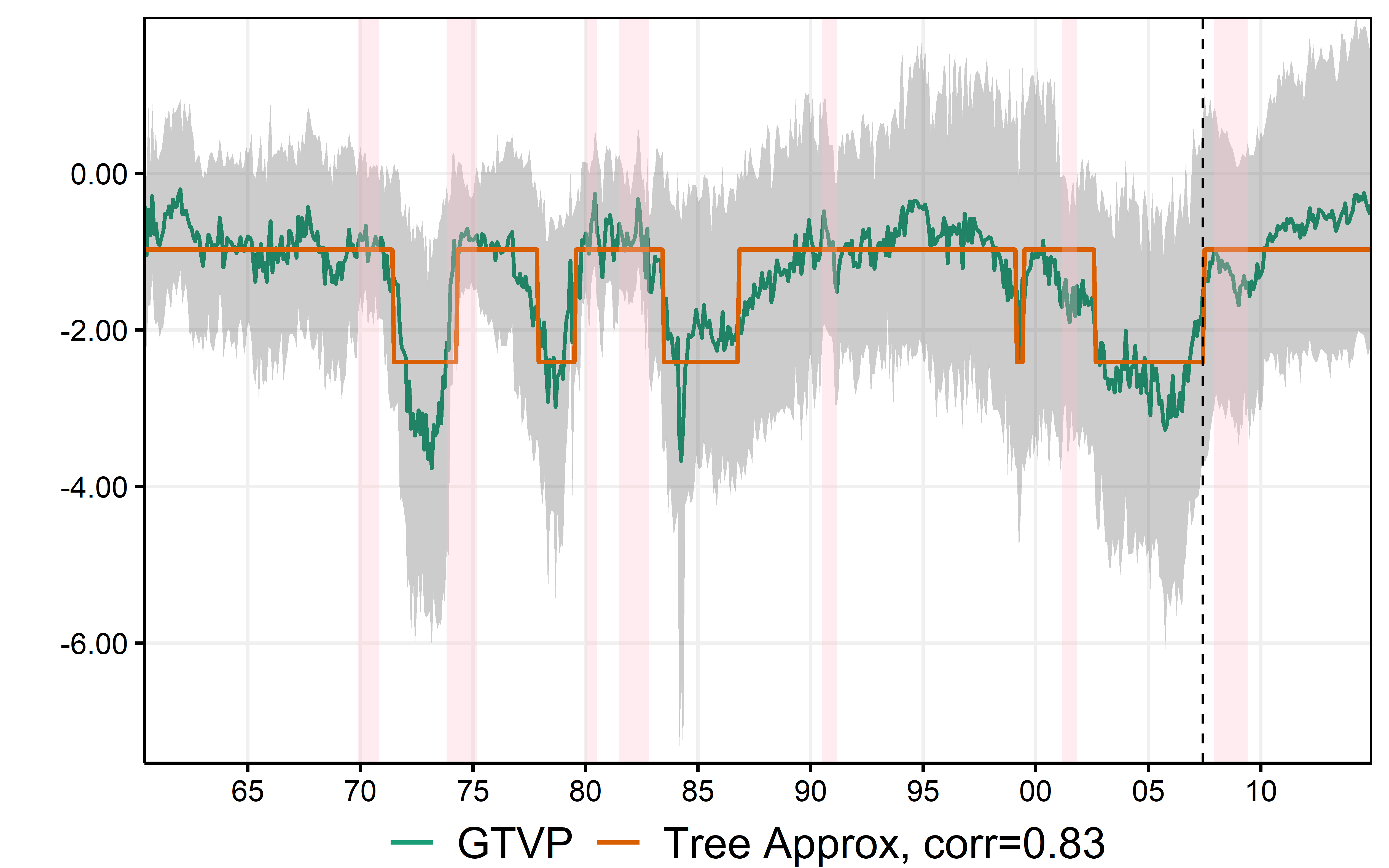}
\caption{$\gamma_{1,t}^{INF,h=1}$: Surrogate Model Replication} 
\label{surro_M_v4b4_ts} 
      \end{subfigure}
  \hspace{1em}
  \begin{subfigure}[b]{0.45\textwidth}
   \includegraphics[trim={0.5cm 2cm 0.5cm 1cm},clip,width=\textwidth]{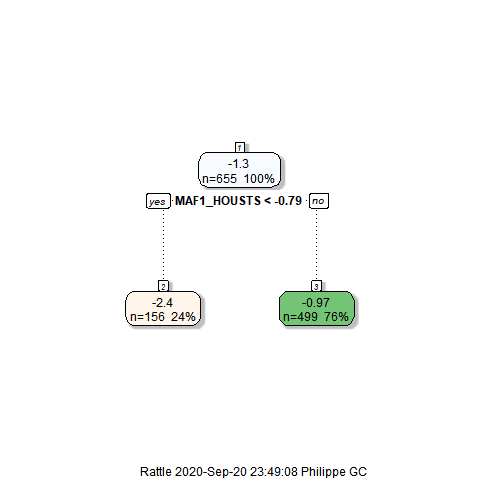}
\caption{$\gamma_{1,t}^{INF,h=1}$: Corresponding Tree}  
\label{surro_M_v4b4} 
      \end{subfigure}
  \begin{subfigure}[b]{0.55\textwidth}
   \includegraphics[trim={0.5cm 0cm 0cm 0cm},clip,width=\textwidth]{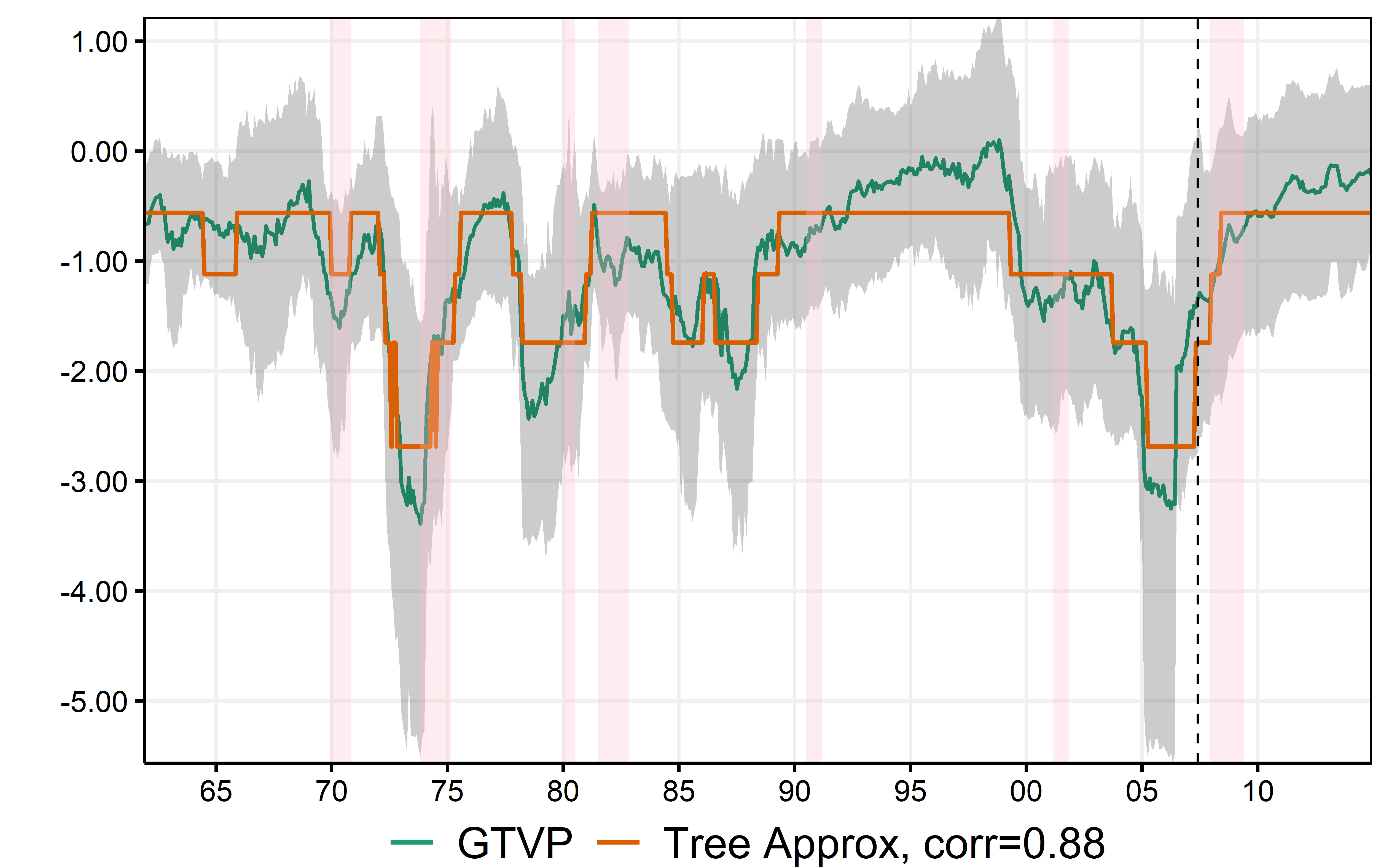}
\caption{$\gamma_{1,t}^{INF,h=12}$: Surrogate Model Replication} 
\label{surro_M_v4b4_ts}  
      \end{subfigure}
  \hspace{1em}
  \begin{subfigure}[b]{0.45\textwidth}
   \includegraphics[trim={0.5cm 2cm 0.5cm 1cm},clip,width=\textwidth]{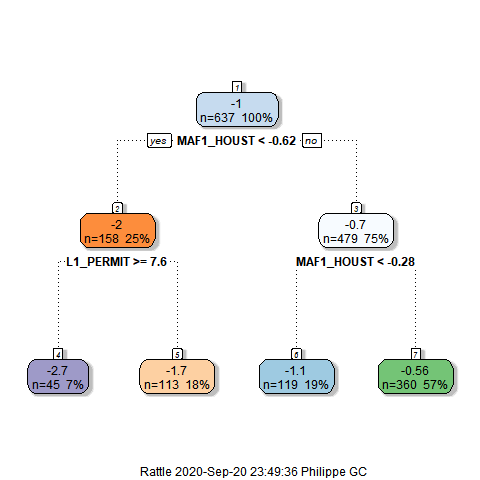}
\caption{$\gamma_{1,t}^{INF,h=12}$: Corresponding Tree} 
\label{surro_M_v4b4}   
  \end{subfigure}
  \caption{\footnotesize Surrogate $\beta_{t,k}$ Trees for Inflation. Shade is 68\% credible region. Pink shading is NBER recessions.}
    \label{surro_M}
\end{figure}

Figures  \ref{surro_M}, \ref{v4h1_betas_M} and \ref{v4h4_betas_M} allow to re-conciliate PC forecasting evidence. For instance, a visible PC death zone spans all of the 90s, which constitutes most of the sample used in \cite{atkeson2001phillips}.\footnote{The decade-long wedge between the OLS estimate and GTVP in Figure \ref{v4h4_betas_M} nicely explains PC failures.} It also includes the post-2008 period, which motivated \cite{blanchard2015inflation}'s inquiry. Most interestingly, for the latter era, $\gamma_{1,t}$ is predicted to head toward 0 \textit{out-of-sample}. To clarify, the parameter is driven by post-2008 data, but the structure itself ($\mathcal{F}$) is not re-evaluated past the dotted line.


By looking at predictive performance results \textit{ex-post}, \cite{stock2008phillips} report that Phillips' curve forecasts usually outperform univariate benchmarks around turning points,  but suffer a reversal of fortune when the output/unemployment gap is close to 0. They note that the finding "cannot yet be used to improve forecasts" because their gap relies on a two-sided filter. More recently, \cite{KLS2019} reinforce this view by showing an ARMA(1,1) is triumphant for inflation \textit{except} in recessionary periods, where a data-rich environment can be helpful. But to capitalize on this, one needs a recession/expansion forecast.  MRF recognize this  potential and relies on leading indicators of the housing market to activate $\gamma_{1,t}$ in a timely manner. This is particularly evident from looking at $\gamma_{1,t}$'s VI measure in Figure \ref{VI_M} and its resulting GTVP in Figure \ref{surro_M}. Overall, we see that the relationship between inflation and economic activity is episodic, as conjectured by \cite{stock2008phillips}, and often prevails before recessions (but not all). Figure \ref{surro_M} proposes a clear-cut answer: inflation responds to the real activity factor when the housing market is booming.

For a long time, housing sector indicators have been known as predictors of future economic activity \citep{stock1998business,leamer2007housing}.  However, when it comes to forecasting inflation itself, including leading indicators (like permits) does not remedy Phillips' curve forecasts failures \citep{stock2007has}. FA-ARRF differs by \textit{not} using housing permits/starts as a replacement and/or additional output gap proxy. Rather, its role is to increase the curvature when the time is right. As mentioned above, one explanation is that housing starts and permits are proxying for future economic activity, resolving the  conundrum posed by  \cite{stock2008phillips}. Overall, this implies a PC which would be highly nonlinear in real activity, as further inquired in section \ref{PC2}. Another hypothesis is that MRF discovers -- through aggregate data -- how to leverage \cite{stock2019slack}'s insights that some components of inflation are much more cyclically sensitive than others. \cite{stock2019slack} show that the most cyclical component of inflation is \textit{housing}, followed closely by food components.  Accordingly, MRF activating $\gamma_{1,t}$ with building permits and housing starts is the algorithm's way of predicting when more cyclically sensitive components take the front stage -- and by doing so, revive the Philipps' curve. In sum, nonlinearities would be a consequence of aggregation.





The predictive PC studied here differs in many aspects from those studied, for instance, in \cite{blanchard2015inflation}. Importantly, $F_1$ summarizes mostly variables in first differences (or growth rates). A typical gap measure, being a deviation from a trend, will be much more persistent. Also, it remains negative for many years following a downturn. In contrast, $F_1$, which is strongly correlated with UR change, will go back up as soon as UR stops growing.  To validate current insights and obtain new ones, I now study a prototypical Phillips' Curve.


\subsection{The Phillips' Curve: Not Dead Yet?}\label{PC2}

The behavior of inflation since the Great Recession -- starting with the missing disinflation and followed by "missing inflation" of recent years -- sparked renewed interest in the Phillips curve. Much attention has been given to its hypothesized flattening  \citep{blanchard2015inflation,galigambetti2019, del2020s}. This body of work supports the view that the PC coefficient (either reduced-form or semi-structural) has substantially declined over the last decades. The focus on slow structural change is operationalized by the modeling strategy -- either random walk TVPs or sample splitting at a specific date.  \cite{coibion2015phillips} show less worry about PC's health. They rationalize post-2008 inflation with a simple OLS PC where expectations are based on consumer survey data rather than lags or professional forecasters. \cite{del2015inflation} demonstrate that a standard DSGE (which encompasses a structural New Keynesian PC) is not baffled by post-2008 inflation since it relies on model-based forward-looking expectations of future marginal cost. More recently, \cite{linde2019resolving} and \cite{mineyama2020downward} articulate theories supporting a nonlinear specification for the reduced-form PC, which could also account for the inflation puzzles punctuating the last 12 years. Given this background and forecasting results reported earlier, a traditional PC must be a fertile ground for MRF-based detective work. 


I contribute to the literature by fitting an MRF which linear part corresponds to an expectations-augmented Phillips' curve. $X_t$ is inspired by what \cite{blanchard2015inflation} (henceforth BCS) considers:
\begin{align}\label{eq:bcs1}
\pi_t = \theta_t \hat{\pi}_t^{LR} + (1-\theta_t) \hat{\pi}_t^{SR}+\phi_t {u}_t^{GAP} + \psi_t {\pi}_t^{IMP}+ \epsilon_t, 
\end{align}
where $\pi_t $ stands for CPI inflation, $\hat{\pi}_t^{LR}$ and $\hat{\pi}_t^{SR}$ respectively for long-run and short-run inflation expectations. ${u}_t^{GAP}$ represents the (negative) unemployment gap and ${\pi}_t^{IMP}$ is import prices inflation. I translate this to the MRF framework by making $\mu_t = \theta_t \hat{\pi}_t^{LR}$ the time-varying intercept, letting $\beta_{t,1} = 1-\theta_t$ and by obtaining ${u}_t^{GAP}$ by means of Hodrick-Prescott filtering.\footnote{Specifically, both this gap and that of BCS get out of negative territory around 2014.}  As in BCS, $\hat{\pi}_t^{SR}$ is the average inflation over the last four quarters. Hence, the estimated equation
\begin{align}\label{eq:bcs2}
\pi_t = \mu_t + \beta_{1,t} \hat{\pi}_t^{SR}+\beta_{2,t} {u}_t^{GAP} + \beta_{3,t}{\pi}_t^{IMP}+ \varepsilon_t 
\end{align}
does not impose the constraint implied by $\theta_t$ in equation \eqref{eq:bcs1}. However, estimation results will desirably have $\beta_{1,t} \in [0,1]$ at almost any point in time. $S_t$ is the same as that considered in the forecasting section. The data set runs up to 2019Q4.
\begin{figure}[ht!]
  \hspace{2em}
\begin{center} 
\hspace{-0.75cm}\includegraphics[scale=.28]{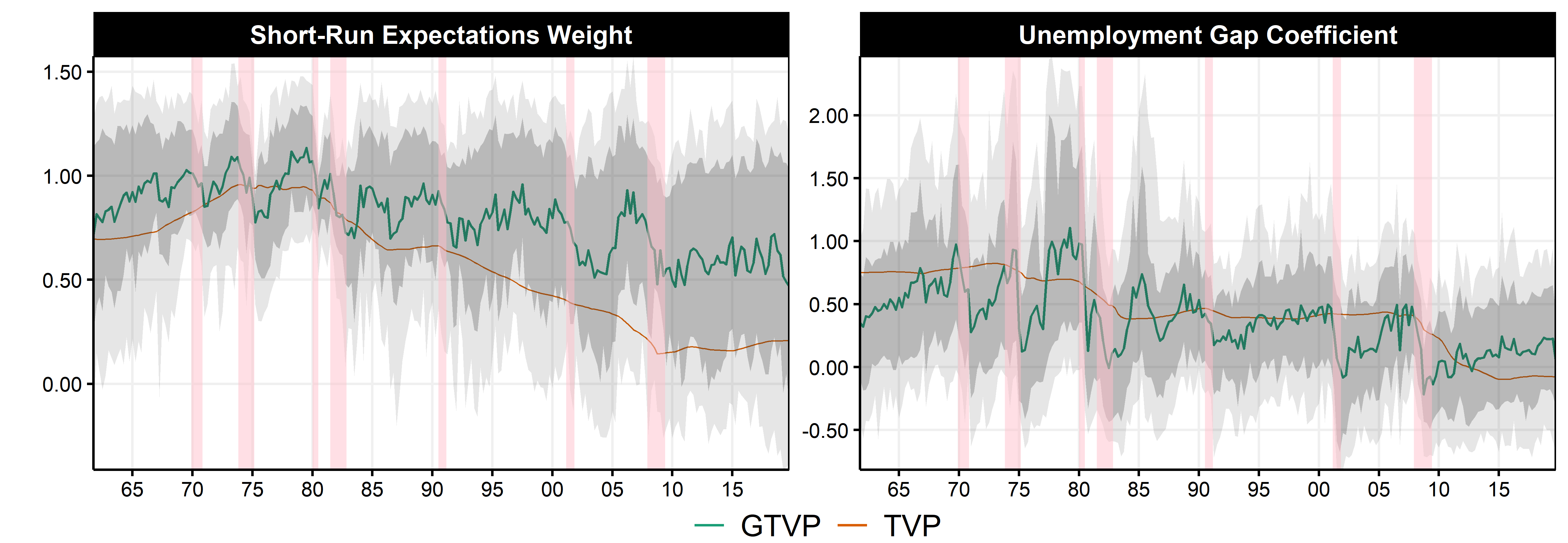}  
\end{center}
\vspace{-0.5cm}
\caption{\footnotesize The gray bands are the 68\% and 90\% credible regions. Pink shading corresponds to NBER recessions.}
\label{beta_bcs}
\end{figure}


Figure \ref{beta_bcs} reports GTVPs of interest: the weight on short-run expectations and the output gap coefficient. Additionally, it contains traditional TVP estimates as means of comparison. The latter convey the usual wisdom: inflation expectations slowly start to be more anchored from the mid 1980s. Around the same time, the unemployment/inflation trade-off begins its slow collapse. The updated data shows that the TVP-based Phillips' curve has further flattened to plain 0 in the last decade. 

For $\beta_{1,t}$, the weight on short-run expectations, both methods agree that it has been decreasing steadily after the 1983 recession.  But GTVPs highlight an additional pattern for the importance of $\hat{\pi}_t^{SR}$: it tends to increase during economic expansions, collapse during recessions then start increasing again until the next downturn. Note that the phenomenon is also observed in Figure \ref{v1h1_betas_arrf} for the simpler ARRF on quarterly inflation. The decrease in the coefficient (usually of about 0.25) is observed for \textit{every} recession and usually last for some additional quarters after the end of it. The linear rise in the coefficient occurs for all expansions except those preceding the early 90s and 2000s recessions, where the pattern is punctuated with additional peaks and troughs. The increased importance of short-run expectations with the age of the expansion is also observed for recent expansionary periods. Hence, the phenomenon is not merely a matter of the 70s and 80s recessions being preceded by a sharp acceleration of inflation. 

From a more statistical point a view, the sharp decline in $\beta_{1,t}$ following every recession suggests that in the aftermath of an important downward shock, the long-run inflation expectation is a more reliable predictor as it is minimally affected by recent events.  As the expansion slowly progress (and recessionary data points get out of the short-run average), $\hat{\pi}_t^{SR}$ becomes a more up to date and reliable barometer of future inflation conditions. This narrative is corroborated by variable importance (Figure \ref{VI_BCS}) for $\beta_{1,t}$, which highlights the importance of the trend, but also recent lags of inflation.


When it comes to the low-frequency movement of the unemployment gap coefficient, both methods agree about a significant decline starting from the 80s. However, GTVPs uncover additional heterogeneity. \textbf{First} and most strikingly, $\beta_{2,t}$ gets very close to 0 following every recession. This suggests a nonlinear Philipps' curve where inflation responds strongly to a very positive ${u}_t^{GAP}$ but not so much to a negative one. \textbf{Second}, the 70s and early 80s are characterized as a series of peaks (preceding the first three recessions of the sample) rather than a sustained high coefficient. Traditional TVPs, by excessive time-smoothing, dissimulate the effects of inflationary spirals on $\beta_{2,t}$. Such pre-recession accelerations still occur during the Great Moderation but in a much milder way. 

\begin{figure}[ht!]
\begin{center} 
\vspace{-0.05cm}
\hspace{-0.5cm}\includegraphics[scale=.3]{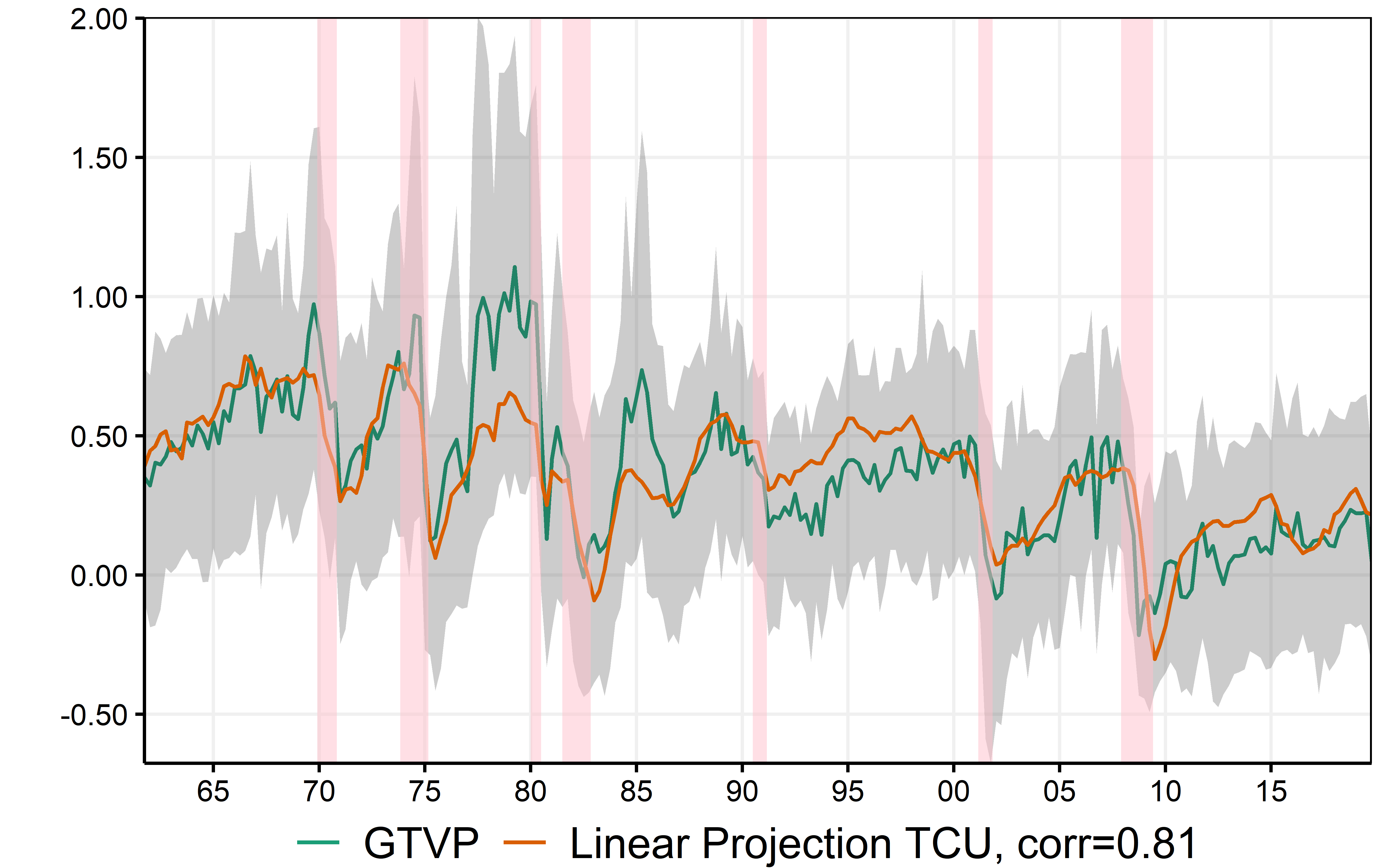}  
\end{center}
\vspace{-0.6cm}
\caption{\footnotesize "What Goes Around Comes Around": Capacity Utilization is substantially correlated with the inflation-unemployment trade-off. The gray band is the 68\% credible region. Pink shading corresponds to NBER recessions.}
\label{beta_bcs_tcu}
\end{figure}

\textbf{Third}, VI measures (in Figure \ref{VI_BCS}) confirm the importance of activity indicators (like Total Capacity Utilization (TCU)) in driving $\beta_{2,t}$ itself. The correlation between $\beta_{2,t}$ and TCU is 0.81, and the correspondence between the two variables is striking in Figure \ref{beta_bcs_tcu}. Many notable increases in $\beta_{2,t}$ are nicely matched (between the two 70s recessions and before 2008). Of course, this simple characterization remains imperfect since it misses some highs (like the end of the 70s) and predicts a higher $\beta_{2,t}$ in the years following the 2008-2009 recession. Generally, given the strong co-cyclicality between TCU and ${u}_t^{GAP}$, this is evidence of a \textit{convex} PC. 

The collapse of $\beta_{2,t}$ following recessions is not unique to 2008: it happened following \textit{every} recession since 1960. As a result, inflation will rise when the economy is running well above its potential, but much more timidly will it go down from economic slack. Recently, \cite{linde2019resolving} have shown that such a phenomenon can be rationalized by a New Keynesian DSGE model. Indeed, by allowing for additional strategic complementarity in firms price- and wage-setting behavior and solving the nonlinear model (rather than considering the linear approximation around the steady state), the authors obtain a state-dependent PC which becomes very flat during large downturns. This can explain both the small coefficient during recessions and its subsequent timid increase.  Theoretically, convexity can also emerge from downward wage rigidities \citep{mineyama2020downward}, but its empirical plausibility for the post-2008 era has been contested \citep{coibion2015phillips}.



This pattern remains when adding controls in the linear part for supply shocks and monetary policy shocks. Those are the usual confounding factors suspected of blurring the relationship by introducing a positive correlation between unemployment and inflation.\footnote{While the time-varying constant can go a long way at controlling for such factors -- being a RF in itself, including them in the linear part makes them "stand out" as everything going through the intercept is inevitably heavily regularized.} The economic suspicion particular to this application is that omitting them could create a downward bias in $\beta_{2,t}$ that only occurs locally, generating the cyclical pattern. As it turns out, controls make cyclicality even more obvious in Figure \ref{beta_bcs_tcu_controls}, especially for the later part of the sample.\footnote{Results being similar for both curves is reminiscent of \cite{galigambetti2019} who report little differences between paths of reduced-form and semi-structural wage PCs (although they focus on long-run change).} However, the overall strength of the coefficient is smaller (especially for the 70s).

Many hypotheses can be accommodated by a model estimated on two disjoint samples, like in \cite{del2020s}. Much fewer of them are compatible with the richer $\beta_{2,t}$ path extracted by MRF. This is important: learning the type of nonlinearity, rather than partially imposing it, helps in discriminating economic suppositions. Figure \ref{beta_bcs_tcu} and recent theoretical developments both suggest that much of the PC's decline is attributable to upward nonlinearities  being less solicited in the last 3 decades. This is in accord with the policy hypothesis: since Paul Volker's chairmanship the monetary authority has responded much more aggressively to inflationary pressures, limiting the spirals that gave rise to high $\beta_{2,t}$'s in the 70s. Two conclusions emerge from this observation. First, exogenous change cannot so simply be ruled out. Second, knowing what were MRF beliefs about PC nonlinearities at different points in time could be enlightening.

\subsubsection{Conditional Coefficient Forecasting}\label{CCF}

$\beta_{2,t}$'s lows are getting lower, and longer. Should we have known? Much of the recent work on PC is directly inspired by Great Recession aftermath, and aims at explaining it. Whether it is theoretical or empirical work, much of it could be overfitting: a model can replicate one or two facts it is trained to replicate, but fails to generalize. That is, even if models are tested out-of-sample (which is itself not so often the case in the literature), the choice of nonlinearity itself is often determined in attempt to match the OOS. Beyond the linear part being a PC, MRF does not assume much --- and its nonlinearities are certainly not "personalized" to the recent inflation experience. Thus, it is interesting to ask: what was MRF "thinking" about $\beta_{2,t}$ in 2007? in 1995? Did it know something we did not, or did it learn (as most economists) of PC's collapse from the post-2008 experience? I conduct a $\beta_{2,t}$ dynamic learning exercise to find out.

To make this operational, MRF is estimated up to 1995, 2007 and 2019, and GTVPs are projected out-of-sample from those dates (when applicable). To be clear, $\hat{\beta}_{2,t|1995}=\hat{\mathcal{F}}_{1995}(S_t)$ means the coefficient \textit{predictive structure} is last estimated in 1995. Coefficients keep moving out-of-sample because $S_t$ does. $\hat{\mathcal{F}}_{1995}(S_t)$ and $\hat{\mathcal{F}}_{2007}(S_t)$ will differ for two main reasons. The first is estimation error -- both in terms of precision and re-evaluating which nonlinearity seems more appropriate.\footnote{The second part has the flavor of model selection "error".} The second is structural change, perhaps completely exogenous or triggered by policy interventions.

\begin{figure}[ht!]
\begin{center} 
\vspace{-0.05cm}
\hspace{-0.5cm}\includegraphics[scale=.3]{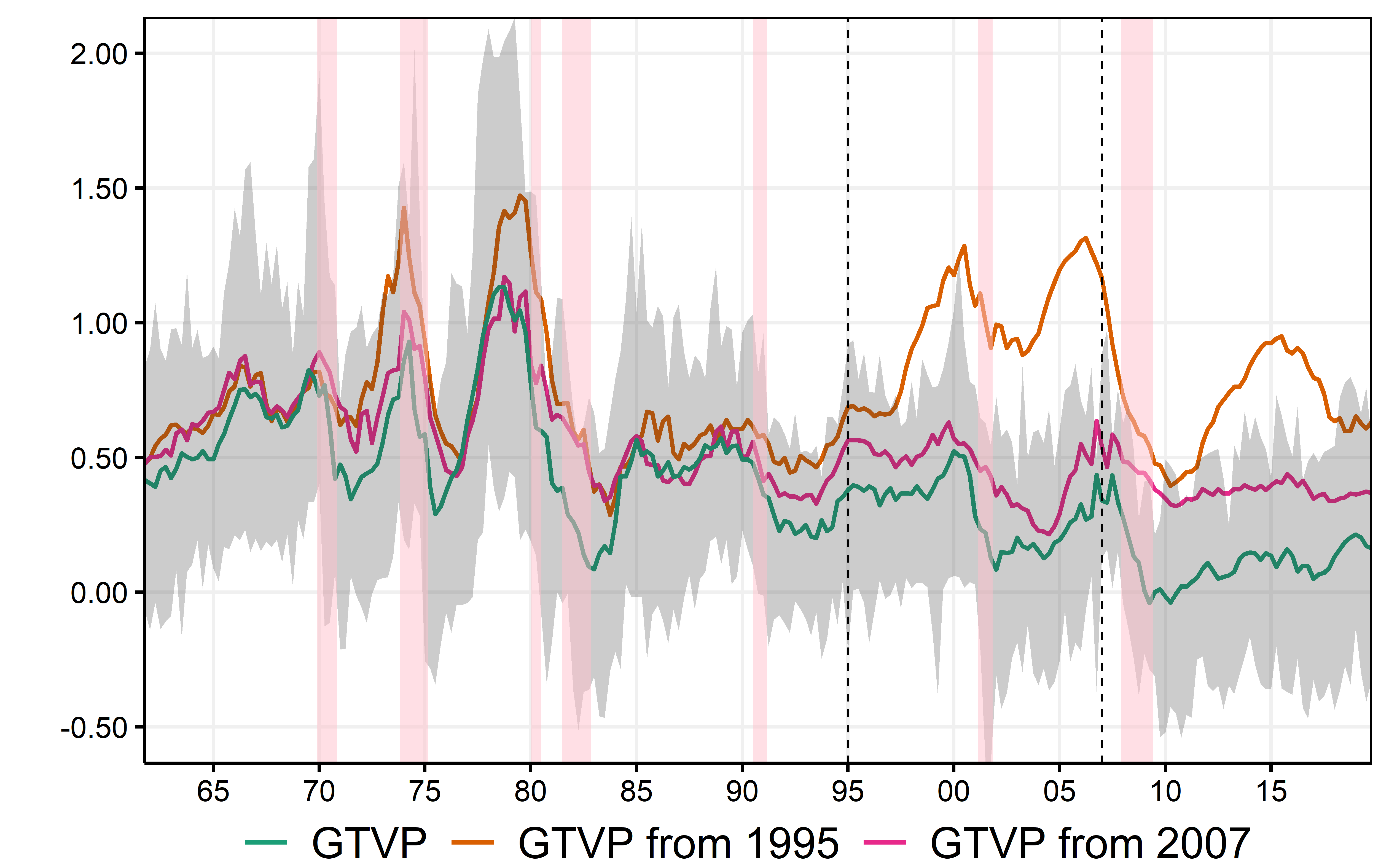}  
\end{center}
\vspace{-0.6cm}
\caption{\footnotesize Conditional $\beta_{2,t}$ Forecasting. The gray band is the 68\% credible region for GTVPs estimated up to 2019Q4. Pink shading corresponds to NBER recessions. For enhanced visibility, GTVPs are smoothed with 1-year moving average. The vertical dotted lines are the end of the training samples.}
\label{beta_bcs_ccf}
\end{figure}

Much can be learned from Figure \ref{beta_bcs_ccf}. First, GTVPs are all very alike for the pre-1995 period, suggesting little was observed post-1995 that made MRF change its reading of the past. Similarly, the green and the magenta line, which both share the 1995-2007 period within their training sets, are close to one another. Overall, this indicates that OOS difference between paths are very unlikely due to a better re-estimation and/or a completely new choice of $\mathcal{F}$. 

Second, unlike what we have seen for the unemployment equation (Figure \ref{v2h1_betas}), there are important disparities between the ex-ante and the ex-post paths \textit{out-of-sample}. Thus, one can rightfully hypothesize that structural change got in the way, making $\hat{\mathcal{F}}_{1995}$'s attempt of replicating the strong nonlinearities of the 70s into the 2000s go wildly off course. An analogous (yet far less noticeable gap) punctuates the post-2007 period. This suggests that while $\beta_{2,t}$ was expected to fall marginally following the crisis and stay low thereafter (according to $\hat{\mathcal{F}}_{2007}$), it was not expected to go \textit{that} low. Indeed, only $\hat{\mathcal{F}}_{2019}$ hits 0 and stays in its vicinity. 

Of course, by design, exogenous structural change cannot be captured out-of-sample -- with the results that we know ($\hat{\mathcal{F}}_{1995}$). This dismal predicament does not apply to cyclical behavior: it has been forecastable at least since 1995. Indeed, $\hat{\mathcal{F}}_{1995}$ propose a  $\beta_{2,t}$ for 2000 and 2008 that is very similar to that of 70s inflation spirals. Moreover, $\hat{\beta}_{2,t|1995}$'s collapse following 2008 is of a magnitude only seen during Arthur Burns' days. Hence, a much weaker PC following large downturns is hardly new. However, what $\hat{\beta}_{2,t|2007}$ and $\hat{\beta}_{2,t|2019}$
tell us is that the overall amplitude (and level) of those variations has evolved exogenously, forcing MRF to update $\mathcal{F}$ repeatedly. 

This exercise may rightfully seem exotic, with no obvious analog in the literature. The simple explanation is that traditional time variations only give "trivial" parameter forecasts by construction, and there is no clear "learning" process to analyze. For example, the "forecasted" random walk TVP would be a straight line over the whole OOS. Doing so with a threshold model would only inform us of the increasing precision of estimation as sample size grows -- i.e., the model itself cannot be re-evaluated. Unlike traditional nonlinear methods, MRF provides non-trivial $\beta_t$ paths out-of-sample --- and discovers exogenous structural change instead of imposing it.  %

\section{Conclusion}\label{sec:con}

I proposed a new time series model that \textbf{(i)} expands multiple nonlinear time series models, \textbf{(ii)} adapts Random Forest for macro forecasting and \textbf{(iii)} can be interpreted as Generalized Time-Varying Parameters. On the empirical front, the methodology provides substantial empirical gains over RF and competing non-linear time series models. The resulting Generalized TVPs have a very distinct behavior vis-à-vis standard random walk parameters. For instance, they adapt nicely to regime-switching behavior that seems pervasive for unemployment -- while not neglecting potential long-run change. This finding is facilitated by the fact that GTVPs lend themselves much more easily to interpretation than either standard RF or random-walk TVPs. Indeed, rather than trying to open the back box of an opaque conditional mean function (like one would with plain RF), MRFs can be compartmentalized in different components of the small macro model. Furthermore, GTVPs can be visualized with standard time series plots and credible intervals are provided by a variant of the Bayesian Bootstrap. 

When looking at Phillips' curves in general, MRF finds both structural change in the persistence and regime-dependent behavior in the economic activity/inflation trade-off. In particular, a recurrent theme across all specifications is that the slowly decaying curve is also much steeper when the economy is overheating -- in line with the convexity/nonlinearity hypothesis. Hence, MRF can be of great help sorting out what is plausible and what is not when it comes to macroeconomic equations with a history of controversy. Since there is no shortage of those, MRF holds many possibilities for future research. 


\clearpage

\bibliographystyle{apalike}

\setstretch{0.75}
\bibliography{C:/Users/Public/Bibtex_files/ref_pgc_v181204}

\clearpage

\appendix
\newcounter{saveeqn}
\setcounter{saveeqn}{\value{section}}
\renewcommand{\theequation}{\mbox{\Alph{saveeqn}.\arabic{equation}}} \setcounter{saveeqn}{1}
\setcounter{equation}{0}
	
\section{Appendix}
\setstretch{1.25}


\subsection{More on Engineering $S_t$}\label{sec:morest}

To appreciate the point that various factors \textit{and} the raw data can both be included together, let us put RF aside for a moment, and look at a high-dimensional linear regression problem. Suppose we define $S_t = [X_t \enskip F_t]$ and by construction the factors are some linear combination of original features ($F_t = X_t R$).\footnote{Note that in this section only, $X_t$ denotes generic raw regressors rather than MRF's linear part. This switch allows for the use of familiar-looking notation.} We can estimate
\begin{align}\label{simplelin}
y_{t+1} = X_t\beta +X_t R \gamma +u_t
\end{align}
using LASSO. Of course, this would not run with OLS because of perfect collinearity, which is the standard motivation for not mixing dense and sparse approaches. By Frisch-Waugh-Lowell theorem and the factor model 
$$X_t = \Lambda F_t+e_t,$$
\eqref{simplelin} above is equivalent to 
$$y_{t+1} = e_t\beta +F_t\gamma +u_t.$$
At first sight, this has more parameters than either the dense or sparse approach. However, with some adequate penalization of $\beta$ and $\gamma$, the model can balance a proper mix of dense and sparse. For instance, activating some $\beta$'s "corrects" the overall prediction when the factor model representation is too restrictive for the effect of a specific regressor $X_k$ on $y_{t+1}$.\footnote{That problem has been documented in \cite{targetedPC} and others.} This representation has been studied in \cite{hahn2013partial} and \cite{hansen2019factor} to enhance hard-thresholding methods' performance (like LASSO) in the presence of highly correlated regressors. Coming back to RF, this means its strong regularization/selection allows for both the original data and its rotation to be included in $S_t$. This also suggests it is relatively costless to explore alternative rotations of $X_t$.

\subsection{Block Bayesian Bootstrap Details}\label{sec:bbb_details}

BBB is a conceptual workaround to reconcile time series data with multinomial sampling. For completeness, I briefly review the \textit{standard} Bayesian Bootstrap. Let all the available data be cast in the matrix $Z_t = [y_t \enskip X_t \enskip S_t]$. $Z$ is considered as a discrete \textit{iid} random variable with $T$ support points. Define $N_{t}=\sum_{\tau=1}^{T} I\left(Z_{\tau}=z_{t}\right)$, which is the number of occurrences of $z_{t}$ in the sample. The goal is to conduct inference on the data weight vector $\theta_{1:T}$, and then obtain credible regions for the posterior functional $\beta_t = \mathcal{T}({\theta}_{1:T})$. To do so, we need to characterize the posterior distribution of vector $\theta$ (stripped of its subscript for readability)
$$
{\pi}(\theta | \mathbf{z})=\frac{f(\mathbf{z} | \theta) \pi(\theta)}{\int f(\mathbf{z} | \theta) \pi(\theta) d \theta}.
$$
Conditional on $\theta$, the likelihood of the data is multinomial. The prior is Dirichlet.  Since Dirichlet is the conjugate prior of the multinomial distribution,  the posterior is also Dirichlet. That is, it can be shown that combining the likelihood 
$$
f(\mathbf{z} | \theta)=\frac{N !}{N_{1} ! \cdots N_{T} !} \prod_{t=1}^{T} \theta_{t}^{N_{t}} \quad \enskip \text{with prior distribution} \quad \enskip {\pi}(\theta)=\frac{1}{B(\alpha_{1:T})}\prod_{t=1}^{T} \theta_{t}^{N_{t}+\alpha_{t}-1}
$$
gives rise to the posterior distribution
$$
{\pi}(\theta | \mathbf{z})= \frac{1}{B(\bar{\alpha}_{1:T})} \prod_{t=1}^{T} \theta_{t}^{N_{t}+\alpha_{t}-1}  \enskip .
$$
where $\bar{\alpha}_{t}=\alpha_{t}+N_{t}$ and $B(\bar{\alpha}_{1:T}) = \frac{\prod_{t=1}^{T} \Gamma\left(\bar{\alpha}_{t}\right)}{\Gamma\left(\sum_{t=1}^{T} \bar{\alpha}_{t}\right)}$. Using the uninformative (and improper) prior $\alpha_t =0 \enskip \forall t$, we can simulate draws from the (proper) posterior using $\theta_t \sim \operatorname{Exp}(1)$. The object of scientific interest is typically not $\theta$ \textit{per se} but rather a functional of it. In \cite{taddy2015bayesian}, the functional of interest is a tree and inference is obtained by computing $\mathcal{T}(\theta_{1:T})$ for each $\theta_{1:T}$ draw. BBB considers a different $Z_t$ so that it is plausibly $iid$ when used with stationary time series data. The derivations above can be carried by replacing $t$ by $\mathfrak{b}$ and $T$ by $\mathfrak{B}$. Practically, this implies drawing $\theta_\mathfrak{b} \sim \operatorname{Exp}(1)$ which means observations within the same block ($\underline{\mathfrak{b}}:\bar{\mathfrak{b}}$) share the same weight. As an alternative to this BBB that would also be valid under dependent data, \cite{cirillo2013urn} provide a more sophisticated urn-based approach with theoretical guarantees. It turns out their approach contains the well-known non-overlapping block bootstrap as a special case, which the above is only its Bayesian rendition.

\subsection{More on Surrogate $\beta_t$ Trees}\label{surro_plus}

The approach described in section \ref{surro} belongs to a family of methods usually referred to as "surrogate models" \citep{molnar2019interpretable}. Attempting to fit the whole conditional mean obtained from a black-box algorithm using a more transparent model is a global surrogate. An obvious critique of this approach is that if the complicated model justifies its cost in interpretability with its predicting gains, it is hard to believe a simple model can reliably recreate its predictions. Conversely, if the surrogate model is quite successful, this casts some doubts about the relevance of the black box itself. In this line of work, a more promising avenue is a local surrogates model as proposed in \cite{ribeiro2016}, which fits interpretable models \textit{locally}. By following \cite{granger2008}'s insights, we already have this: by looking at the $\beta_t$ paths directly, we effectively have a local model -- in time.  The purpose of surrogate models is to learn about the model, not the data. The former is much easier in MRF than in standard RF since the vector $\beta_t$ fully characterizes the prediction at a particular point in time.\footnote{More generally, any partially linear model in the spirit of MRF has a potential for local surrogate analysis along the linear regression space rather than the observations line.} Moreover, the coefficients are attained to predictors that can have themselves a specific economic meaning. Considering this and the earlier discussion of section \ref{macroastrees}, it is natural in a macro time series context to fit surrogate models to time-varying parameters themselves -- a blatant divide-and-conquer strategy.

\subsubsection{About $VI_{OOB}$, $VI_{OOS}$ and $VI_{\beta}$}
I now explain the motivation and mechanics behind the different VI measurements. The first measure, $VI_{OOB}$, is the standard out-of-bag (hence OOB) VI permutation measure widely used in RF applications \citep{wei2015variable}. It consists of randomly permuting one feature $S_j$ and comparing predictive accuracy to the full model on observations that were not used to fit the tree.\footnote{This is thought as the equivalent for a black-box model to setting a specific coefficient to 0 in a linear regression and then comparing fits. However, VI as implemented here (and in most applications) does not re-estimate the model after dropping $S_j$. This differs from a t-test since it is well known that the latter is equivalent to comparing two $R^2$'s -- the original one and that of a re-estimated model, under the constraint.}  This pseudo evaluation set is convenient because it is a direct byproduct of the construction of the forest. Under a well-specified model that includes enough lags of $y_t$, autocorrelation of residuals will not be an issue. This condition is likely to be met here since the analysis focuses on results for $h=1$. \footnote{Notwithstanding, at longer horizons, $VI_{OOB}$ could paint a distorted picture in the presence of autocorrelation -- the same way K-fold cross validation can be inconsistent for time series data \citep{bergmeir2018note}. This worry can be alleviated by using a block approach like in section \ref{bayes}.} $VI_{OOS}$ considers a different testing set more natural for time series data: the real OOS, which in this section spans from 2007q2 to the end of 2014. By construction, this measure focuses on finding variables which contribution paid off during a specific forecasting experiment, rather than throughout the whole sample. This is not bad \textit{per se} but is a different concept that can be of independent interest.  Finally, both $VI_{OOB}$ and $VI_{OOS}$ focus on overall fit. $VI_{\beta}$ implements the same idea as $VI_{OOB}$ but is calculated using a different loss function. That is, $VI_{\beta_{k,j}}$ reports a measure of how much the path of $\beta_k$ is altered (out-of-bag) when variable $S_j$ is randomly permuted in the forest part. Finally, I use the various VI measurements as devices to narrow down the set of predictors for the construction of intuitive trees.

I restrict the number of considered variables (for the next step) to be 20 for each VI criteria. When VI suggest that a parsimonious set of variables matter, it is very rarely more than 3 or 4 variables. Thus, restricting it to 20 is a constraint that only binds if all variables contribute, but marginally, in the spirit of a Ridge regression \citep{ESL}. When it comes to that, the cut-off is simply the natural reflection of a trade-off between interpretability and fit.

\subsection{On Tuning Parameters}\label{CV_section}

The bulk of the discussion on the algorithm's specifics is deferred to the \href{https://philippegouletcoulombe.com/code}{{\texttt{R} package}}. None of the RFs reported in the text were tuned. This is not heresy, as minuscule performance gains from doing so (like optimizing $\texttt{mtry}$) are the norm rather than the exception. Additionally, restraining the terminal nodes size can only alter performance very mildly and it is now clear why \citep{MSoRF}. Nonetheless, reviewing some of those untuned tuning parameters can be insightful about MRFs inner workings. "Algorithm" \ref{mrf_algo_full} below summarizes when and where those enter the MRF procedure. 

\begin{itemize} 
  \setlength\itemsep{0.001cm}
  		\item \texttt{RWR}: stands for Random Walk Regularization strength as discussed in \ref{RWR}. It is the $\zeta$ in equation (\ref{mrf_algo_rw}).
		\item \texttt{RL}: stands for Ridge Lambda ($\lambda$) in equation (\ref{mrf_algo}). Prior means are OLS estimates.
											\item \texttt{Minimal Node Size}: Minimal parent leaf size to consider a new split. Set to 10 for quarterly data and 15 for monthly.
		\item \texttt{MLF}: stands for Minimum Leaf Fraction. It is the parameter in MRF that has a role complementary to that of minimum node size. The so-called "fraction" is the ratio of parameters in the linear part to that of observations in any node (which includes most importantly the terminal ones). Here is an example. Set $\texttt{MLF}=2$, the linear part has 3 parameters, and we are trying to split a subset of 15 observations. This setting implies that any split that results in having less than 6 observations in the children note will not be considered. This specific setting ensures that the ratio of parameters to observations never exceeds $\sfrac{1}{2}$ in any node. This ensure stability, especially if the two aforementioned HPs are set to 0. However, when \texttt{RWR} and \texttt{RL} are active, it is possible to consider $\texttt{MLF}=1$ or even lower. 
The extra regularization allows in the latter case to have base regressions that have parameters/observations ratio exceeding 1 (high-dimensional setting). This is desirable with quarterly data because setting $\texttt{MLF}>2$ or higher seriously restricts the potential depth of the trees.  

				\item \texttt{mtry}: how many $S_j$'s do we consider as potential "splitter" at each split? It is easier to think about it as a fraction of the total number of predictors. For regression settings, the suggested value is $\sfrac{1}{3}$. The lower it gets, the more random tree generation gets, and better diversification may ensue. Moreover, \texttt{mtry} directly impacts computational burden. It is often found, in a macro context, that lowering \texttt{mtry} to 0.2 does not alter performance noticeably, while reducing appreciably computations. In fact, running RF-MAF with \texttt{mtry}$\in \{0.1,0.2,0.33,0.5 \}$ delivers nearly identical performance for all variable/horizon pairs of the quarterly exercise. This is likely attributable to macro data having a factor structure. If $S_j$ is "not available" for a split when it would in fact maximize fit locally, there is another strongly correlated $S_{j'}$ ready for the task. For instance, if the unemployment rate is discarded by \texttt{mtry}, then there are more than 20 other labor indicators that can possibly substitute for it. If those 20 variables are all a noisy representation of the same latent variable the model wants to split on, then the probability of having none to split with at a given point is $\left(1-\frac{\texttt{mtry}}{\#\text{regressors}}\right)^{20}\approx 0$.
				\item \texttt{Trend Push}: Some minorities may end up being underrepresented as a result of \texttt{mtry}'s discriminating action. While there are 20+ labor indicators in the data base, there is only one trend. Since exogenous change should most certainly not be underrepresented, its "personalized" probability of inclusion can be pushed beyond what \texttt{mtry} suggests.
							\item \texttt{Subsampling Rate}: is set at 75\%.

\end{itemize}

A scaled down quarterly forecasting exercise was conducted to see whether tuning any of those could help. Precisely, horizons 1, 2, and 4 quarters were considered and models (ARRF,FA-ARRF,VARRF) were estimated once at the beginning of the OOS period (2002). Tuning parameters were optimized targeting 1998-2002 data as an artificial test set. Possible values were \texttt{RWR}$\in \{0,0.5,0.95\}$, \texttt{RL}$\in \{0.1,0.5\}$, \texttt{mtry}$\in \{0.2,0.33,0.5\}$ and \texttt{min.node.size}$\in \{10,40\}$. It is found that results are largely invariant to pre-optimized HPs. As mentioned earlier, what matters most in the linear part. It is observed that optimizing tuning parameters can help reduce marginally RMSEs of MRFs that were sometimes struggling (like VARRF). Results are available upon request.

\begin{algorithm}
\caption{How the key tuning parameters enter MRF, and other practical aspects \label{mrf_algo_full}}
\begin{algorithmic}[1] \itemsep 1em
  \small
\STATE Draw blocks of some size (8 for quarterly, 24 monthly), that makes for \texttt{Subsampling Rate}\% of the sample. To simply get the mean prediction, 100 trees are usually more than enough. To get credible regions to stabilize, 200-300 trees are typically needed.
\STATE 
\begin{itemize} \itemsep -0.5em
\item For each subsample: run \eqref{mrf_algo_rw} recursively on that sample given $\lambda$ and $\zeta$ values until each (potential) parent nodes are smaller than \texttt{Minimal Node Size}.
\item A total of \texttt{mtry} predictors are considered at each splitting step $\mathcal{J}^-$ is randomly picked out of $\mathcal{J}$. Those probabilities are all $\sfrac{1}{\text{dim}(\mathcal{J})}$ by default. \texttt{Trend Push} pushes that of the trend further if judged appropriate for a given data set. 
\item When evaluating potential splits, discard those that would not meet \texttt{MLF}'s requirements on resulting children nodes.
\item This outputs one tree structure $\mathcal{T}$.
\end{itemize}
 \STATE When inputted with new observations of $X_t$ and $S_t$, each tree produces a forecast. MRF forecast is the mean of the those. 
 \STATE Same goes for $\beta_t$: each tree predicts its own $\beta_{t}$ out-of-sample and the posterior mean is the average of all those.
\STATE In-sample $\beta_t$'s need an extra step: only draws that did not use observation $t$ to construct the tree (that is, for which $t$ was left out of the subsample) are used to characterize the distribution of $\beta_t$.
\end{algorithmic}
\end{algorithm}

\subsection{Additional Simulations Results}\label{sec:moresims}

\begin{figure}[p!]
  \begin{subfigure}[b]{\textwidth}
  \begin{center}
         \includegraphics[trim={1.5cm 0cm 0.25cm 0cm},clip,width=0.94\textwidth]{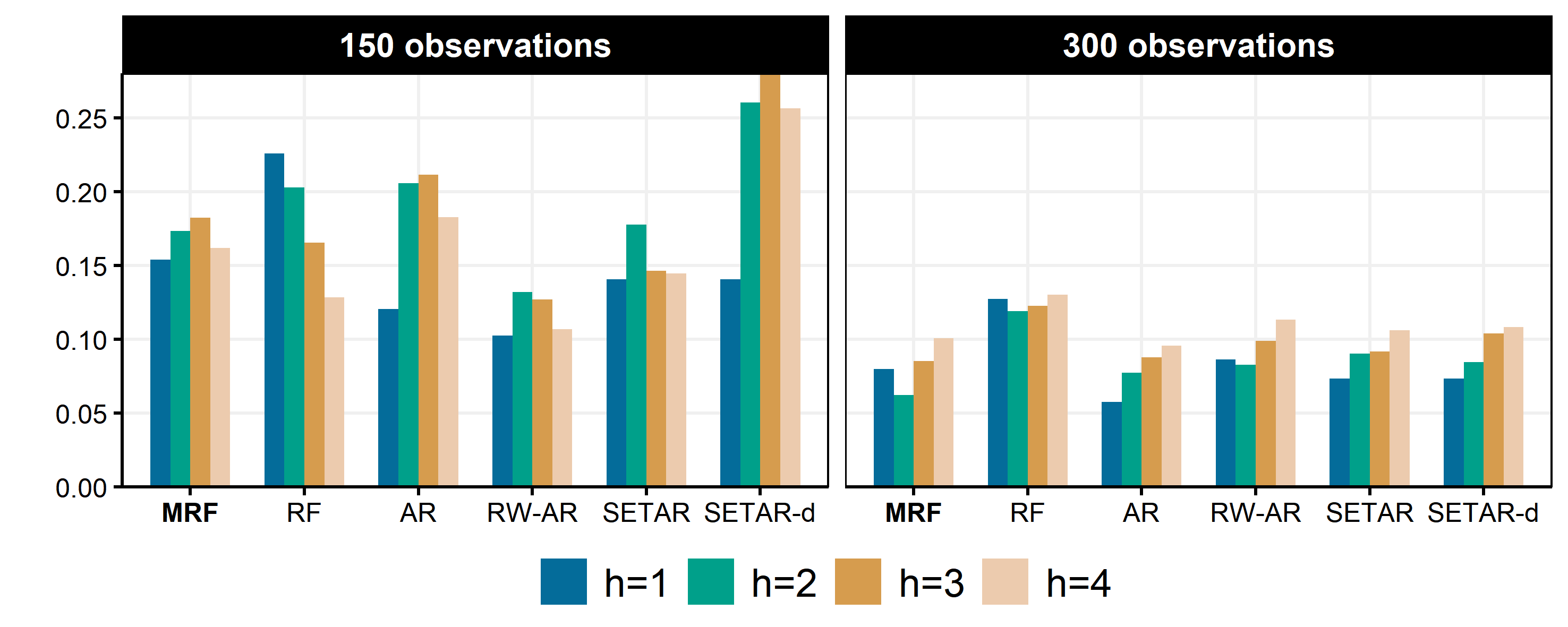}
    \caption{DGP is SETAR.}
        \label{v4_graph}
                    \end{center}
  \end{subfigure}
  \hspace{2em}
  \begin{subfigure}[b]{\textwidth}
    \begin{center}
         \includegraphics[trim={1.5cm 0cm 0.25cm 0cm},clip,width=0.94\textwidth]{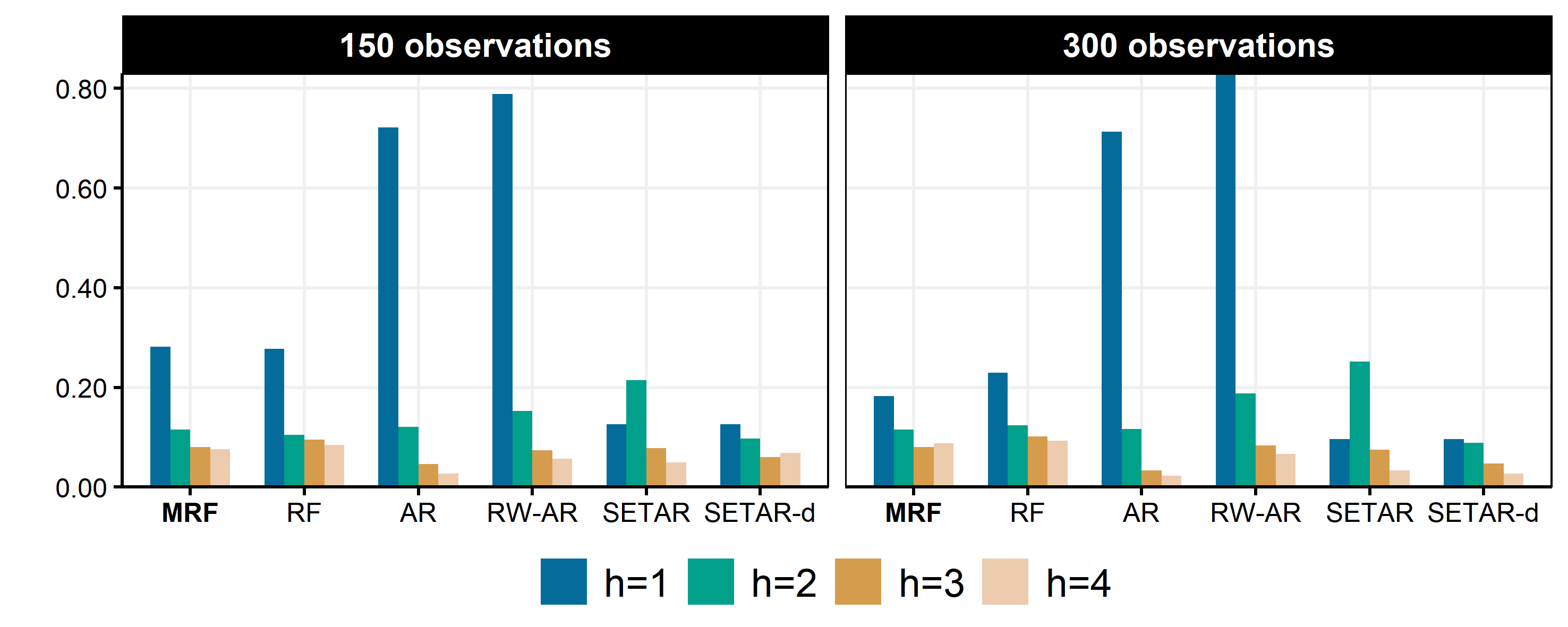}
    \caption{DGP is AR(2) with structural break.}
        \label{v5_graph}
                    \end{center}
      \end{subfigure}
  \begin{subfigure}[b]{\textwidth}
    \begin{center}
         \includegraphics[trim={1.5cm 0cm 0.25cm 0cm},clip,width=0.94\textwidth]{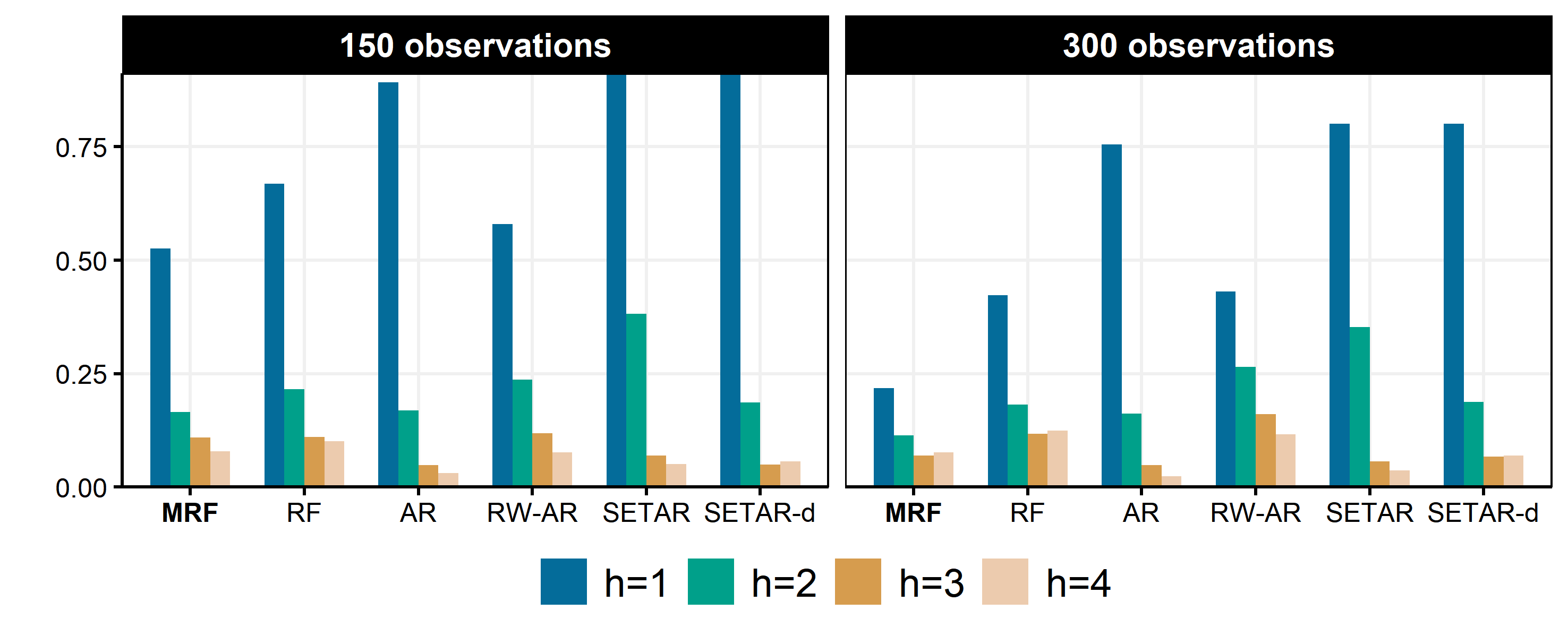}
    \caption{DGP is SETAR with structural break.}
        \label{v6_graph}
                    \end{center}
      \end{subfigure}
  \caption{Displayed are increases in relative RMSE with respect to the oracle.}
    \label{oldsimul_all1}
\end{figure}

\vskip 0.2cm
{\noindent \textbf{DGP 4: SETAR.}} In this second SETAR example 
\begin{align*}
y_{t}&=X_{t}{\beta_{t}}+{\epsilon_{t}}, \quad {\epsilon}_{t}\sim{N(0,0.5^2)} \\
{\beta_{t}}&=
\begin{cases}
[2 \enskip 0.8 \enskip -0.2] ,& \text{if } y_{t-1}\geq 1  \\
[0 \enskip 0.4 \enskip -0.2] ,& \text{otherwise},
\end{cases} 
\end{align*}
AR models are doing badly by not capturing the change in mean and dynamics. It is noteworthy that in this DGP, predictive power quickly vanishes after $h=1$, which is why we observe little performance heterogeneity at longer horizons in Figure \ref{v4_graph}: those are dominated by the unshrinkable prediction error. Specifically tailored for this class of DGPs, the two SETARs are offering the best performance. A less trivial observation is that MRF and RF, while much more general, perform only marginally worse than SETARs. The tie between MRF and RF is attributable the importance of the switching constant in the current DGP, which both models allow for. 

\vskip 0.2cm
{\sc \noindent \textbf{DGP 5: AR(2) with a break.}} Results for 
\begin{align*}
y_{t}&=X_{t}{\beta_{t}}+{\epsilon_{t}}, \quad {\epsilon}_{t}\sim{N(0,0.3^2)} \\
{\beta_{t}}&=
\begin{cases}
[0 \enskip 0.7 \enskip -0.35] ,& \text{if } t<T/2  \\
[0.15 \enskip 0.6 \enskip 0] ,& \text{otherwise}
\end{cases}
\end{align*}
are reported in Figure \ref{v5_graph}. In this setup, RW-AR is expected to have an edge, with the estimation window excluding pre-break data. At horizon 1, both RW-AR and ARRF are the best model, beating the robust AR by a thin margin. For $h>1$, ARRF emerges as the best model at both 150 and 300 sample sizes. Naturally, RW-AR is always close behind.\footnote{Although not reported here, I considered a simple linear model where I search for a single break (in time) and use the data after the break for forecasting. This option does as well as ARRF for this particular DGP.}  As expected, the two models are better than the remaining alternatives by allowing for exogenous structural change (which SETARs and AR do not) and explicitly modeling the autoregressive part (which RF does not).  

\vskip 0.2cm
{\sc \noindent \textbf{DGP 6: SETAR with a Structural Break.}} This is slight complications of DGP 1. Again, SETARs are expected to fail because they are not designed to catch breaks. RW-AR is also expected to fail because it does not model switching. RF is general enough, but is anticipated to be inefficient. All these heuristics for
\begin{align*}
\text{DGP 4}=
\begin{cases}
\text{DGP 2} ,& \text{if } t<T/2  \\
\text{DGP 3} ,& \text{otherwise}
\end{cases} 
\end{align*}
are verified in Figure \ref{v6_graph}: MRF is the better model followed closely by RW-AR and RF for short horizons. With 300 observations, the lead of ARRF, as well as the second position of RF, are both strengthened. At longer horizons, all models perform poorly (including the oracle) due to the fundamental unpredictability of the law of motion for $\beta_t$. For these horizons, misspecification only plays a minor role in total forecast error variance, explaining the small and homogeneous decrease in performance with respect to the oracle. 

\begin{figure}[!p]
  \begin{subfigure}[b]{0.4997\textwidth}
   \includegraphics[width=\textwidth]{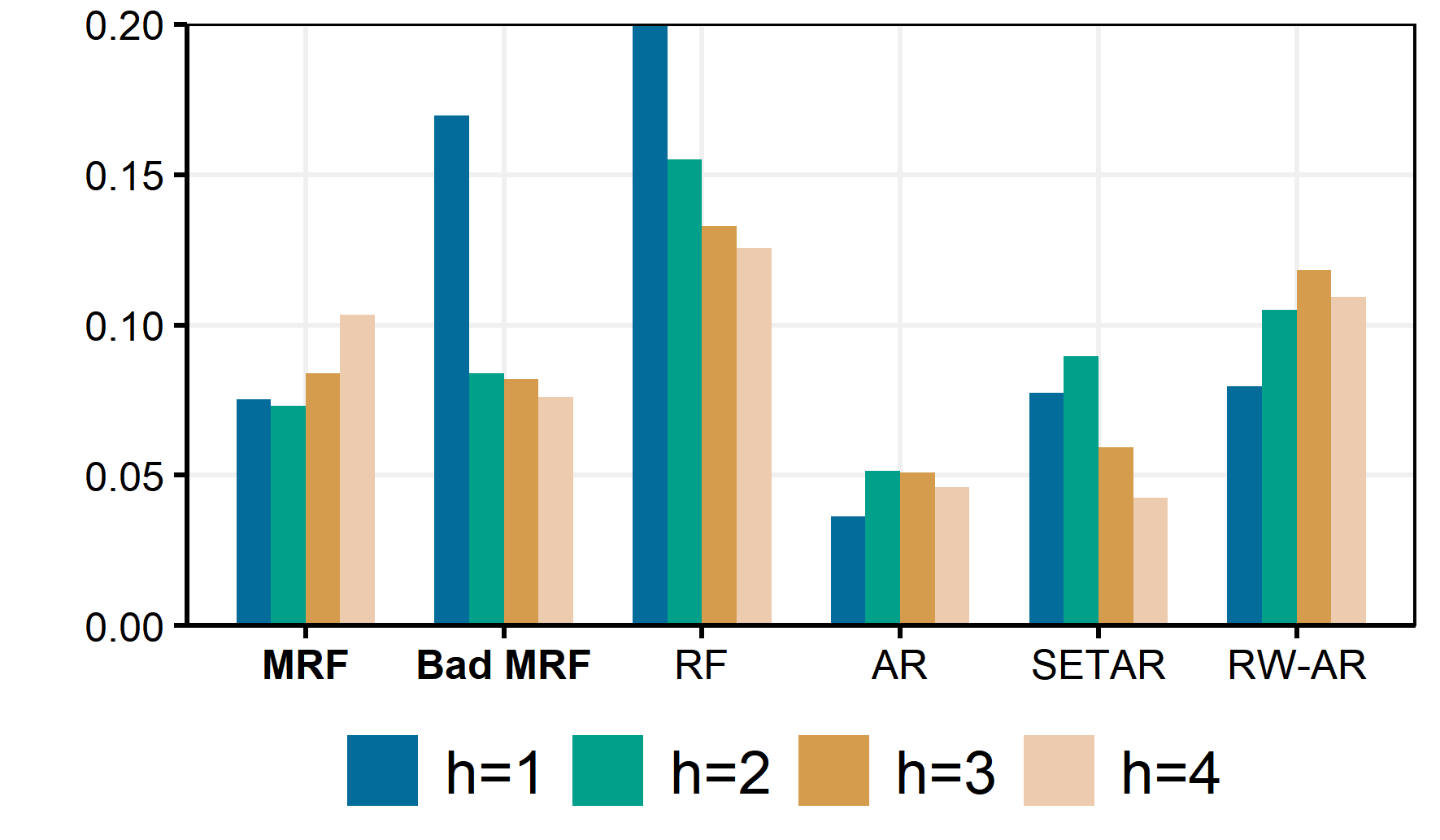}
    \caption{DGP 1}
  \end{subfigure}
  \begin{subfigure}[b]{0.4997\textwidth}
   \includegraphics[width=\textwidth]{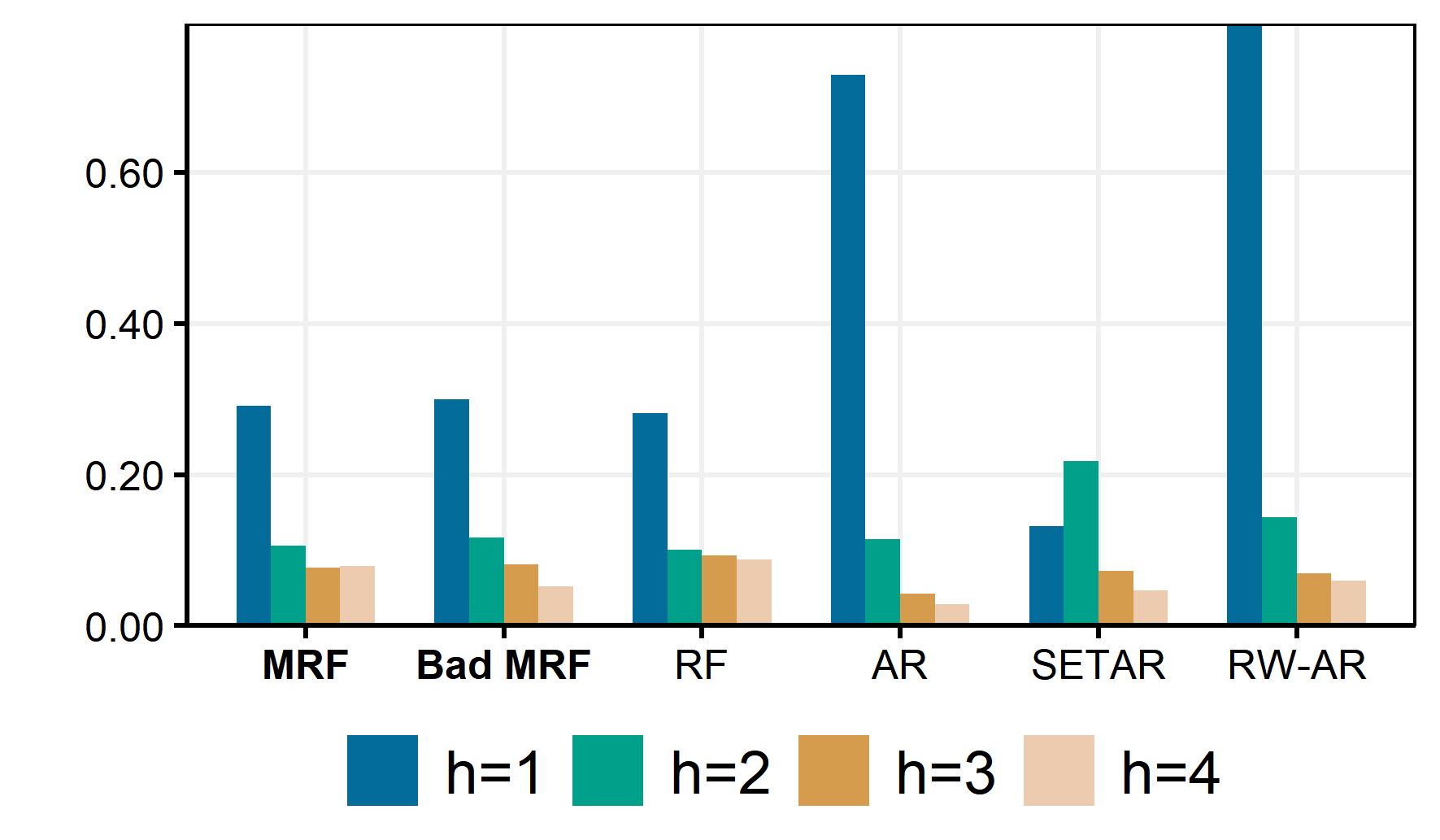}
    \caption{DGP 2}
      \end{subfigure}
  \begin{subfigure}[b]{0.4997\textwidth}
   \includegraphics[width=\textwidth]{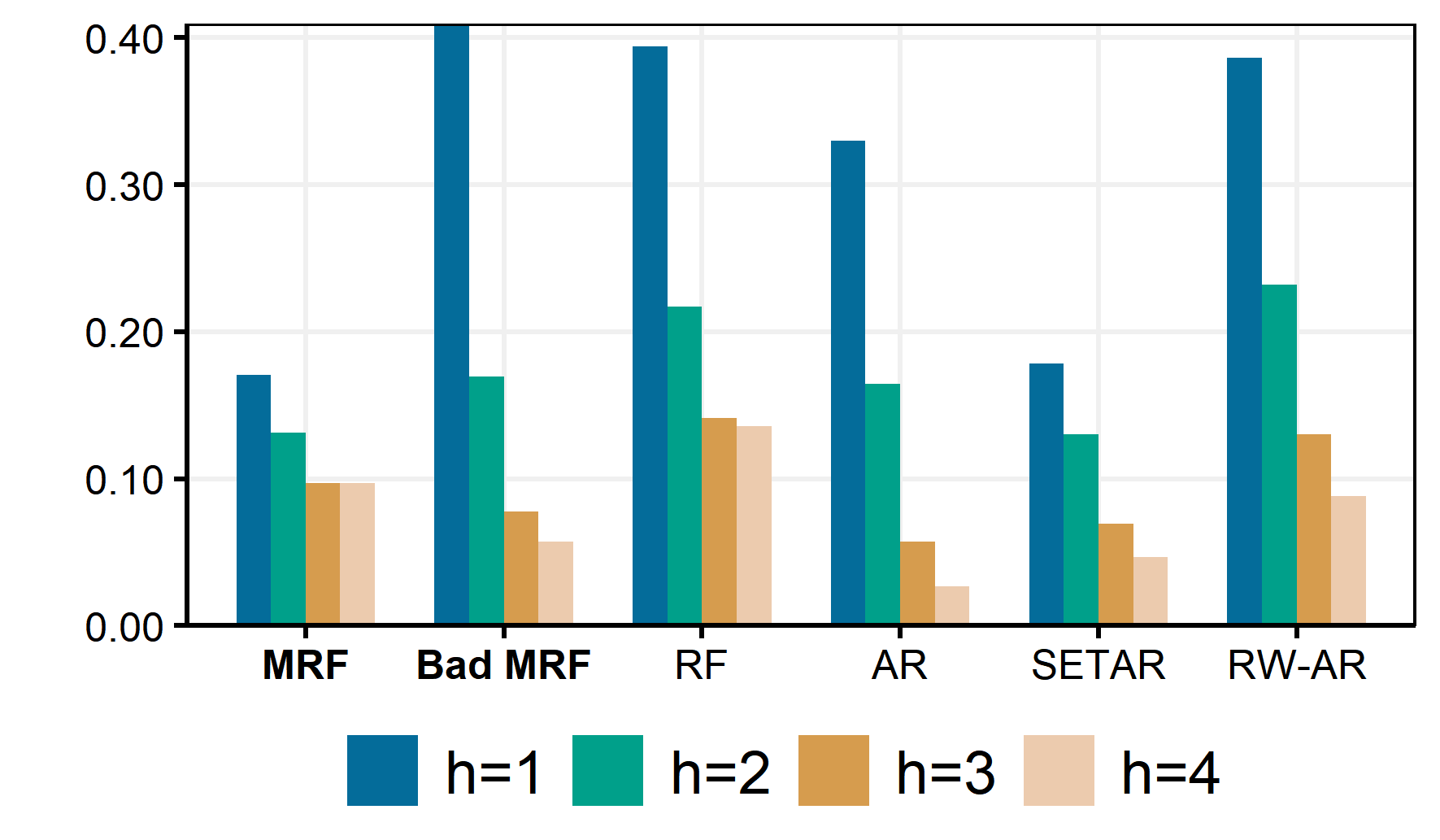}
    \caption{DGP 3}
      \end{subfigure}
  \begin{subfigure}[b]{0.4997\textwidth}
   \includegraphics[width=\textwidth]{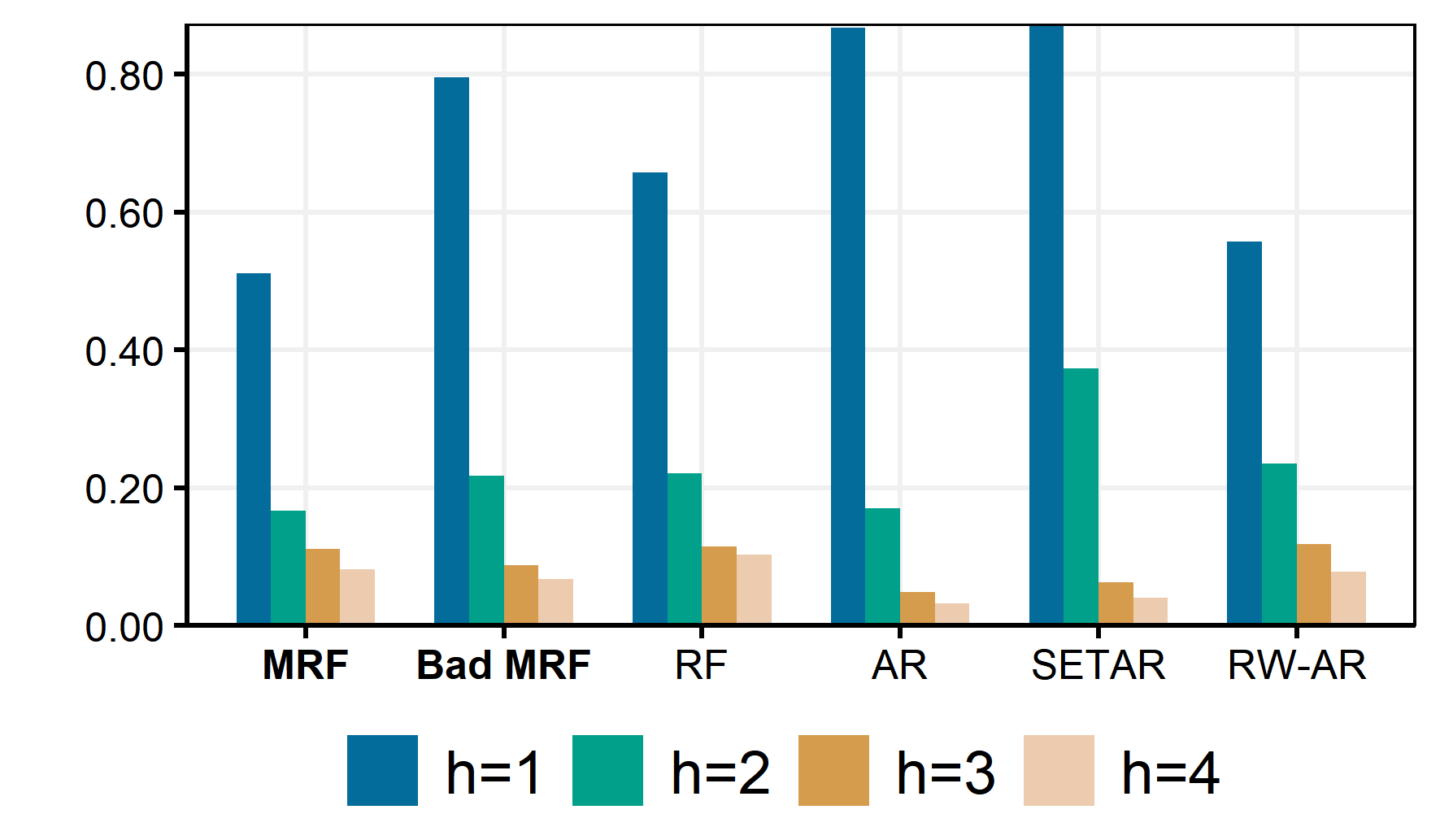}
    \caption{DGP 4}
      \end{subfigure}
  \begin{subfigure}[b]{0.4997\textwidth}
   \includegraphics[width=\textwidth]{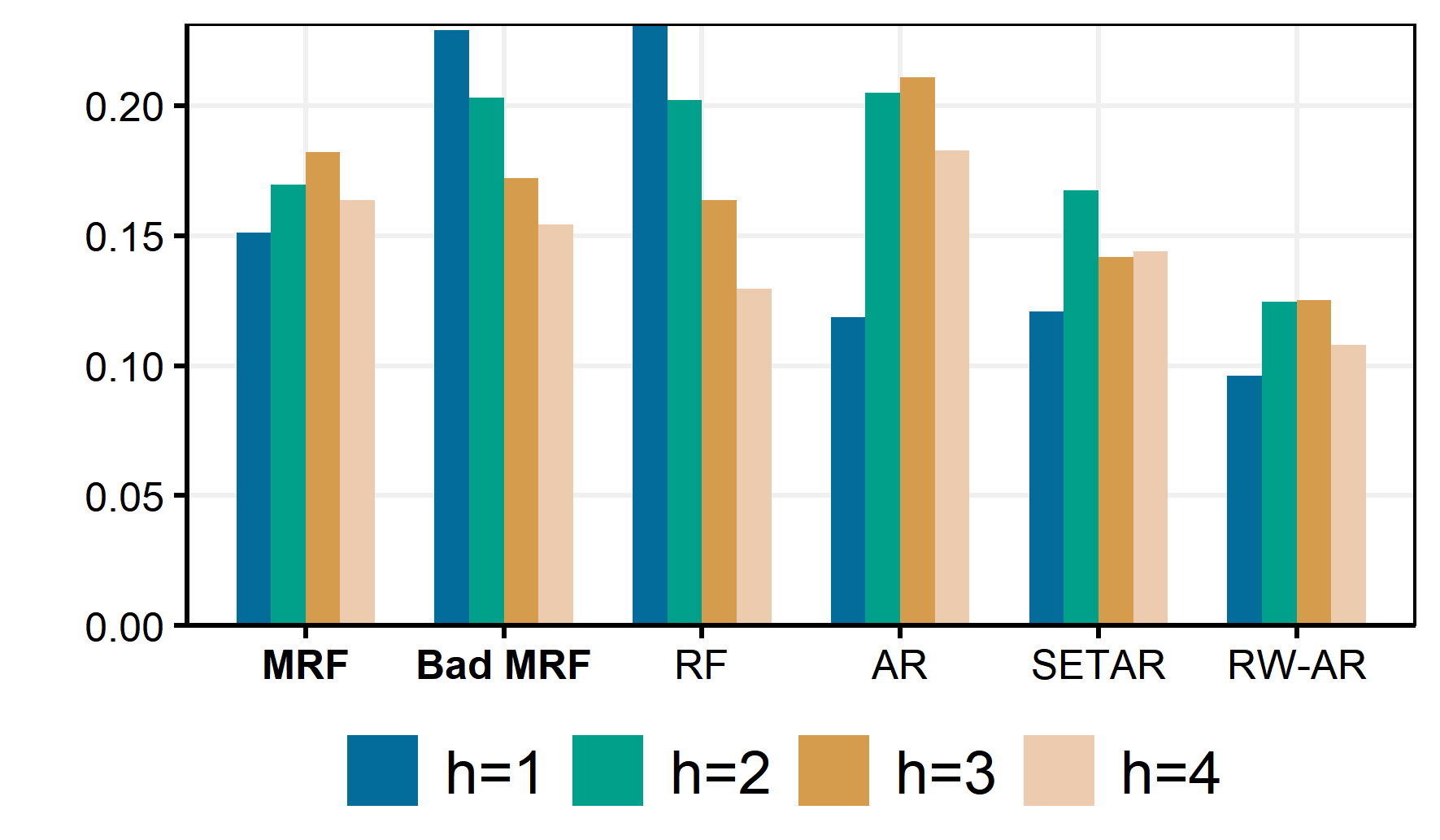}
    \caption{DGP 5}
      \end{subfigure}
  \begin{subfigure}[b]{0.4997\textwidth}
   \includegraphics[width=\textwidth]{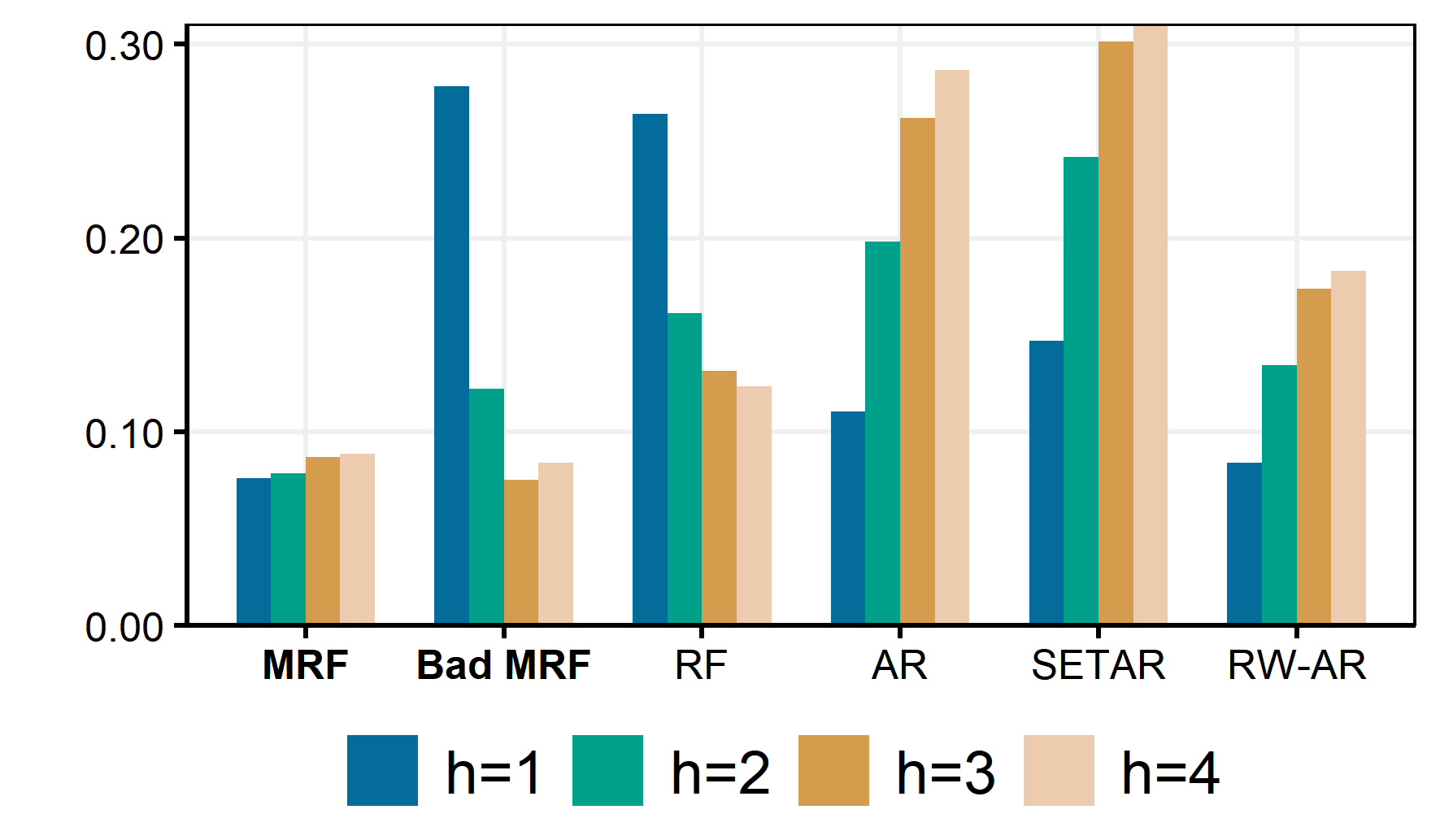}
    \caption{DGP 6}
  \end{subfigure}
  \caption{\footnotesize Investigation of the consequences of $X_t$'s misspecification, as exemplified by "Bad ARRF". Instead of the first two lags of $y_t$, $X_t$ is replaced by randomly generated \textit{iid} (normal) variables. Total number of simulations is 50, and the total number of squared errors is thus 2000.} 
    \label{MISS_graph}
\end{figure}

\begin{figure}[p!] 
  \begin{subfigure}[b]{\textwidth}
  \begin{center}
   \includegraphics[trim={1.6cm 1.8cm 0.3cm 0.5cm},clip,width=0.9\textwidth]{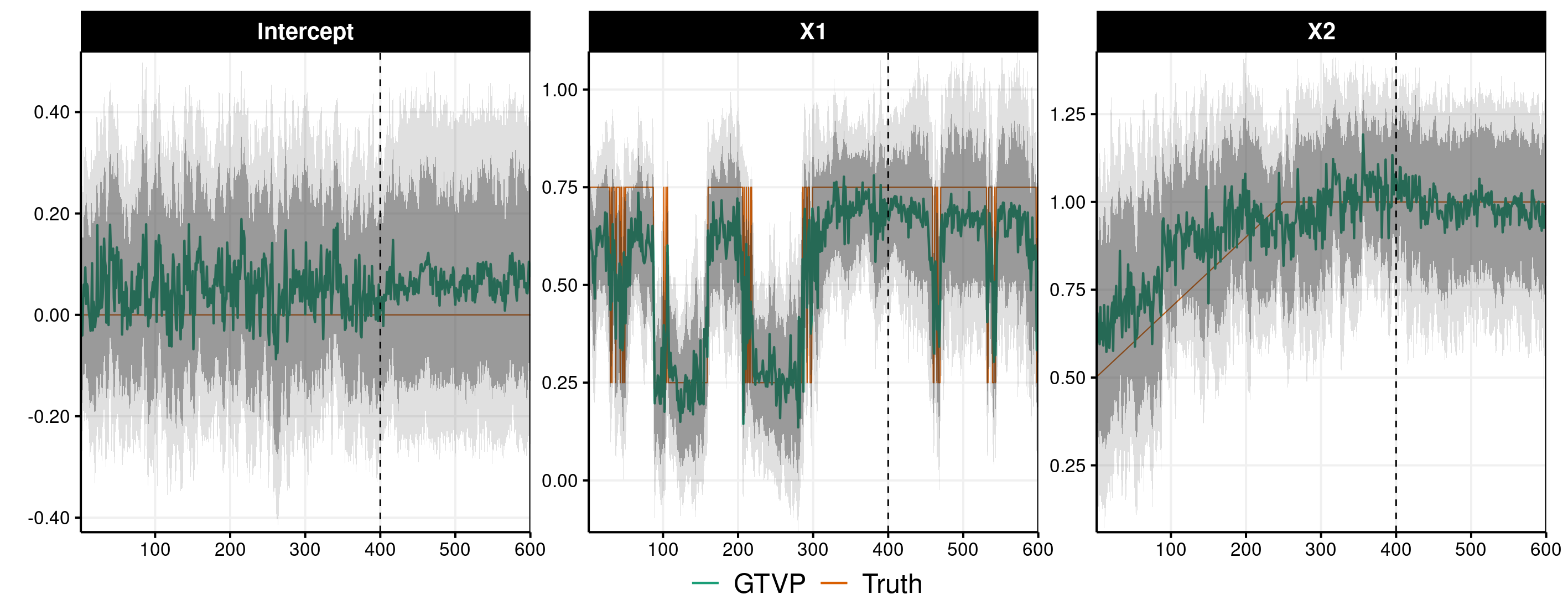}
    \caption{DGP 4}
        \label{newsimul_v4}
          \end{center}
  \end{subfigure}
  \hspace{2em}
  \begin{subfigure}[b]{\textwidth}
      \begin{center}
   \includegraphics[trim={1.6cm 1.8cm 0.3cm 0.5cm},clip,width=0.9\textwidth]{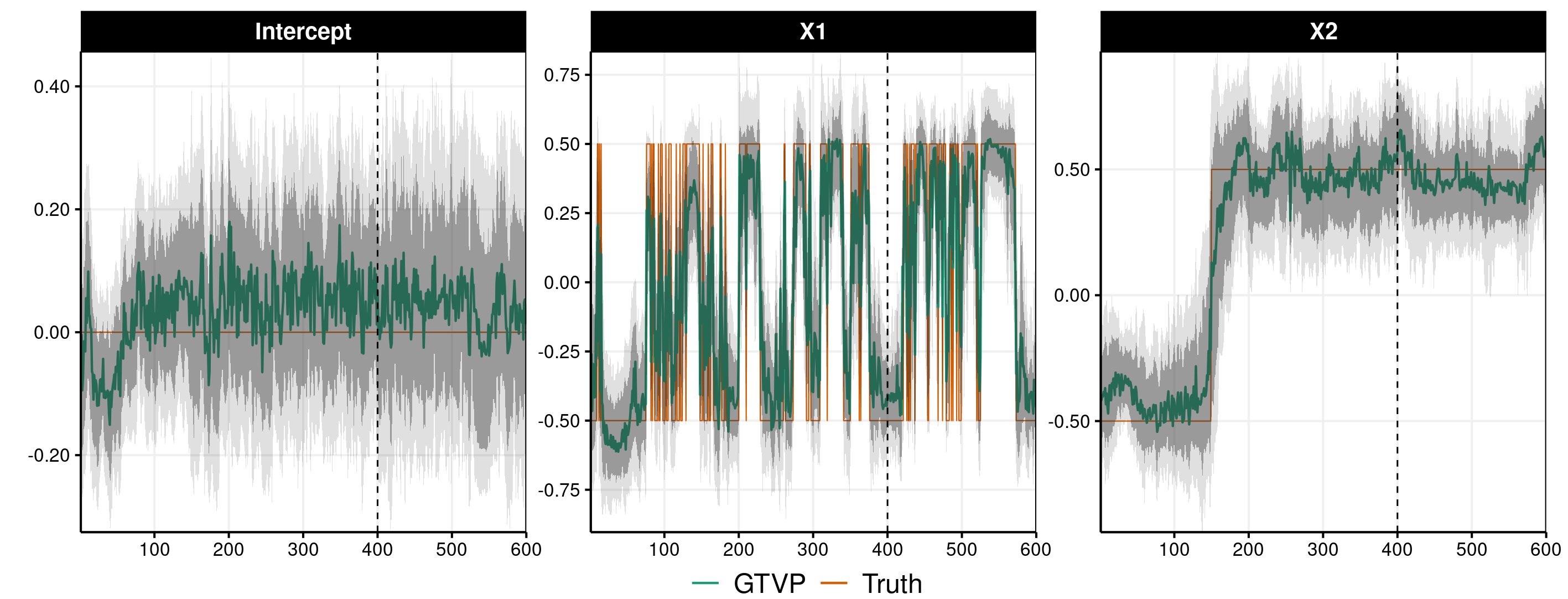}
    \caption{DGP 5}
        \label{newsimul_v5}
                  \end{center}
      \end{subfigure}
  \begin{subfigure}[b]{\textwidth}
    \begin{center}
   \includegraphics[trim={1.6cm 1.8cm 0.3cm 0.5cm},clip,width=0.9\textwidth]{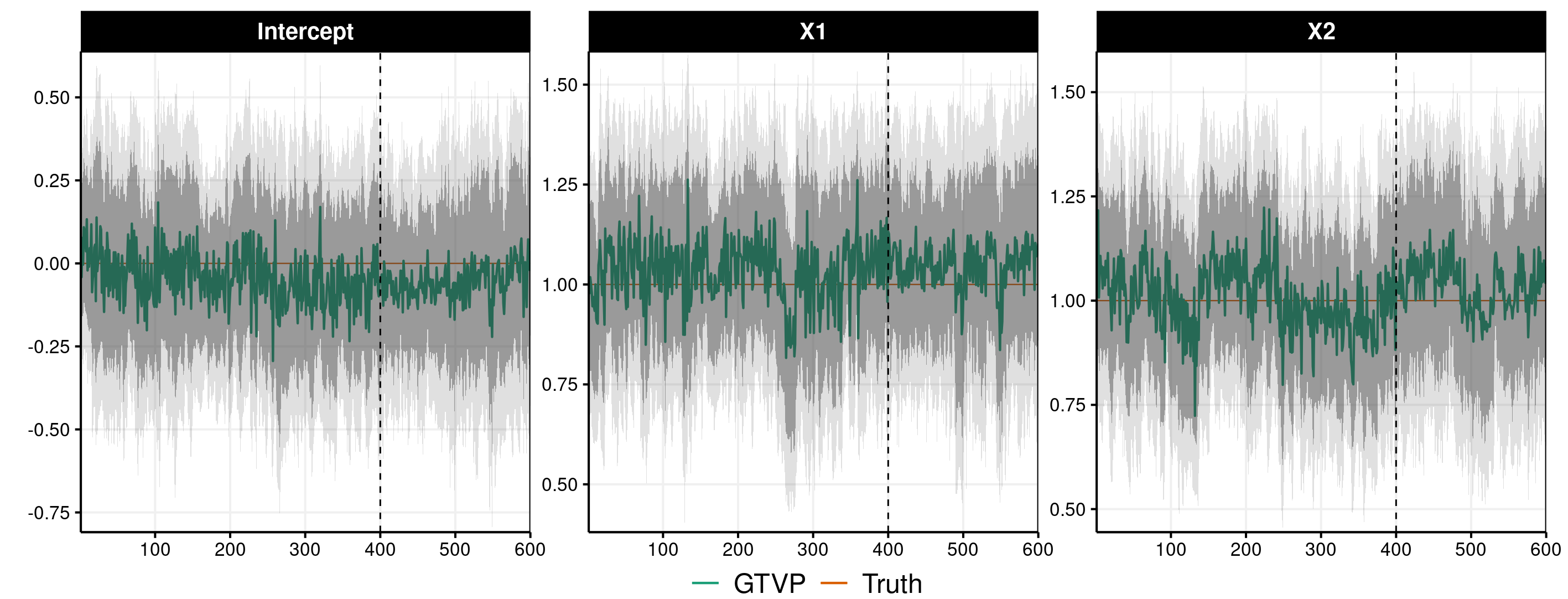}
    \caption{DGP 6}
        \label{newsimul_v6}
                  \end{center}
      \end{subfigure}
  \caption{\footnotesize The grey bands are the 68\% and 90\% credible region. After the blue line is the hold-out sample. Green line is the posterior mean and orange is the truth. The plots include only the first 400 observations.}
    \label{newsimul_graph}
\end{figure}

\begin{figure}[h!]
\center
   \includegraphics[width=\textwidth]{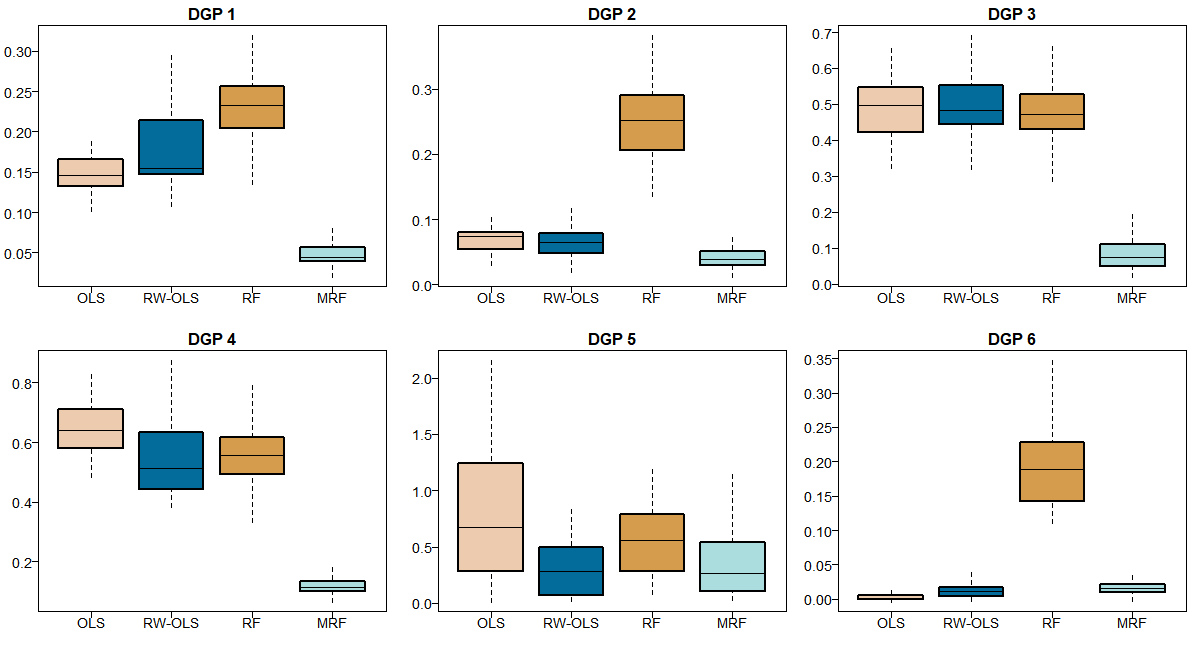}      
\caption{\footnotesize The distribution of RMSE dis-improvements with respect to the oracle's forecast for 4 models: OLS, Rolling-Window OLS, plain RF, MRF. 50 simulations of 750 OOS forecasts each.}  
\vspace{-0.3cm}
  \label{ns_rmse}
\end{figure}

\subsection{Monthly Forecasting Results}\label{sec:month}

I run a similar exercise as in \cite{GCLSS2018} which is very close to what has been precedently conducted for quarterly data. FRED-\textbf{M}D is now used. It contains 134 monthly US macroeconomic and financial indicators observed from 1960M01 \citep{mccrackenng}. To match the experimental design of \cite{GCLSS2018} for ML methods, Industrial Production (IP) replaces GDP and IR is dropped. The horizons of interest are $h=1,3,9,12,24$ months. The forecast target is the average growth rate 
$\sfrac{\sum_{h'}^h y_{t+h'}^v}{h}$
which is much less noisy than the monthly growth rate. For example, for inflation 24 months ahead, I target the average inflation rate over the next two years -- rather than the monthly inflation rate in 2 years. The OOS period is the same as before. 

\begin{figure}[h!]
\begin{center} 
\hspace*{-.3cm}\includegraphics[trim={1.5cm 1.5cm 0.25cm 0cm},clip,scale=.3]{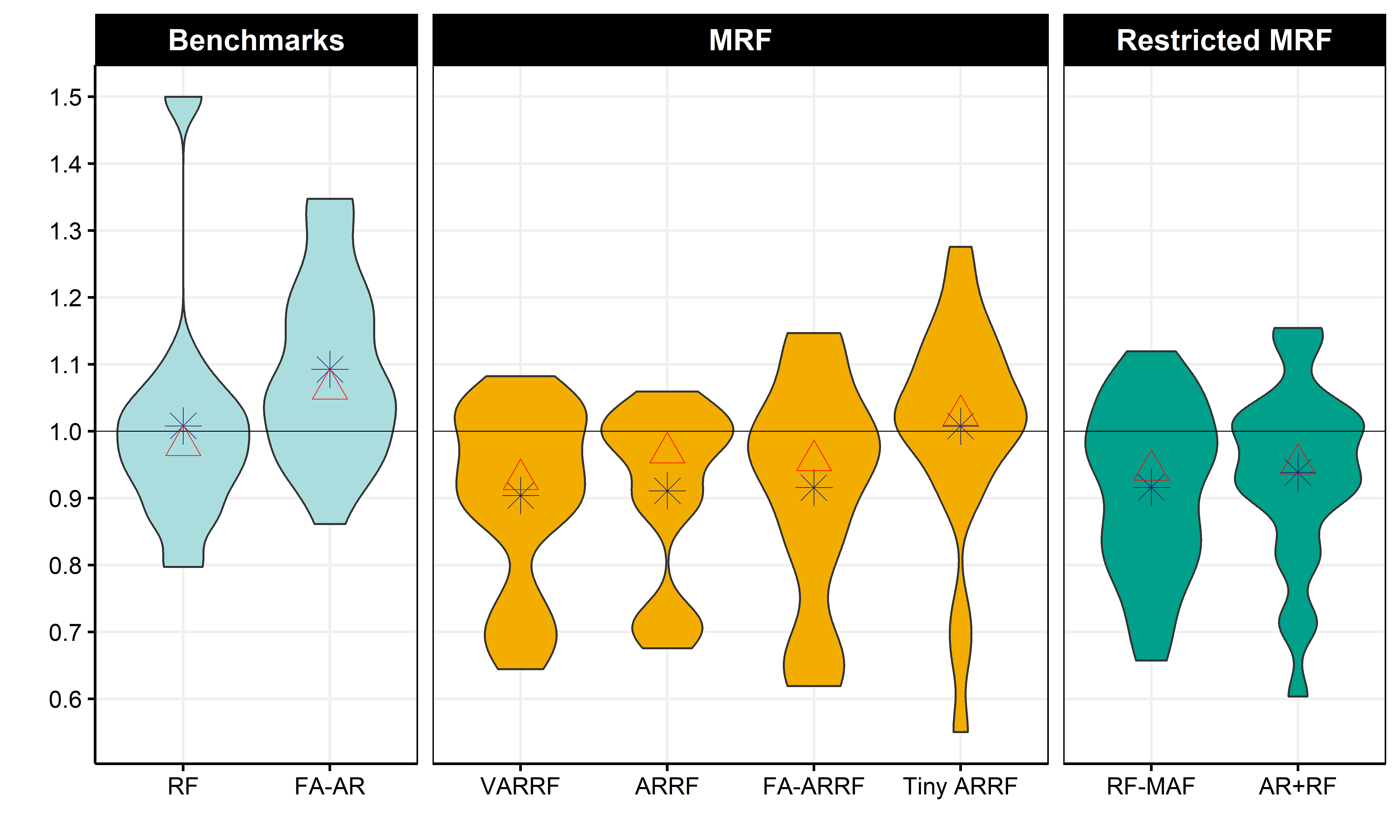}
\caption{\footnotesize The distribution of $RMSE_{v,h,m}/RMSE_{v,h, AR}$ for monthly data. The star is the mean and the triangle is the median.}\label{monthly_urn}
\end{center}
\end{figure} 

In Figure \ref{monthly_urn}, VARRF is now doing much better on average, ranking first in terms of mean improvement over AR. ARRF still provides great insurance against doing worse than a plain AR counterpart (here AR(12)).\footnote{This is also true for the more parsimonious AR, see Table \ref{M_table}.} FA-ARRF remains very competitive. The models that do not have the MAFs (benchmarks) are clearly outperformed by the rest that do. This  unsurprisingly indicates that lag polynomial compression can be of even greater use at the monthly frequency. 

Table \ref{M_table} reports specific $RMSE_{v,h,m}/RMSE_{v,h, AR}$'s with Diebold-Mariano tests. Broadly, they show that (i) MAFs are without any doubt the major improvement for the first three variables (IP, UR, SPREAD), (ii) simpler approaches like RF-MAF and AR+RF do well (\textit{except} for INF) (iii) \textit{all} MRFs do very well for inflation. Particularly, for (iii), ARRF and Tiny ARRF provide significant gains of 33\% and 45\% over the benchmark at $h=12$ and $h=24$, respectively. It is clear from this evidence, and that of the quarterly section, that forcing time-invariant inflation dynamics is costly in terms of RMSPE. GTVPs will confirm that, in accord with classic evidence on the matter \citep{cogley2001evolving}. 

Gains for INF are miles ahead from the usual competition. Table \ref{M_table} includes forecasts inspired by the contribution of \cite{atkeson2001phillips}: 1, $h$ and 12 months moving averages are considered (where $h$ is the targeted horizon). As in the original paper, the "AO-12" forecasts prove remarkably resilient, but are bested with sizable margins at each horizons by ARRF, Tiny ARRF, and FA-ARRF. For instance, at $h=24$, the next best non-MRF forecast delivers 16\% gains over the benchmark AR, whereas the worst MRF provides a gain of 27\%. Tiny ARRF supremacy at longer horizons is sensible given that restricting $S_t$ emphasizes long-run exogenous change, a usual suspect for INF.

Another interesting observation emerges from MRFs successes with monthly inflation. FA-ARRF is often close to the best model, and that, at all horizons. Naturally, this is intriguing as FA-ARRF can be thought of as a Phillips' curve forecast, which recurrent failures are well documented \citep{atkeson2001phillips,stock2007has}. Moreover, it is reported that FA-AR, in contrast, does really bad. To sort this out, FA-ARRF's GTVPs are studied in section \ref{PCanal}.


\subsubsection{Non-US Data}

Much attention has been paid to the prediction of US economic aggregates. An even greater challenge is that of forecasting the future state of a small open economy. Such an application is beyond the scope of this paper but is considered in \cite{rapportMFQ}. The study considers the prediction of more than a dozen key economic variables for Canada and Québec using the large Canadian data base of \cite{dalibordata}. Forecasts from about 50 models and different averages of them are compared, with ARRF and FA-ARRF among them. MRFs generate substantial improvements especially at the one-quarter horizon for numerous real activity variables (Canadian GDP, Québec GDP, industrial production, real investment). In such cases, ARRF or FA-ARRF provide reductions (with respect to autoregressive benchmark) that are sizable and statistically significant, going up to 32\% in RMSE. That performance is sometimes miles ahead from the next best option (among Complete Subset Regression, Factor models, Neural Networks, Ridge, Lasso, plain RF and different model averagaging schemes). \cite{rapportMFQ}'s results suggest that MRFs forecasting abilities generalize beyond the traditional exercise of predicting US aggregates. 

More recently, \cite{GCMS} uses MRF (along with a plethora of ML models) with a newly-built large UK macro data base, and finds that it can provide substantial gains during the Pandemic Recession. One of the reasons for that is the capacity of MRF to be nonlinear \textit{and} extrapolate, which off-the-self tree-based methods (like RF) lack.

\subsection{Additional Figures and Tables}\label{sec:addigraphs}


\setstretch{0.75}

\begin{figure}[h!]
  \begin{subfigure}[b]{0.49\textwidth}
\hspace{-0.5cm}\includegraphics[scale=.18]{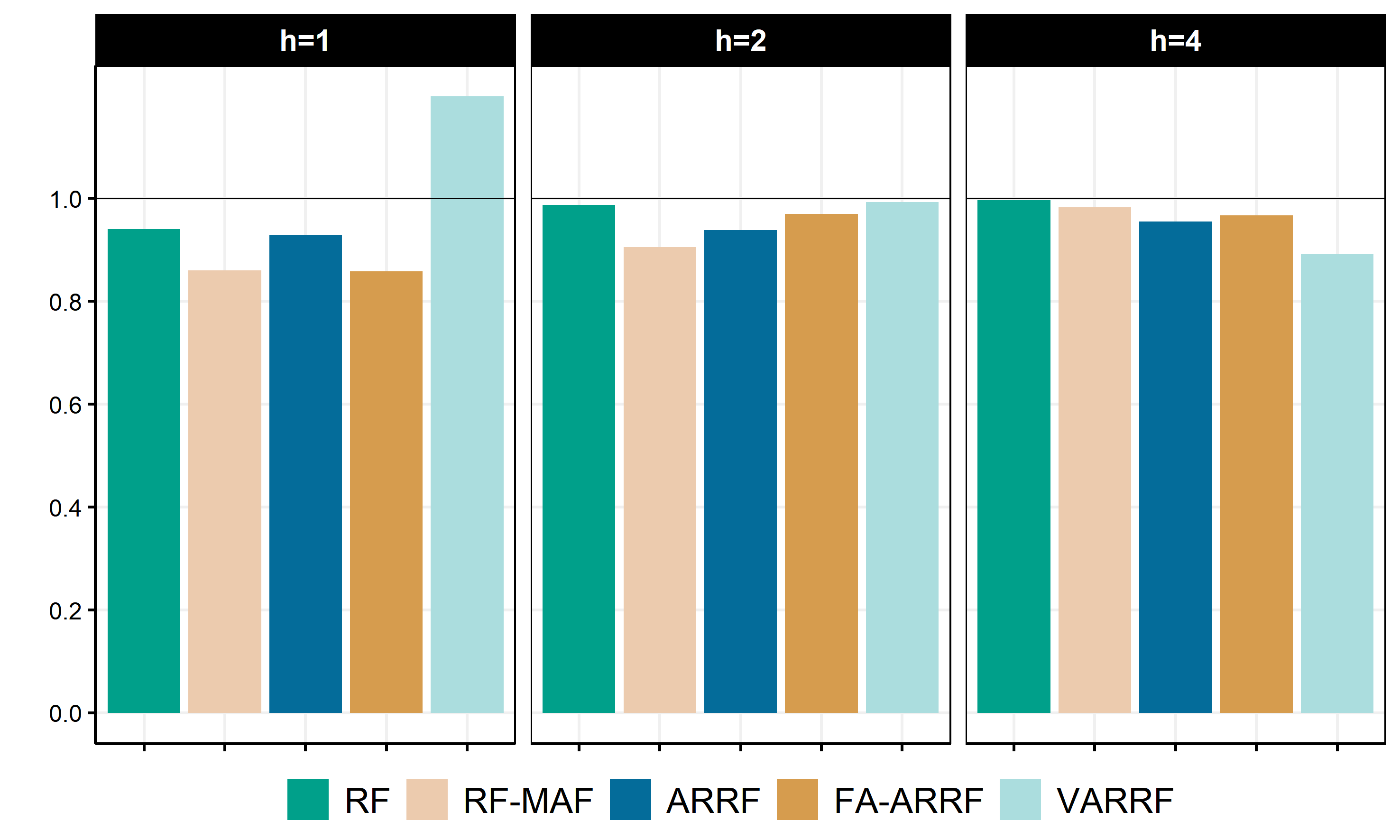}
\caption{$RMSE_{\textcolor{black}{GDP},h,m}/RMSE_{\textcolor{black}{GDP},h, AR}$ }  
  \end{subfigure}
  \hspace{2em}
  \begin{subfigure}[b]{0.49\textwidth}
\hspace{-1cm}\includegraphics[scale=.18]{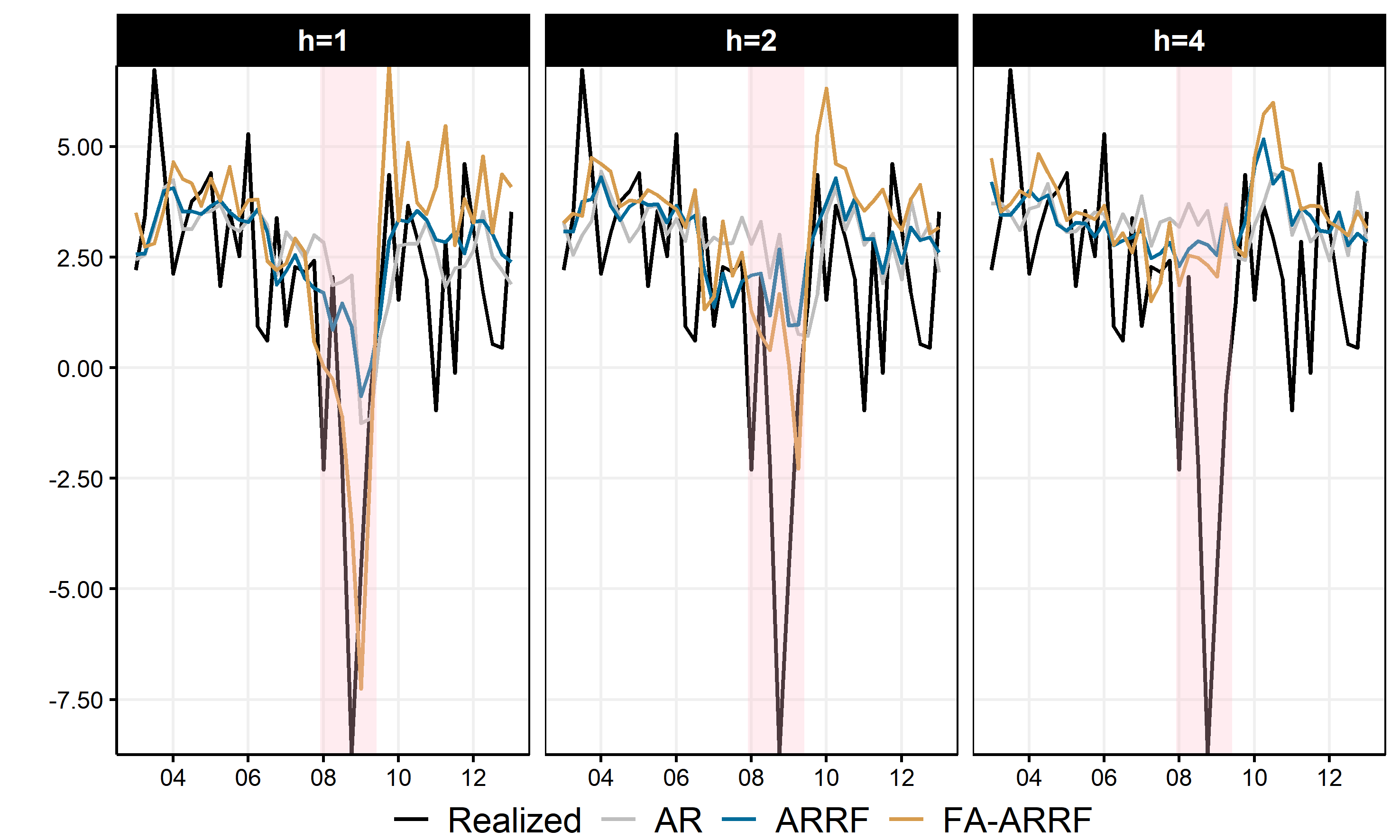}
\caption{A look at forecasts}  
  \end{subfigure}
  \caption{GDP results in detail}  
\label{GDP_detail}
\end{figure}

\begin{figure}[h!]
  \begin{subfigure}[b]{0.49\textwidth}
\hspace{-0.5cm}\includegraphics[scale=.18]{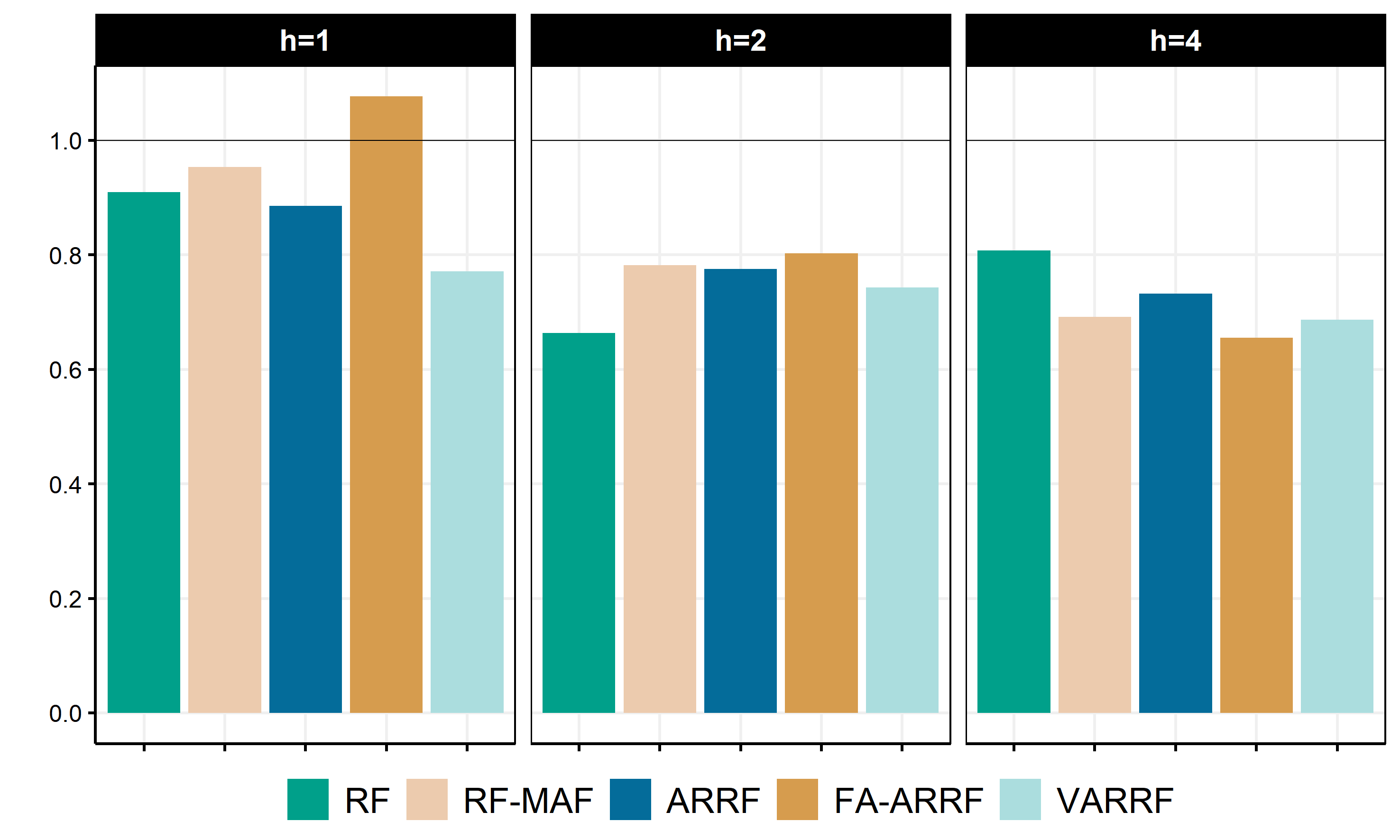}
\caption{$RMSE_{\textcolor{black}{SPREAD},h,m}/RMSE_{\textcolor{black}{SPREAD},h, AR}$ }  
  \end{subfigure}
  \hspace{2em}
  \begin{subfigure}[b]{0.49\textwidth}
\hspace{-1cm}\includegraphics[scale=.18]{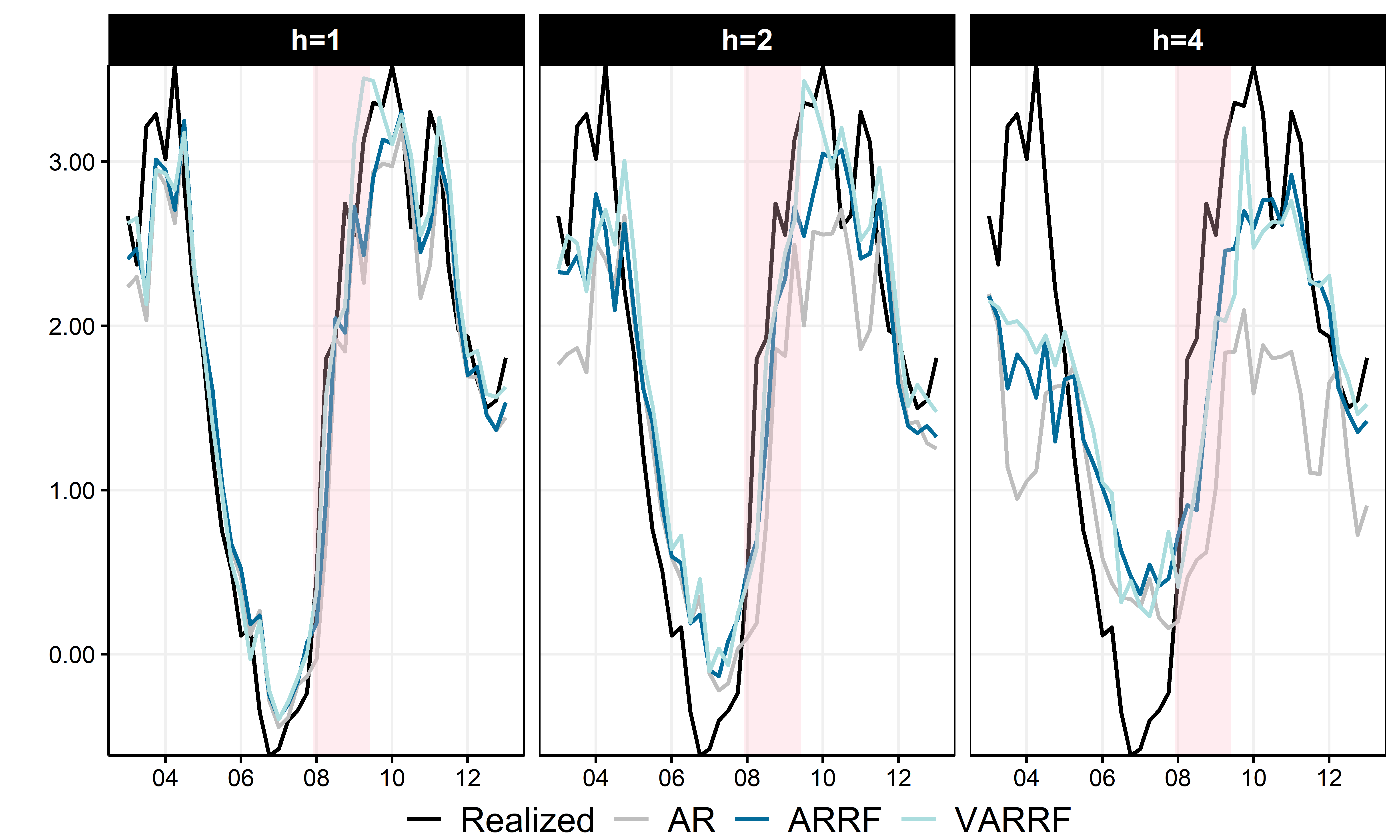}
\caption{A look at forecasts}  
  \end{subfigure}
  \caption{SPREAD results in detail}  
\label{SPREAD_detail}
\end{figure}

\begin{figure}[h!]
  \begin{subfigure}[b]{0.49\textwidth}
\hspace{-0.5cm}\includegraphics[scale=.18]{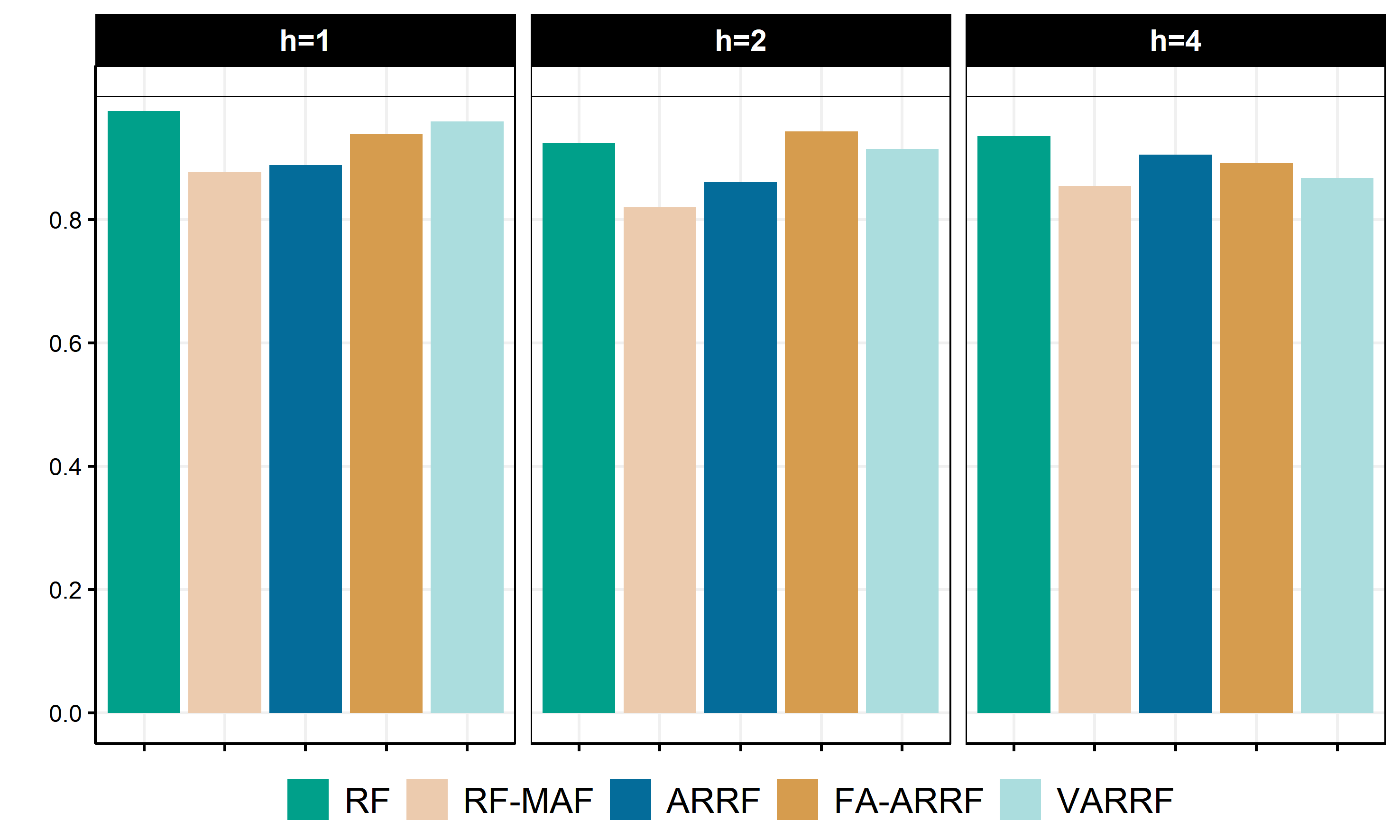}
\caption{$RMSE_{\textcolor{black}{INF},h,m}/RMSE_{\textcolor{black}{INF},h, AR}$ }  
  \end{subfigure}
  \hspace{2em}
  \begin{subfigure}[b]{0.49\textwidth}
\hspace{-1cm}\includegraphics[scale=.18]{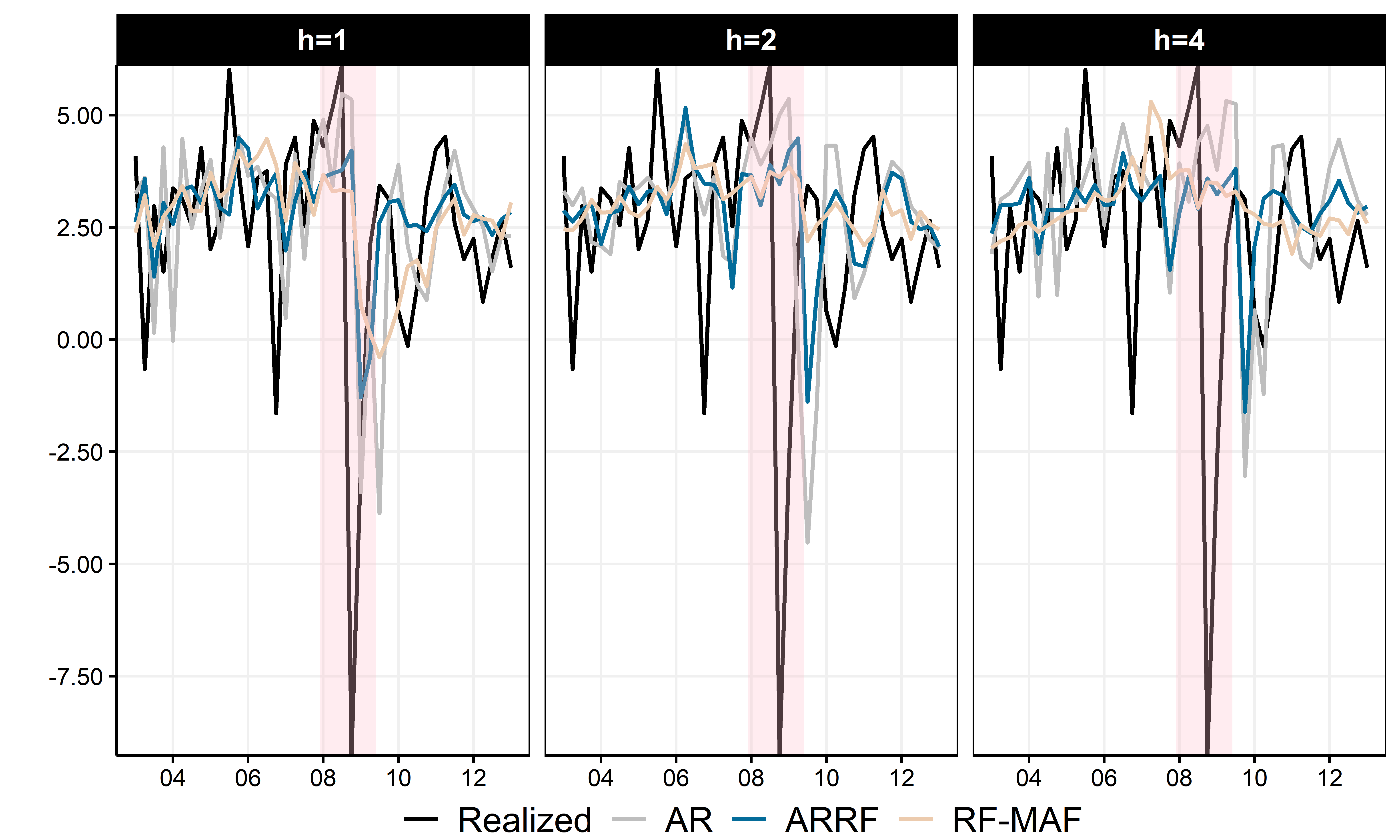}
\caption{A look at forecasts}  
  \end{subfigure}
  \caption{INF results in detail}  
\label{INF_detail}
\end{figure}

\begin{figure}[ht!]
\begin{center} 
\hspace{-0.5cm}\includegraphics[scale=.25]{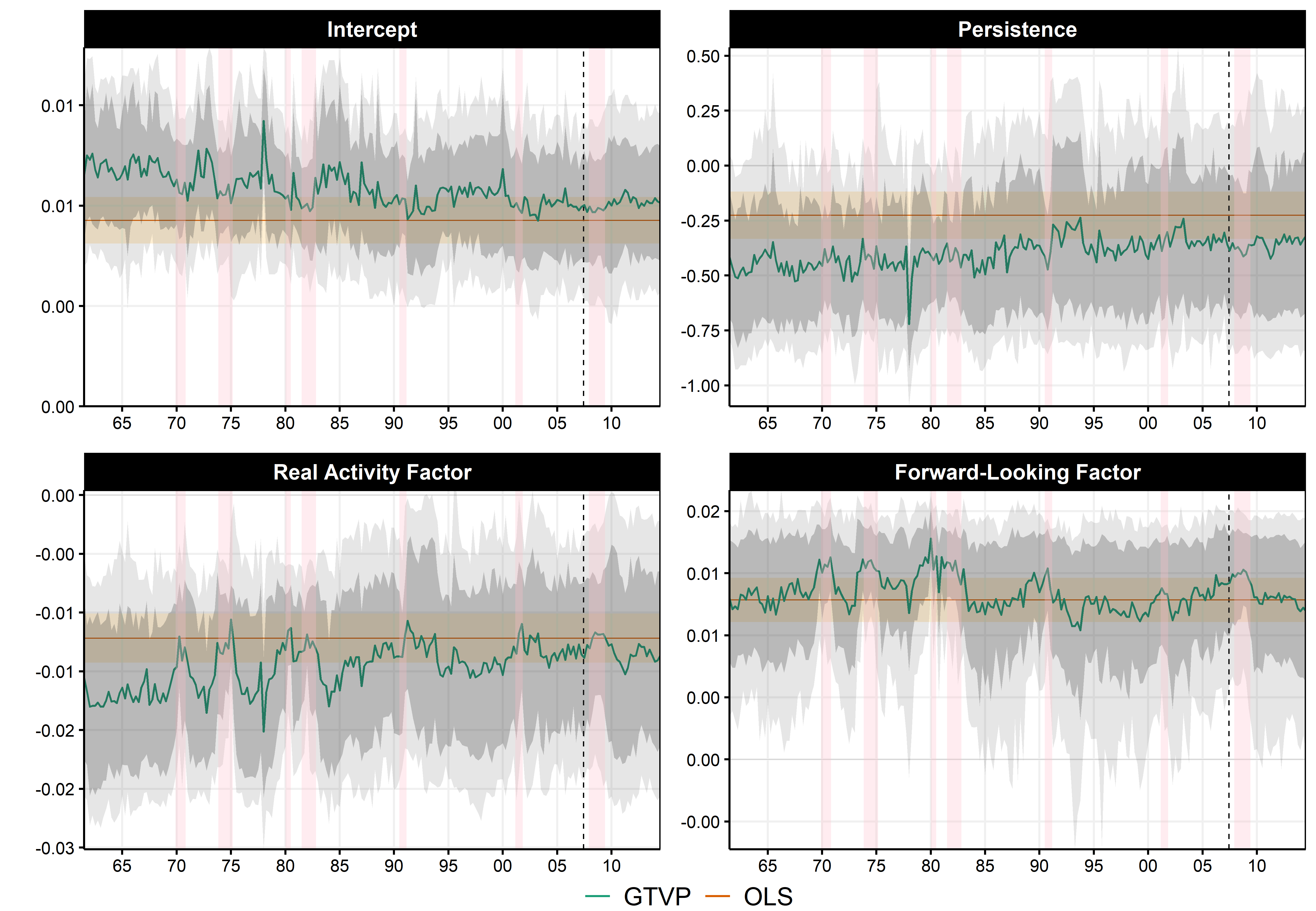}  
\caption{\footnotesize GTVPs of the one-quarter ahead GDP forecast. Persistence is defined as $\phi_{1,t}+\phi_{2,t}$. The grey bands are the 68\% and 90\% credible region. The pale orange region is the OLS coefficient $\pm$ one standard error. The vertical dotted blue line is the end of the training sample. Pink shading corresponds to NBER recessions.}
\label{v1h1_betas}

\end{center}
\end{figure}

\begin{figure}[ht!]
\begin{center} 
\hspace*{-1cm}\includegraphics[scale=.25]{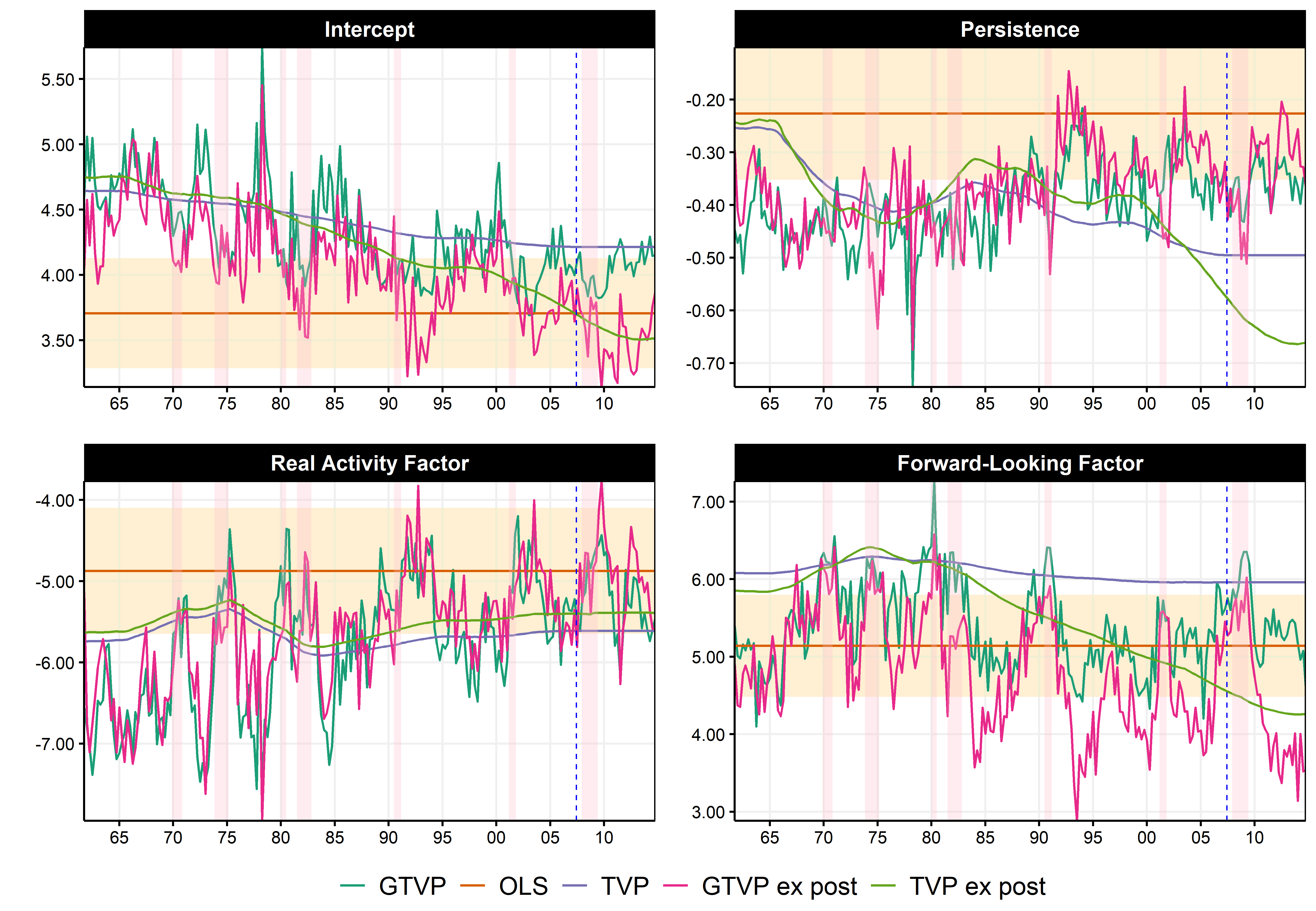}  
\caption{\footnotesize GDP equation $\beta_t$'s obtained with different techniques. Persistence is defined as $\phi_{1,t}+\phi_{2,t}$. TVPs estimated with a ridge regression as in \cite{GC2019} and the parameter volatility is tuned with k-fold cross-validation. Ex Post TVP means using the full sample for estimation and tuning as opposed to only using pre-2002 data as for GTVPs. The pale orange region is the OLS coefficient $\pm$ one standard error. Pink shading corresponds to NBER recessions.}
\label{v1h1_tvpcomp}
\end{center}
\end{figure}

\begin{figure}[ht!]
  \begin{subfigure}[b]{\textwidth}
   \includegraphics[width=\textwidth]{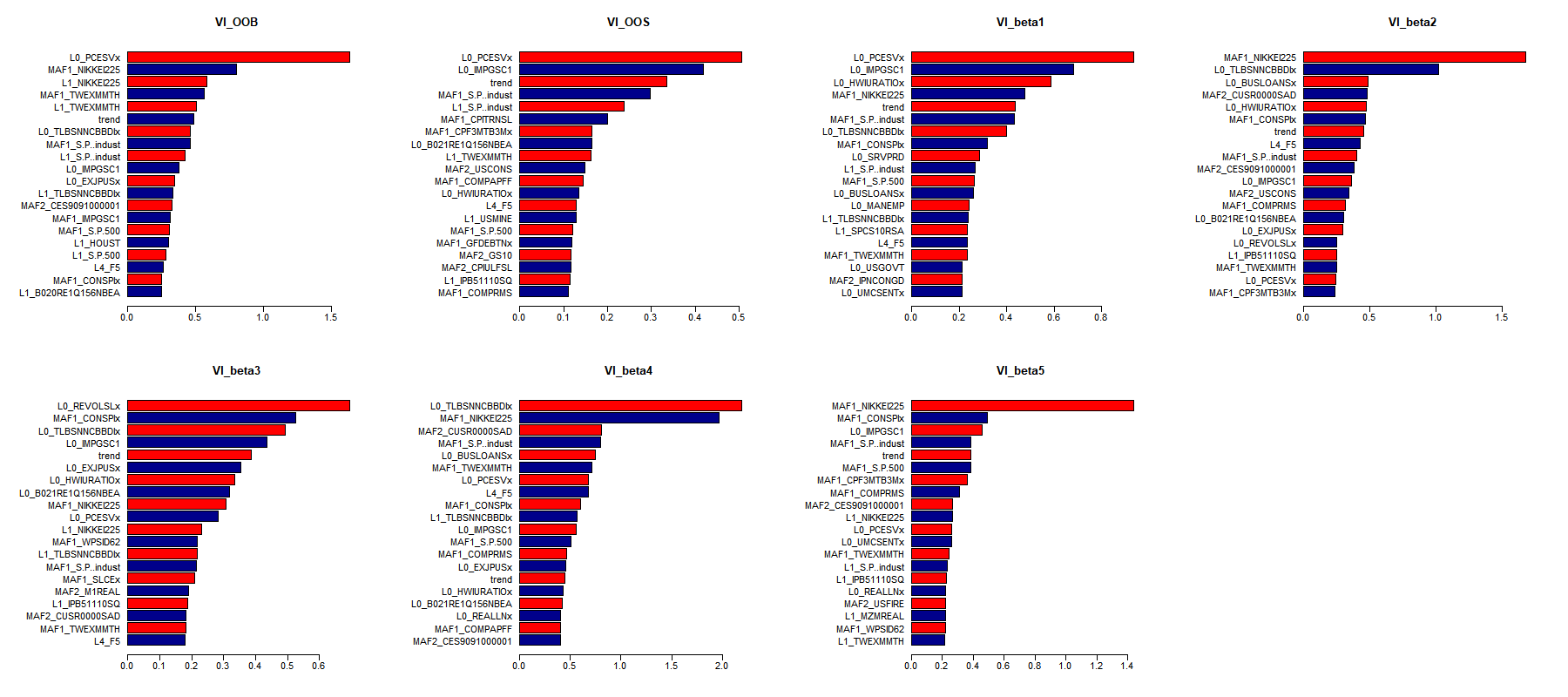}
    \caption{GDP horizon 1}
        \label{VI_Q_v1h1}

  \end{subfigure}
  \hspace{2em}
  \begin{subfigure}[b]{\textwidth}
   \includegraphics[width=\textwidth]{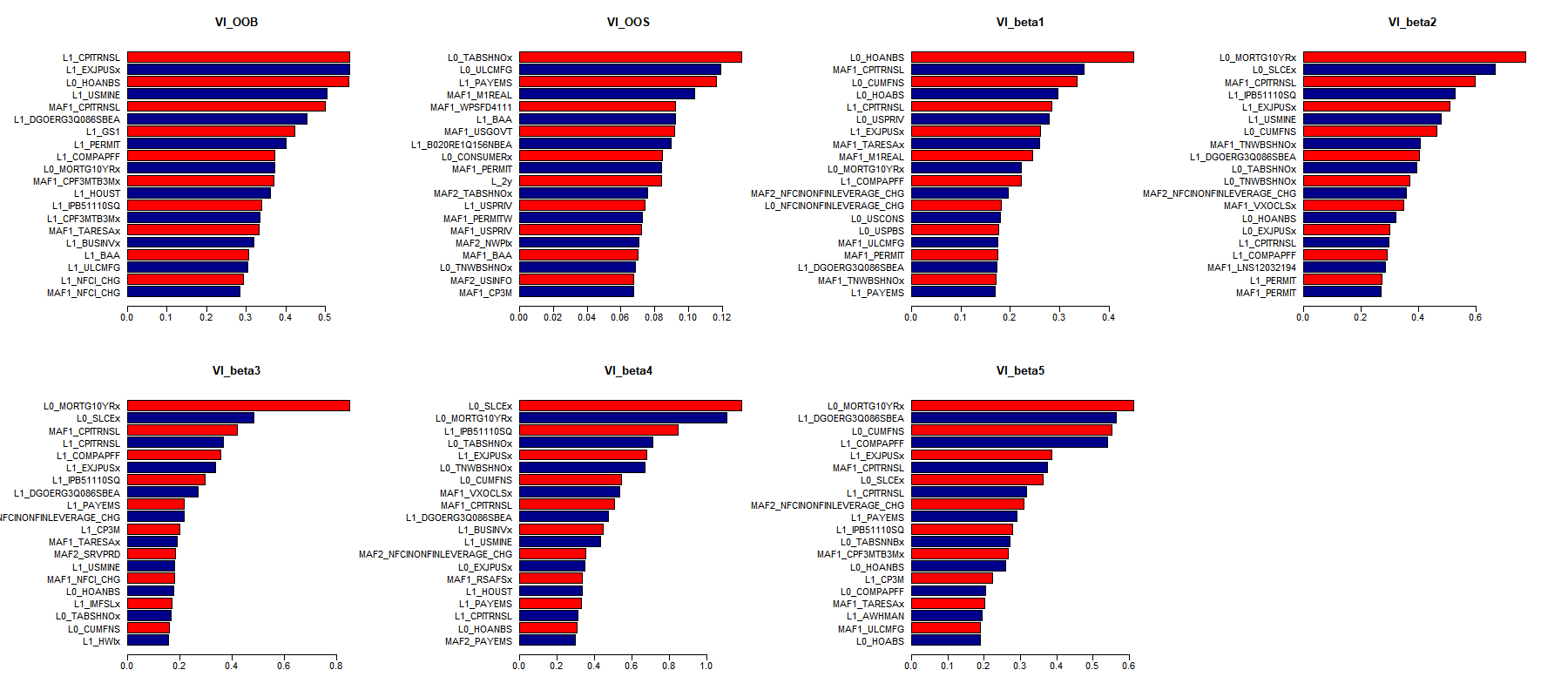}
    \caption{UR horizon 1}
        \label{VI_Q_v2h1}

      \end{subfigure}
  \caption{20 most important series according to the various variable importance (VI) criteria. Units are relative RMSE gains (in percentage) from including the specific predictor in the forest part. $VI_{OOB}$ means VI for the out-of-bag criterion. $VI_{OOS}$ is using the hold-out sample. $VI_{\beta}$ is an out-of-bag measure of how much $\beta_{t,k}$ varies by withdrawing a certain predictor.}  
    \label{VI_Q}
\end{figure}

\clearpage

\begin{figure}[ht!]
  \begin{subfigure}[b]{\textwidth}
   \includegraphics[width=\textwidth]{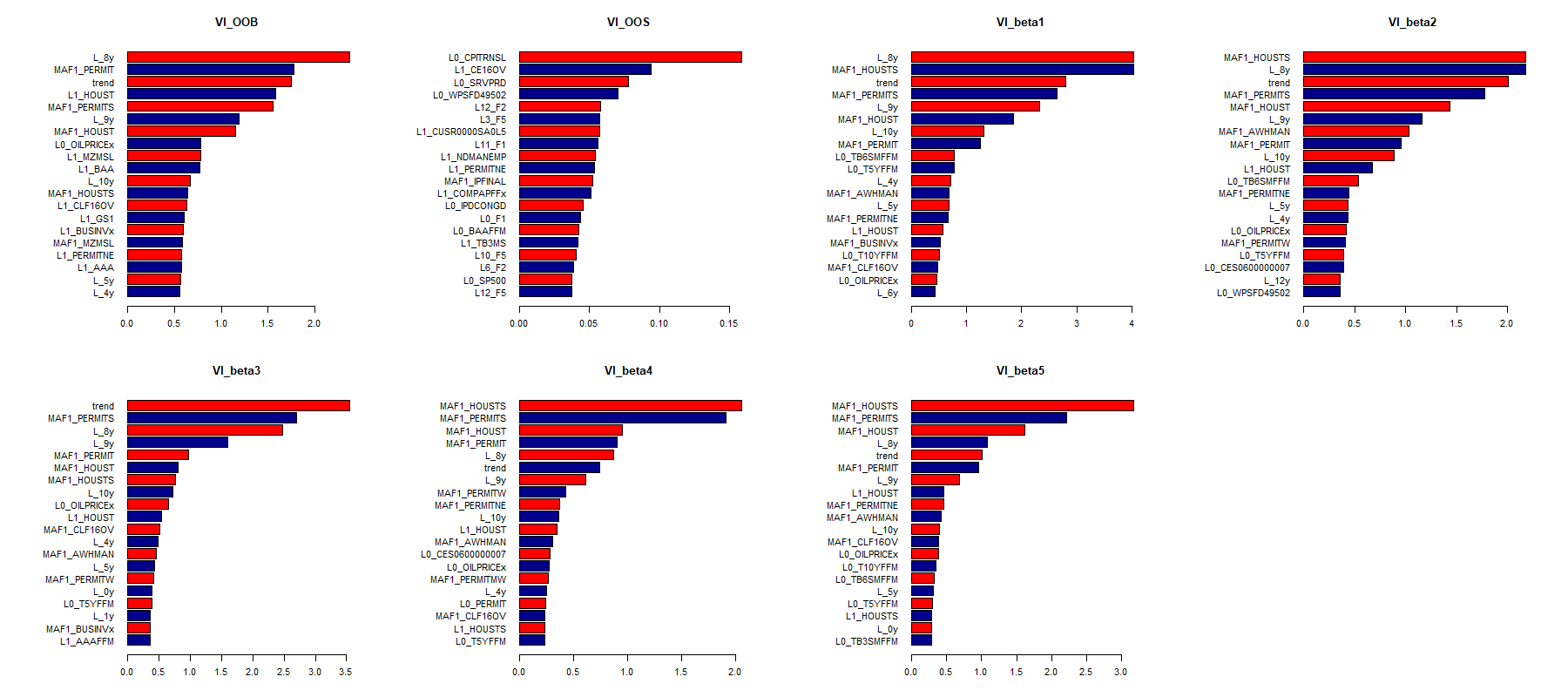}
    \caption{One month ahead inflation forecast}
        \label{VI_M_v4h1}

  \end{subfigure}
  \hspace{2em}
  \begin{subfigure}[b]{\textwidth}
   \includegraphics[width=\textwidth]{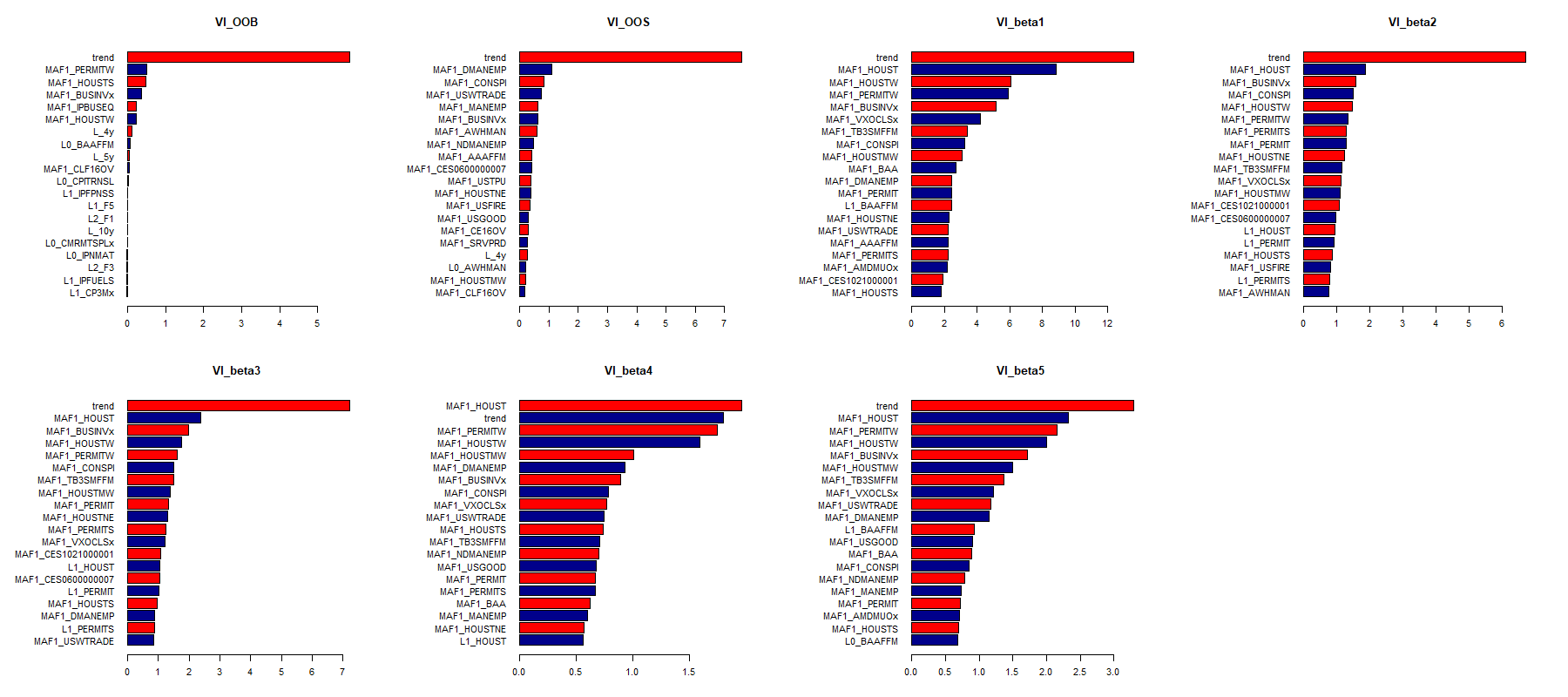}
    \caption{Average inflation over the next 12 months}
        \label{VI_M_v4h4}

      \end{subfigure}
  \caption{20 most important series according to the various variable importance (VI) criteria. Units are relative RMSE gains (in percentage) from including the specific predictor in the forest part. $VI_{OOB}$ means VI for the out-of-bag criterion. $VI_{OOS}$ is using the hold-out sample. $VI_{\beta}$ is an out-of-bag measure of how much $\beta_{t,k}$ varies by withdrawing a certain predictor.}  
    \label{VI_M}
\end{figure}

\begin{figure}[ht!]
\begin{center} 
  \begin{subfigure}[b]{\textwidth}
  \begin{center}
   \includegraphics[width=0.85\textwidth]{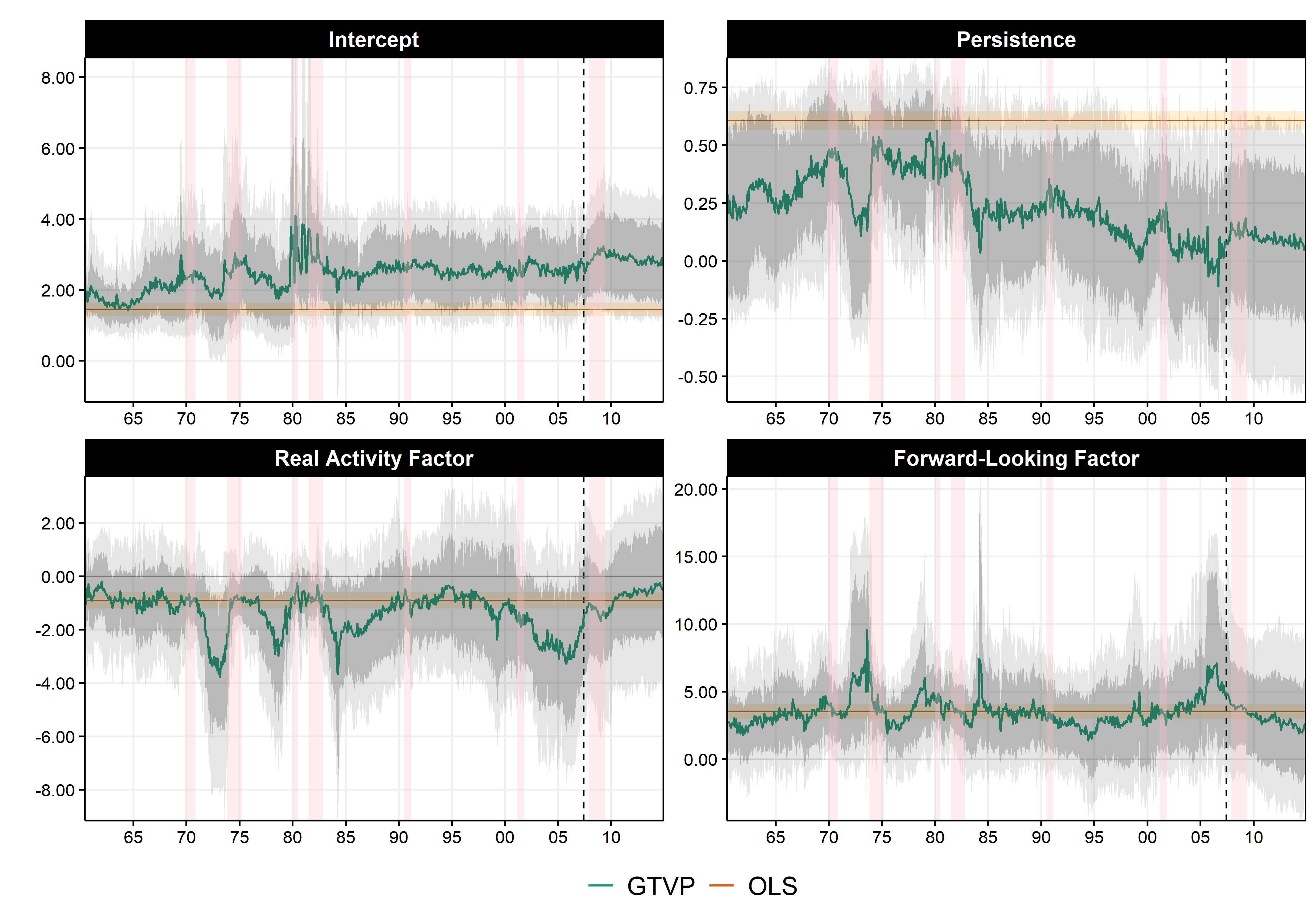}
    \caption{One-month ahead}
        \label{v4h1_betas_M}
    \end{center}
  \end{subfigure}
  \hspace{2em}
  \begin{subfigure}[b]{\textwidth}
  \begin{center} 
   \includegraphics[width=0.85\textwidth]{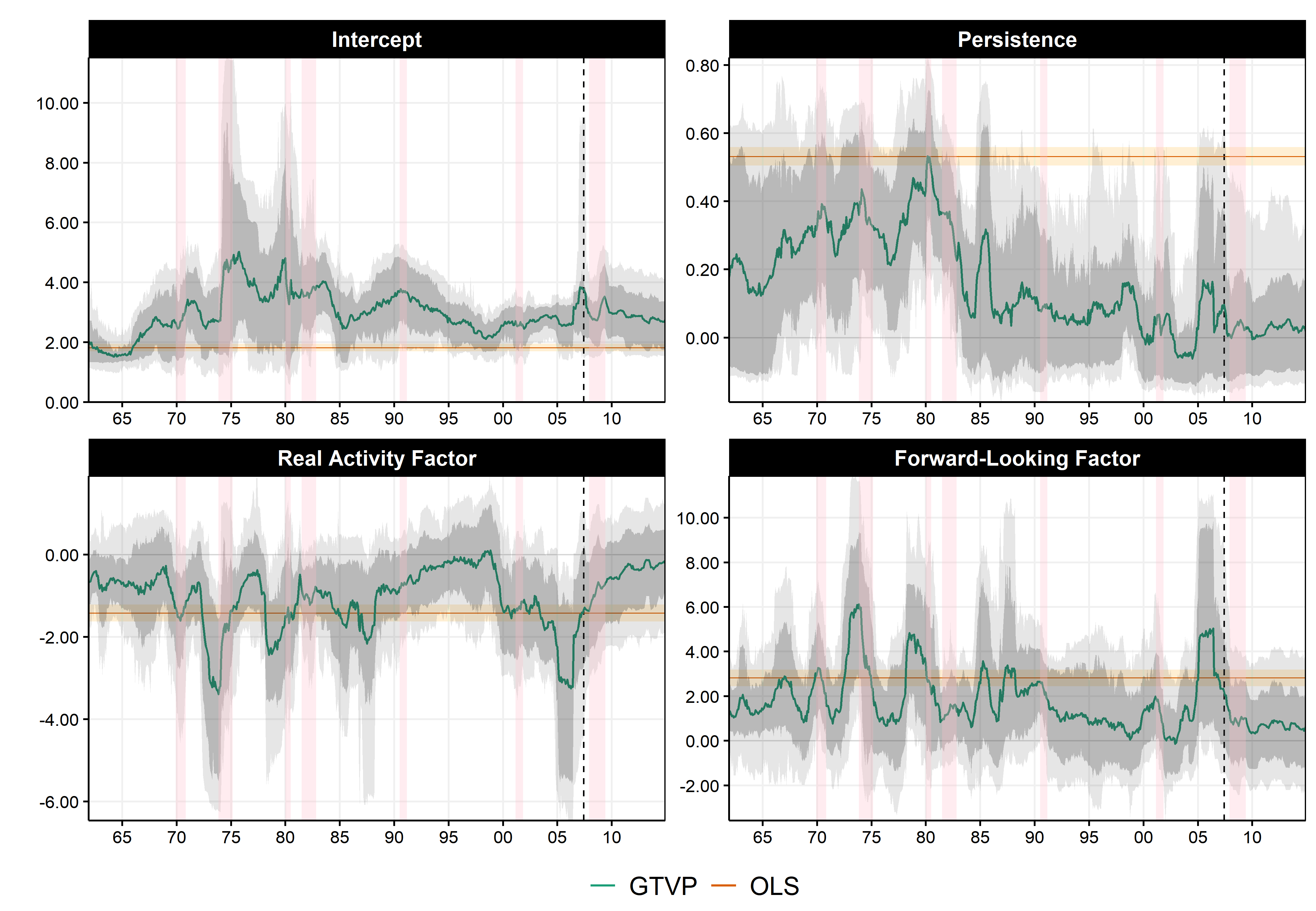}
   \caption{12-months ahead}
        \label{v4h4_betas_M}
    \end{center}
      \end{subfigure}
  \caption{\footnotesize GTVPs of monthly inflation forecast. The grey bands are the 68\% and 90\% credible regions. The pale orange region is the OLS coefficient $\pm$ one standard error. The vertical dotted line is the end of the training sample. Pink shading corresponds to NBER recessions.}  
    \label{v4_betas_M}
    \end{center}
\end{figure}

\begin{figure}[ht!]
\begin{center} 
  \begin{subfigure}[b]{\textwidth}
\begin{center} 
\hspace*{-0.5cm}\includegraphics[width=0.8\textwidth]{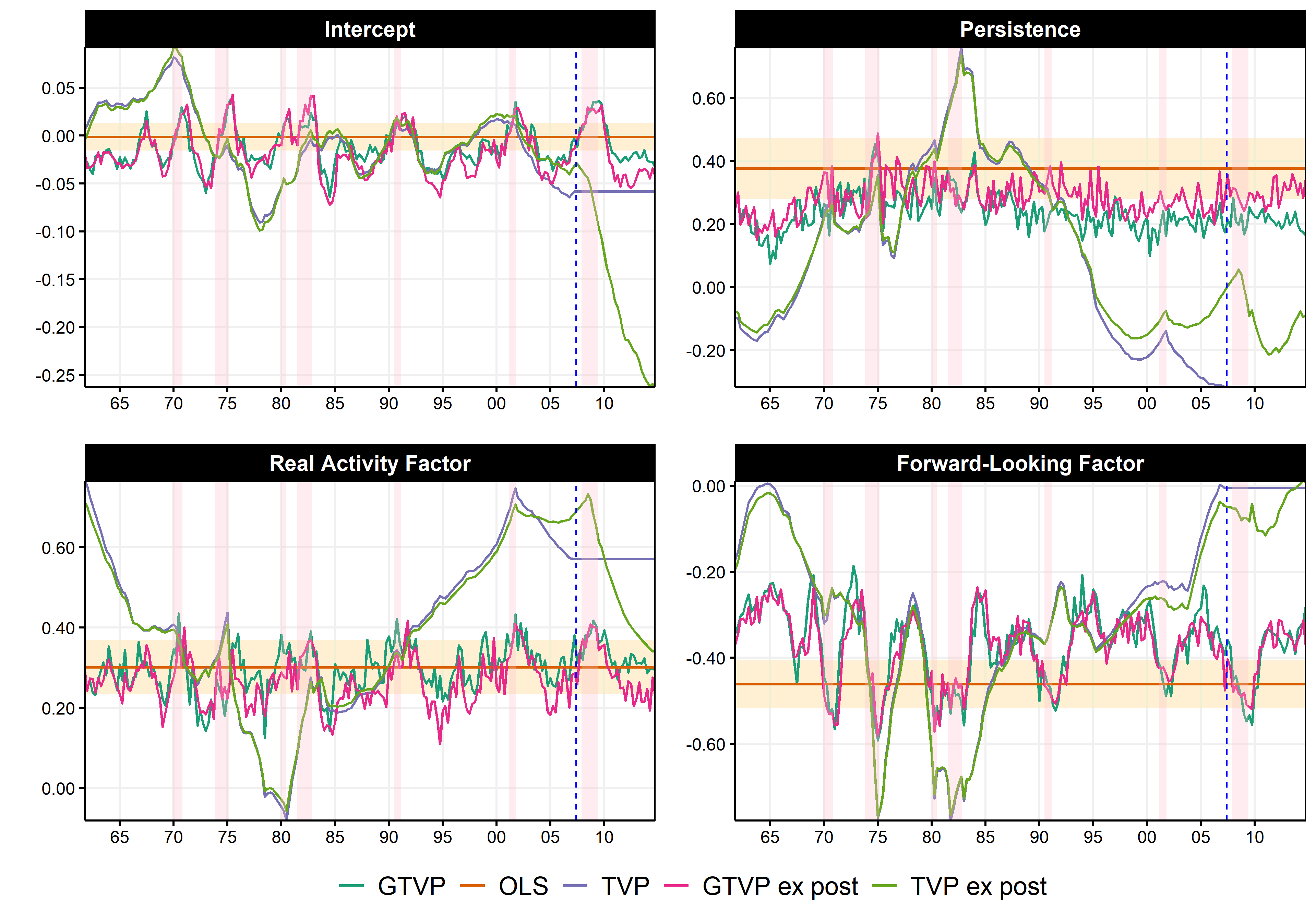}  
\caption{UR equation}
\label{v2h1_tvpcomp_lambdatresdoux}
\end{center}
  \end{subfigure}
  \hspace{2em}
  \begin{subfigure}[b]{\textwidth}
\begin{center} 
\hspace*{-0.5cm}\includegraphics[width=0.8\textwidth]{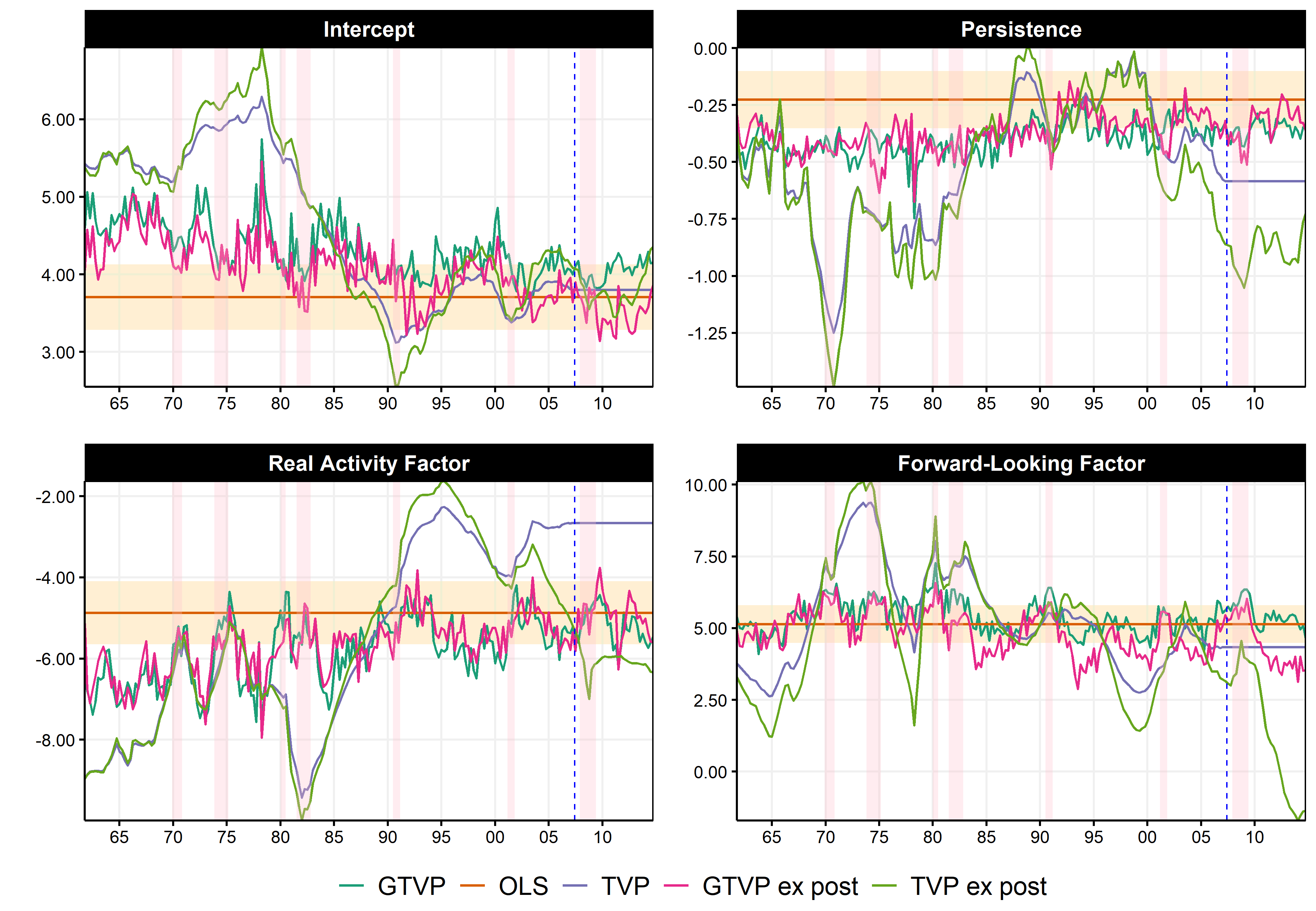}  
\caption{GDP equation}
\label{v1h1_tvpcomp_lambdatresdoux}
\end{center}
      \end{subfigure}
  \caption{\footnotesize $\beta_t$'s obtained with different techniques. TVPs estimated with a ridge regression as in \cite{GC2019} and the parameter volatility $\lambda$ is tuned with k-fold cross-validation, \textbf{then divided by 100}. This means the standard deviation of parameters shocks is allowed to be about 10 times higher than what CV recommends. Ex Post TVP means using the full sample for estimation and tuning as opposed to only using pre-2002 data as for GTVPs. The pale orange region is the OLS coefficient $\pm$ one standard error.}  
    \label{tvpcomp_lambdatresdoux}
    \end{center}
\end{figure}

\begin{figure}[ht!]
   \includegraphics[width=\textwidth]{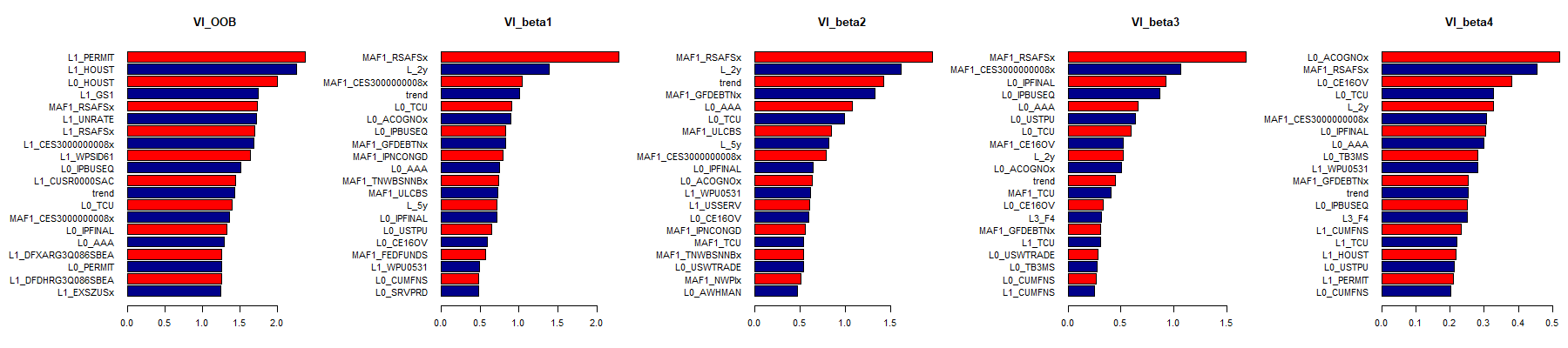}
    \caption{20 most important series according to the various variable importance (VI) criteria. Units are relative RMSE gains (in percentage) from including the specific predictor in the forest part. $VI_{OOB}$ means VI for the out-of-bag criterion. $VI_{OOS}$ is using the hold-out sample. $VI_{\beta}$ is an out-of-bag measure of how much $\beta_{t,k}$ varies by withdrawing a certain predictor.}
        \label{VI_BCS}
\end{figure}

\begin{figure}[ht!]
  \hspace{2em}
\begin{center} 
\hspace{-0.5cm}\includegraphics[scale=.3]{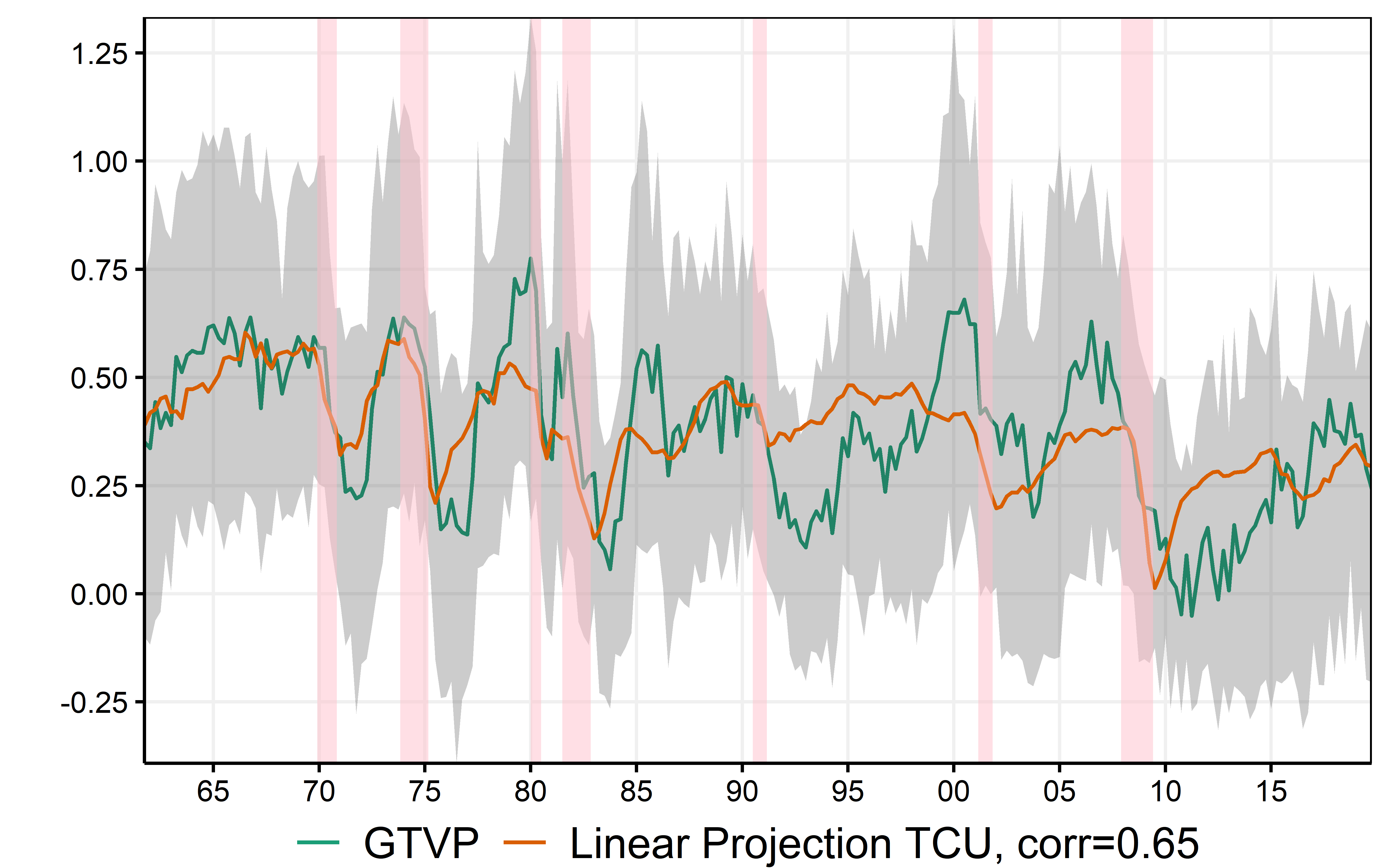}  
\end{center}
\caption{\footnotesize $\beta_{3,t}$ in \eqref{eq:bcs2} with additional controls for supply and monetary policy shocks. Capacity Utilization is still substantially correlated with the inflation-unemployment trade-off. The grey band is the 68\% credible region. Pink shading corresponds to NBER recessions.}
\label{beta_bcs_tcu_controls}
\end{figure}




\tiny
\input{tables/FCST_tableQ_200929}

\input{tables/FCST_tableM_200929}

\pagebreak

\normalsize

\end{document}

%% file: tables/FCST_tableQ_200929.tex
\begin{landscape}\begin{table}[!tbp]
\caption{Main Quarterly Results \label{Q_table}} 
\begin{center}
\footnotesize\addtolength{\tabcolsep}{-2.5pt}
\begin{tabular}{lllllllllllllll}
\hline\hline
\multicolumn{1}{l}{}&\multicolumn{1}{c}{FA-AR}&\multicolumn{1}{c}{LASSO-MAF}&\multicolumn{1}{c}{Ridge-MAF}&\multicolumn{1}{c}{RF}&\multicolumn{1}{c}{RF-MAF}&\multicolumn{1}{c}{AR+RF}&\multicolumn{1}{c}{Tiny RF}&\multicolumn{1}{c}{FA-ARRF}&\multicolumn{1}{c}{ARRF}&\multicolumn{1}{c}{Tiny ARRF}&\multicolumn{1}{c}{VARRF}&\multicolumn{1}{c}{SETAR}&\multicolumn{1}{c}{STAR}&\multicolumn{1}{c}{TV-AR}\tabularnewline
\hline
{\bfseries GDP}&&&&&&&&&&&&&&\tabularnewline
~~h=1&1.02&0.96&0.89**&0.94&0.86&0.89&1.03&\textbf{0.86}&0.93&1.04&1.20&1.01&1.03&0.99\tabularnewline
~~h=2&0.96&0.98&0.98&0.99&\textbf{0.91}&0.93&1.01&0.97&0.94**&1.03&0.99&0.97&0.98&1.03\tabularnewline
~~h=4&1.03&0.98&0.99***&1.00&0.98&0.99&1.03&0.97&0.95&0.98&\textbf{0.89}&0.97***&0.96***&0.96\tabularnewline
~~h=6&1.36&0.98&0.98&0.98&1.00&1.00&1.08&1.01&0.97&0.98&1.00&0.98&\textbf{0.95}&0.98\tabularnewline
~~h=8&1.37&1.00&0.99&0.99&0.99&\textbf{0.96}&1.15&1.06&1.00&1.01&1.04***&1.00&0.97&1.00\tabularnewline
\hline
{\bfseries UR}&&&&&&&&&&&&&&\tabularnewline
~~h=1&0.83&0.99&0.99&1.00&0.85*&0.84&1.24**&\textbf{0.72}&0.90***&1.00&1.24&1.18&1.10&1.00\tabularnewline
~~h=2&0.80&0.98&0.92*&0.98&0.85&0.84&1.15*&\textbf{0.76}&0.90&0.96&0.89&1.03&0.97&0.99\tabularnewline
~~h=4&0.88&0.96***&0.94**&0.96*&0.87*&0.84*&1.37&\textbf{0.79}&0.87&0.92&0.91&1.02&1.01&1.34\tabularnewline
~~h=6&1.18*&0.98&0.98&1.01&0.94&0.90&1.60*&\textbf{0.89}&0.95&0.97&0.95&1.07&1.04&1.14\tabularnewline
~~h=8&1.25&0.98&1.01&1.01&\textbf{0.95}&0.95&1.57&1.01&0.98&0.98&1.04&1.09&1.06&1.11**\tabularnewline
\hline
{\bfseries SPREAD}&&&&&&&&&&&&&&\tabularnewline
~~h=1&1.28&2.16***&0.93&0.91&0.95&0.79**&0.96&1.08&0.89**&1.06&\textbf{0.77}**&1.51***&1.53***&0.98\tabularnewline
~~h=2&1.13&1.20&0.77&\textbf{0.66}**&0.78&0.72***&0.93&0.80&0.78**&1.11&0.74**&1.19&1.20&1.04\tabularnewline
~~h=4&0.86&0.95&1.01&0.81&0.69**&\textbf{0.61}**&1.48*&0.66**&0.73**&1.07&0.69**&1.04&1.06&1.30\tabularnewline
~~h=6&1.51&0.80*&1.13&0.98&0.80&0.80&1.43&\textbf{0.72}**&0.82&1.05&0.74*&1.03&1.06&1.19\tabularnewline
~~h=8&1.28&\textbf{0.76}**&0.96&0.92&0.83&0.89&1.36&0.82&0.88&0.99&0.85&1.11&1.14&0.99\tabularnewline
\hline
{\bfseries INF}&&&&&&&&&&&&&&\tabularnewline
~~h=1&1.01&0.93&0.95&0.98&0.88&1.23&0.90&0.94&0.89&\textbf{0.87}*&0.96&1.05&1.00&0.93\tabularnewline
~~h=2&1.01&0.96&0.92&0.92&\textbf{0.82}&1.00&0.88&0.94&0.86&0.87&0.91&0.86*&0.86&0.89\tabularnewline
~~h=4&1.08&0.92&0.87&0.94&\textbf{0.85}**&0.96&0.86&0.89&0.91*&0.95*&0.87*&0.90*&0.87*&0.91\tabularnewline
~~h=6&1.32&0.96&0.90&1.01&0.88&1.00&0.86&0.91&\textbf{0.85}&0.92**&0.87&0.94&0.89&0.98\tabularnewline
~~h=8&1.21&0.98&1.27&1.44&\textbf{0.88}*&0.94&0.88&0.91*&0.92&0.94&0.91*&0.96&0.92&0.98\tabularnewline
\hline
{\bfseries HOUST}&&&&&&&&&&&&&&\tabularnewline
~~h=1&1.13&1.04&0.94*&\textbf{0.92}*&1.00&1.01&1.24***&1.08&0.94**&0.95&1.09&1.01&0.99&1.00\tabularnewline
~~h=2&1.13&0.99&0.94**&0.95*&1.01&1.02&1.10*&1.06&1.00&1.02&0.99&\textbf{0.94}&0.97&1.01\tabularnewline
~~h=4&1.11&0.98**&0.97*&0.97&1.01&1.03&1.12&1.02&1.00&1.02&1.02&\textbf{0.95}&0.96&1.08\tabularnewline
~~h=6&1.40&0.96&0.96&0.96&0.96***&1.01&1.16&0.97***&0.99&1.00&0.98&\textbf{0.95}&0.96&0.99\tabularnewline
~~h=8&1.04&0.95&0.95&0.95&0.99&1.02&1.44&0.96&0.99&1.01&1.00&\textbf{0.95}&0.95&1.03\tabularnewline
\hline
{\bfseries IR}&&&&&&&&&&&&&&\tabularnewline
~~h=1&1.85&1.02&1.55&1.17&1.11&0.97&0.99&1.29&0.94&\textbf{0.92}&1.43&1.39&1.20&0.97\tabularnewline
~~h=2&1.49&0.96&1.01&1.00&0.93&0.98&1.29***&1.22&0.93&\textbf{0.92}&1.10&1.15&1.11&1.04\tabularnewline
~~h=4&\textbf{0.96}&1.00&1.03&1.03&1.04&0.99&1.39*&0.99&0.97&1.12&0.97&1.08&1.07&1.09\tabularnewline
~~h=6&1.87&0.95&0.99&1.00&\textbf{0.93}&0.93&1.23*&0.98&0.95*&1.07&1.12&1.19&1.14&1.06**\tabularnewline
~~h=8&1.58&0.98&1.02&1.03&\textbf{0.96}&0.96&1.20&1.04&0.96&1.10&0.98&1.25**&1.20**&1.06\tabularnewline
\hline
\end{tabular}
    \vspace{0.25ex}
    
     \raggedright Notes: This table report the root MSPE of the model $m$ with respect to the root MSPE the AR(4). Best forecast of the row is in bold.  Diebold-Mariano test is conducted for each model against the AR(4). "*", "**" and "***" means p-values of below 10\%, 5\% and 1\%.
\end{center}
\end{table}\end{landscape}

%% file: tables/FCST_tableM_200929.tex
\begin{landscape}\begin{table}[!tbp]
\caption{Monthly Results \label{M_table}} 
\begin{center}
\footnotesize\addtolength{\tabcolsep}{-1.5pt}
\begin{tabular}{llllllllllll}
\hline\hline
\multicolumn{1}{l}{}&\multicolumn{1}{c}{AR4}&\multicolumn{1}{c}{AO-12}&\multicolumn{1}{c}{AO-$$h$$}&\multicolumn{1}{c}{FAAR}&\multicolumn{1}{c}{RF}&\multicolumn{1}{c}{RF-MAF}&\multicolumn{1}{c}{AR+RF}&\multicolumn{1}{c}{ARRF}&\multicolumn{1}{c}{FA-ARRF}&\multicolumn{1}{c}{Tiny ARRF}&\multicolumn{1}{c}{VARRF}\tabularnewline
\hline
{\bfseries IP}&&&&&&&&&&&\tabularnewline
~~h=1&1.00&1.11*&1.14&0.96&1.03&\textbf{0.94}*&0.97&0.99&0.96&1.02&1.02\tabularnewline
~~h=3&1.02&1.17*&1.02&0.99&1.12&0.98&\textbf{0.96}&1.03&1.01&1.02&1.08\tabularnewline
~~h=9&1.01&1.04&1.03&1.06&1.02&1.06&1.02&1.04&1.10&1.09&1.03\tabularnewline
~~h=12&1.01&1.00&1.00&1.05&0.99&0.97&\textbf{0.91}&0.97&1.05&1.13&0.96\tabularnewline
~~h=24&1.00&\textbf{0.84}&0.84&1.17&0.92&0.86&0.86&0.88&0.95&1.11&0.89\tabularnewline
\hline
{\bfseries UR}&&&&&&&&&&&\tabularnewline
~~h=1&1.01&1.03&1.09&0.95&0.97&\textbf{0.87}***&0.95&0.91***&0.90**&0.98&0.94**\tabularnewline
~~h=3&1.00&1.10&1.05&0.86&1.05&\textbf{0.81}***&0.92&0.89**&0.82*&1.03&0.89***\tabularnewline
~~h=9&0.99&1.11&1.10&0.92&1.02&0.96&\textbf{0.91}&0.97&0.98&1.16*&0.97\tabularnewline
~~h=12&0.99&1.07&1.07&0.96&0.97&0.96&\textbf{0.91}&0.99&0.94&1.17&0.96\tabularnewline
~~h=24&1.02**&1.02&1.03&1.06&0.91*&0.84&\textbf{0.81}&0.91&0.97&1.28&0.87\tabularnewline
\hline
{\bfseries SPREAD}&&&&&&&&&&&\tabularnewline
~~h=1&0.99&2.88***&1.23***&1.21**&3.52***&1.07&\textbf{0.91}***&0.99&0.98&0.96&0.93**\tabularnewline
~~h=3&1.01&1.68***&1.07&1.25&1.69***&0.82**&\textbf{0.81}***&1.06&0.85**&1.00&0.88**\tabularnewline
~~h=9&1.01&1.36&1.27&1.06&0.94&0.73**&0.72**&0.70***&\textbf{0.62}***&1.07&0.67***\tabularnewline
~~h=12&1.02&1.28&1.28&1.05&0.80***&0.66***&\textbf{0.60}***&0.68***&0.65***&1.07&0.64***\tabularnewline
~~h=24&1.03&1.34*&1.34*&0.96&0.80*&0.70*&0.71*&0.69**&\textbf{0.63}***&0.90&0.70**\tabularnewline
\hline
{\bfseries INF}&&&&&&&&&&&\tabularnewline
~~h=1&1.02&1.11*&1.18*&0.99&1.07&1.06*&1.01&0.95&0.96&0.95&\textbf{0.93}**\tabularnewline
~~h=3&1.04&1.02&1.24*&1.04&0.93&\textbf{0.88}&1.05&0.90&0.88&0.90&0.88\tabularnewline
~~h=9&1.07&0.92&1.01&1.16&0.86&0.78&1.15*&\textbf{0.72}&0.82&0.73&0.76\tabularnewline
~~h=12&1.09*&0.91&0.91&1.21&0.88&0.79&1.15*&0.73&0.67&\textbf{0.67}*&0.70\tabularnewline
~~h=24&1.04&0.90**&0.86**&1.35&1.00&1.12&1.12&0.71&0.69&\textbf{0.55}**&0.73\tabularnewline
\hline
{\bfseries HOUST}&&&&&&&&&&&\tabularnewline
~~h=1&\textbf{1.00}&1.10**&1.35***&1.07&1.08**&1.02&1.00&1.01&1.02&1.02&1.01\tabularnewline
~~h=3&\textbf{0.96}**&1.06&1.34***&1.15&1.03&1.07&1.03&1.04&1.03&1.01&1.04\tabularnewline
~~h=9&\textbf{0.98}&1.05&1.12&1.35&0.98&1.02&1.01&1.02&1.14&1.03&1.03\tabularnewline
~~h=12&0.98&1.05&1.05&1.32&\textbf{0.95}&1.00&1.01&1.00&1.12&1.11&1.03\tabularnewline
~~h=24&0.95&1.09&1.07&1.17&\textbf{0.87}&0.94&0.95&1.00&1.15&1.23&1.06\tabularnewline
\hline
\end{tabular}
    \vspace{0.25ex}
    
     \raggedright Notes: This table report the root MSPE of the model $m$ with respect to the root MSPE the AR(4). Best forecast of the row is in bold.  Diebold-Mariano test is for each model against the AR(4). "*", "**" and "***" means p-values of below 10\%, 5\% and 1\%. "AO-$i$" means $i$-months moving average forecasts à la \cite{atkeson2001phillips}. 
\end{center}
\end{table}\end{landscape}